\documentclass[floatfix,twocolumn,aps,pra,superscriptaddress,showpacs]{revtex4-1}
\usepackage{graphicx}
\usepackage{amsmath}
\usepackage{amssymb}
\usepackage{soul}

\usepackage[usenames]{color}

\newcommand{\ket}[1]{|#1\rangle}

\newcommand{\ev}[1]{\langle #1\rangle}

\newcommand{\vcre}[2]{\hat{\mathbf{#1}}^{\dagger}_{#2}}
\newcommand{\vdes}[2]{\hat{\mathbf{#1}}_{#2}}

\newcommand{\op}[1]{ \hat{#1} }
\newcommand{\cre}[2]{\hat{#1}^{\dagger}_{#2}}
\newcommand{\des}[2]{\hat{#1}_{#2}}

\renewcommand{\vec}[1]{\mathbf{#1}}

\newcommand{\beq}{\begin{equation}}
\newcommand{\eeq}{\end{equation}}

\begin{document}
\unitlength = 1mm

\title{Optical signatures of antiferromagnetic ordering of fermionic atoms in an optical lattice}
\author{Francisco \surname{Cordobes Aguilar}}
\email{francisco.cordobesaguilar.2009@live.rhul.ac.uk}
\affiliation{Department of Physics, Royal Holloway, University of London, Egham, Surrey TW20 0EX, United Kingdom}
\affiliation{Mathematical Sciences, University of Southampton,
Southampton, SO17 1BJ, United Kingdom}
\author{Andrew \surname{F. Ho}}
\affiliation{Department of Physics, Royal Holloway, University of London, Egham, Surrey TW20 0EX, United Kingdom}
\author{Janne \surname{Ruostekoski}}
\affiliation{Mathematical Sciences, University of Southampton,
Southampton, SO17 1BJ, United Kingdom}

\date{\today}

\date{\today}
\begin{abstract}
We show how off-resonant light scattering can provide quantitative information on antiferromagnetic ordering of a two-species fermionic atomic gas in a tightly-confined two-dimensional optical lattice. We analyze the emerging magnetic ordering of atoms in the mean-field and in random phase approximations and show how the many-body static and dynamic correlations, evaluated in the standard Feynman-Dyson perturbation series, can be detected in the scattered light signal. The staggered magnetization reveals itself in the magnetic Bragg peaks of the individual spin components. These magnetic peaks, however, can be considerably suppressed in the absence of a true long-range antiferromagnetic order. The light scattered outside the diffraction orders can be collected by a lens a with highly improved signal-to-shot-noise ratio when the diffraction maxima are blocked.
The collective and single-particle excitations are identified in the spectrum of the scattered light. We find that the spin-conserving and spin-exchanging atomic transitions convey information on density, longitudinal spin, and transverse spin correlations. The different correlations and scattering processes exhibit characteristic angular distribution profiles for the scattered light and, e.g., the diagnostic signal of transverse spin correlations could be separated from the signal by the scattering direction, frequency, or polarization.
We also analyze the detection accuracy by estimating the number of required measurements, constrained by the heating rate that is determined by inelastic light scattering events.
The imaging technique could be extended to the two-species fermionic states
in other regions of the phase diagram where the ground state properties
are still not fully understood.
\end{abstract}
%03.75.Ss	Degenerate Fermi gases
%42.50.Ct Quantum description of interaction of light and matter; related experiments
%37.10.Jk	Atoms in optical lattices
%Ultracold degenerate Fermi gases, 67.85.Lm
% Lattice fermion models, 71.10.Fd
%%%%Strongly correlated electron systems, 71.27.+a
%37.10.Vz	Mechanical effects of light on atoms, molecules, and ions

\maketitle

\section{Introduction}

The two-species fermionic Hubbard model is one of the most studied models with strong correlation
effects in condensed matter physics, particularly since Anderson proposed in 1987~\cite{Anderson1987}
that the Hubbard model (or its strong coupling limit, the t-J model) is the minimum model to describe the physics
of the high temperature superconductors~\cite{RevModPhys.78.17}. It is well understood that at half-filling, i.e., at an average of
one spin-1/2 fermion per site, the Hubbard model has the ground state which is a Mott insulator with
antiferromagnetic (AFM) ordering. Away from half-filling, quantitatively accurate predictions
are scarce, but it is generally believed that a $d$-wave superconductor can be present in some regions of the phase diagram, hence the relevance of the Hubbard model for the high temperature superconductors.

With the rapid advance in cooling and trapping technologies of neutral atoms in optical lattices,
there is the exciting prospect that the phase diagram of the fermionic Hubbard model can be explored
via quantum simulation (rather than numerical simulation)
in an optical lattice set up with ultracold two-species fermionic atomic gas~\cite{ISI:000302557600011,ISI:000259377800004,PhysRevA.79.033620}. Such a system has been cooled down in a laboratory to
the Mott insulator regime~\cite{ISI:000259090800042,U.Schneider12052008} and recent experiments have demonstrated evidence of AFM correlations~\cite{Greif14062013,HuletDamop}, owing to exchange coupling between the atomic spin states. Magnetic ordering has also been observed in doublon formation in tilted lattice systems~\cite{spinchain,PhysRevLett.111.053003}.
Accurate diagnostic tools for ultracold atoms in lattices form key elements in emulating strongly-correlated physics of condensed-matter systems. Currently, advanced imaging provides a microscopic scanning technology of atoms with a single-site resolution~\cite{Bloch2DMicroscope,2009Natur.462...74B,PhysRevLett.102.053001,Bakr30072010,2011Natur.471..319W,Endres14102011,2013arXiv1303.5652E,2007NatPh...3..556N}
in which case each atom may resonantly scatter thousands of photons while being detected. The analogy between x-ray (or neutron) diffraction of crystalline structure in solids and off-resonant light scattering from an ultracold atomic gas in an optical lattice provides an alternative route. The periodic crystalline order leads to constructive interference and the emergence of sharp diffraction maxima, or Bragg peaks, revealing the underlying lattice structure. Importantly, diffuse background of scattered light in between the diffraction maxima can convey information on {\em fluctuations} of atomic positions. The far-field off-resonant imaging therefore provides a powerful probe of strongly-correlated ultracold atom systems in optical lattices, being sensitive, e.g., to correlations, excitations, and temperature.

Spontaneously scattered off-resonant light was first proposed as a diagnostic tool for correlations of ultracold atoms in an optical lattice in Ref.~\cite{PhysRevLett.91.150404}. This study considered a two-species gas where the hopping element of the atoms between adjacent sites acquired an artificially constructed phase factor from the Peierls substitution. It was then shown that the resulting topological properties of edge states, which originate from nontrivial spin correlations, can be mapped onto fluctuations of scattered light and directly detected~\cite{PhysRevLett.91.150404,PhysRevA.77.013603}. Several other studies have addressed signatures of atom statistics in an optical lattice from the scattered light with~\cite{PhysRevLett.98.100402,PhysRevA.76.053618} and without~\cite{ruostekoski:170404,PhysRevA.80.043404,PhysRevA.81.013404,PhysRevA.82.033434, PhysRevA.84.033637,PhysRevA.81.063618,PhysRevA.84.053608,PhysRevA.83.051604,PhysRevA.86.023607,PhysRevA.84.043825} an additional optical cavity.

A two-species fermionic Hubbard model exhibits a Mott insulator state where the fluctuations of the total on-site atom number are suppressed~\cite{ISI:000259090800042,U.Schneider12052008}.
At sufficiently low temperatures the effective exchange coupling can generate magnetic ordering of the atomic spins within the Mott state. At half-filling with equal spin populations this leads to alternating checkerboard pattern of atom densities of individual spin components. The resulting doubling of periodicity of atom densities leads to the emergence of additional magnetic Bragg peaks of scattered light that have been proposed as a detection mechanism of AFM ordering~\cite{PhysRevA.81.013415}. The extra Bragg peaks in the optical signal of a two-species system were experimentally observed using an artificially constructed stripe pattern of atomic densities~\cite{PhysRevLett.106.215301}.

Here we analyze the far-field diffraction pattern for off-resonantly scattered light from a two-species fermionic atomic gas in an optical lattice. We consider a tightly-confined two-dimensional (2D) Hubbard model at half-filling with equal spin populations. We show how optical diagnostics can provide quantitative information on properties of strongly correlated states in a lattice. The time-ordered correlation functions that can be calculated using the Feynman-Dyson perturbation series are mapped onto the fluctuations of the scattered light and detected in the optical signal. We evaluate the relevant correlation functions in the mean-field theory (MFT) and in the random phase approximation (RPA) for the AFM ordered state. MFT provides information on single-particle excitations but fails to capture the effect of quantum fluctuations and collective excitations that are approximately incorporated in RPA.
In the RPA calculations we follow the approach of Ref.~\cite{PhysRevB.39.11663}.

The scattered light intensity may be separated into elastic and inelastic components. In the elastic scattering process the atom scatters back to its original state. The elastic part produces a diffraction pattern from a nonfluctuating atom density, analogous to that of $N_s\times N_s$ diffracting apertures when the lattice has $N_s\times N_s$ sites. The overall envelope of the diffraction pattern is determined by the lattice site wave function of the atoms. The magnetic ordering specifies the relative atom densities of the two spin components and appears as additional Bragg peaks. The measurement of the magnetic Bragg peaks is possible,
provided that the lattice spacing $a>\lambda/\sqrt{2}$, where $\lambda$ denotes the probe wavelength. Our analysis confirms that detecting the magnetic peaks by scattering light from a single spin component alone constitutes the most accurate observable for the staggered magnetization of the AFM order. However, the absence of true long-range AFM order can significantly suppress the magnetic Bragg peaks. We demonstrate this by considering a phenomenological Ising model for the staggered magnetization when the system exhibits a finite correlation length.

The inelastic scattering processes, on the other hand, are those in which an atom scatters between two {\em different} quasimomentum states. The inelastically scattered light conveys information on correlations between the atoms and results in diffuse scattering of light outside the diffraction orders, generating  fluctuations of the diffraction pattern.
We collect the light in the near-forward direction with a lens and block the diffraction maxima, so that the amount of elastically scattered light entering the detector is suppressed~\cite{ruostekoski:170404}. This method is then extended to detection of light in the direction approximately perpendicular to the propagation direction of the incident field. In Ref.~\cite{ruostekoski:170404} detecting the near-forward light was shown to provide an experimentally feasible technique for measurements of temperature of fermionic atoms in a lattice. We find that in an AFM ordered two-species state the scattered light in the near-forward direction is not only sensitive to the temperature of the atoms but provides also a suitable probe of density and longitudinal spin correlations (from spin conserving atomic transitions). The scattering processes in which the atomic spin state changes, on the other hand, are prominent in the scattered light perpendicular to the light propagation direction  and can be employed in detection of transverse spin correlations. Furthermore, we estimate the detection accuracy of the magnetic ordering in different measurement configurations. This is done by calculating the number of required experimental realizations of the lattice system to detect the order parameter above the shot noise of light at a desired accuracy. In each experimental realization the scattered light heats up the atomic sample and the total number of inelastic scattering events is constrained to be a small fraction of the total atom number. We find that a strong lattice and trap confinement is beneficial for the measurements in suppressing scattering of atoms to higher energy bands as compared with the inelastic lower energy band scattering.

We also calculate the spectrum of the scattered light and show how it reveals the excitation of the system that could be measured using optical heterodyne techniques~\cite{Hoffges1997170}. The differences between the MFT and RPA treatments are especially prominent in the spectrum of transverse spin correlations: The low-energy collective mode excitations manifest themselves as a well-separated peak (at sufficiently strong interaction energy) from the gapped single-particle excitations.

The optical setup that blocks the Bragg diffraction maxima in the measured signal can considerably reduce the scattered light that is insensitive to correlations, therefore improving the signal-to-noise efficiency. Similar separation of less important part of the signal, for instance, in atom shot-noise correlation measurements \cite{2005Natur.434..481F,2006Natur.444..733R} would be challenging.
We estimate the detection accuracy of the AFM ordering in the scattered light intensity by calculating the number of experiments needed to achieve a given measurement accuracy of the magnetic order parameter, when the heating rate of the atoms limits the total number of possible scattering events.

The proposed technique differs from the simple angle-resolved imaging of the diffraction peaks, since we collect  the inelastically scattered light with a lens that has a relatively large numerical aperture (NA). By analyzing the spectrum of the scattered light using the realistic optics setup, we show that the essential features of the excitation spectrum can be captured even when the light is collected over a range of scattering angles. A large lens provides a stronger signal even when the number of inelastically scattered photons remains a small fraction of the total atom number. In less demanding detection scenarios (e.g., in large lattices and at high temperatures) even a single experimental realization of the lattice system can then be sufficient to determine, e.g., the approximate temperature of the atoms in the lattice. This contrasts with the single-site microscopy~\cite{Bloch2DMicroscope,Physics.4.41}
in which case typically several thousands of photons are scattered from each atom and the light is also simultaneously used to cool down the atoms. The microscopy also requires a scanning of the lattice sites that in very large lattice systems may become less practical. The off-resonant light scattering, on the other hand, becomes more efficient a method when the size of the system increases. Other advantages of the off-resonant imaging include the possibility for spectral measurements of excitations and the access to correlation functions that include combinations of different spin states via spin-exchanging scattering processes. Finally, the diffractive far-field imaging does not need to be limited to probing the atoms only by light, but also matter-wave probes are possible~\cite{ISI:000305970400014}.

The remainder of the paper is organized as follows:
Section~\ref{sec:key_results} presents a summary of the key results.
We then start in Sec.~\ref{sec:optical_lattices} by introducing the basic formalism of the lattice system. We continue in Sec.~\ref{sec:hubbard_model}, where the MFT and RPA results for the lattice system are presented. The scattered light as a diagnostic tool is introduced in Sec.~\ref{sec:quantum_optics}.
In Sec.~\ref{sec:optical_signatures} we present results for the scattered light intensity for $^{40}\text{K}$ atoms. The specific experimental setups for the optical detection and the estimates for the measurement accuracy are considered in Sec.~\ref{sec:detection}.
The scattered spectrum is studied in Sec.~\ref{sec:scattered_spectrum} and some concluding remarks are made
in Sec.~\ref{sec:conclusions}. Finally, a diagrammatic description of the RPA susceptibilities is presented in Appendix~\ref{App:RPA-Feynman} and the finite temperature MFT susceptibilities are given in Appendix~\ref{app:mfa_susceptibilities}.

\section{Summary of Key Results}
\label{sec:key_results}
In this Section we briefly highlight the main findings of the paper. We study a two-species fermionic atomic gas trapped in a tightly-confined 2D optical lattice [Eq.~\eqref{eq:optical_lattice_potential}].
We assume equal spin populations and that on average, there is one atom per site in the lattice. The atoms are probed by incoming laser light propagating perpendicular to the lattice. The scattered light is collected  by a lens positioned at various angles relative to the propagation direction, see Fig.~\ref{fig:setup}.

The Hubbard Hamiltonian for the atoms reads (see Sec.~\ref{sec:optical_lattices},  Eq.~\eqref{eq:hubbard-direct-space})
\begin{align*}
\notag
\mathcal{H}=&
        -J\sum\limits_{
                        \langle\vec{j}_1\vec{j}_2\rangle,g
                        }
                {\left(
                                \cre{c}{\vec{j}_1g}\des{c}{\vec{j}_2g} +\text{H.c.}
                \right)}
        \\&+U\sum\limits_{\vec{j}}{\op{n}_{\vec{j}\uparrow}\op{n}_{\vec{j}\downarrow}}
        -\mu\sum\limits_{\vec{j}g}{\op{n}_{\vec{j}g}} \; ,
\end{align*}
where  $\op{n}_{\vec{j}g}=\cre{c}{\vec{j}g}\des{c}{\vec{j}g}$ and $\cre{c}{\vec{j}g}$($\des{c}{\vec{j}g}$) is the fermionic creation (annihilation) operator
in the lowest band for hyperfine state $g=\uparrow, \downarrow$ at site $\vec{j}=(j_x,j_y)$ with
$j_i=1,\ldots,N_s$ for $i=x,y$. Here
$J$ denotes the hopping amplitude [Eq.~\eqref{eq:hopping_amplitude}], $U$ the on-site interaction strength [Eq.~\eqref{eq:onsite_interaction}] and $\mu$ is the chemical potential.

We study the Mott insulator state of the Hubbard model
where fluctuations of total on-site atom number are suppressed.
We focus on the AFM ground state,
 characterized by a checkerboard pattern  of ordering (period doubling).
The associated AFM order parameter is the staggered magnetization $m$, defined as
[Eq.~\eqref{eq:order_parameter}]
\begin{equation*}
m = \frac{1}{2} \: e^{i \: \vec{Q}\cdot \vec{r}_{\vec j}} \; \ev{\op{S^z_{\vec{j}}}} \;,
\end{equation*}
where
$\op{S^z_{j}}=(\cre{c}{\vec{j}\uparrow}\des{c}{\vec{j}\uparrow}- \cre{c}{\vec{j}\downarrow}\des{c}{\vec{j}\downarrow})$ is the real space spin operator.
The magnetic ordering wavevector is $\vec{Q}=(\pi/a,\pi/a)$ ($a$ denotes the lattice spacing) and $\vec{r}_{\vec j}$ is the coordinate of the center of the lattice site $\vec j$. We employ MFT in Sec.~\ref{sec:mean_field_hamiltonian} to diagonalize the Hamiltonian around the AFM order parameter $m$ using the Bogoliubov transformation. This leads to the gap equation and description of single-particle excitations of the system. However, MFT  does not
include quantum fluctuations (specifically, spin waves).
Crucially, such  quantum fluctuations modify {\em qualitatively}
the correlation functions of spin and density operators, which then leads to large changes to
the optical response  (see Secs.~\ref{sec:optical_signatures} and~\ref{sec:scattered_spectrum}).
Hence, in  Sec.~\ref{sec:RPA-Chi} and Appendix~\ref{App:RPA-Feynman}, we describe the calculation of the correlation functions by a partial summation of the Feynman-Dyson perturbation series in RPA that incorporates the correct spin wave physics at strong coupling.

We show in Sec.~\ref{sec:intensity} how quantum statistical correlations of the atoms can be mapped onto fluctuations of the scattered light.
Measurements on the scattered light therefore convey information about the correlated phases of the ultracold atoms in the lattice. The formalism describing the relationship between optical signal (the intensity and spectrum
of the scattered light) and the atomic correlations is described in Sec.~\ref{sec:quantum_optics}:
see Eqs.~\eqref{eq:intensity-elastic+inelastic}-\eqref{eq:intensity-inelastic-general} for the scattered intensity, and
Eqs.~\eqref{eq:spectrum-elastic+inelastic}-\eqref{eq:spectrum-inelastic-general} for the scattered spectrum.

In the calculations of the optical response we specialize to a level scheme and transitions of $^{40}$K, see Fig.~\ref{fig:sigmatransition} and Sec.~\ref{sec:40K:intensity}, where
the two electronic ground states (labeled as spin up and down) are
$\ket{\uparrow}=\ket{4S_{1/2},F_g=9/2,m_F=-7/2}$ and $\ket{\downarrow}=\ket{4S_{1/2},F_g=9/2,m_F=-9/2}$ [Eq.~\eqref{eq:states:g_levels}]. These are coupled to electronically excited states $\ket{1}=\ket{4P_{3/2},F_e=11/2,m_F=-11/2}$ and $\ket{2}=\ket{4P_{3/2},F_e=11/2,m_F=-9/2}$ [Eq.~\eqref{eq:states:e_levels}] by the $\sigma^-$ polarized incident light. One of the states, $\ket{\downarrow}$, undergoes a cycling transition, but the atoms in $\ket{\uparrow}$ may either scatter back to $\ket{\uparrow}$ (spin-conserving transition) or to $\ket{\downarrow}$ (spin-exchanging transition).

The scattered light intensity contains contributions from the {\em elastic} and {\em inelastic}
scattering events.  The elastic part
generates the diffraction pattern of the atomic lattice structure. If the detected signal cannot distinguish the two spin
components, the light provides almost no information about the AFM order. But with {\it spin-specific imaging, the emerging AFM order and the period doubling can be identified as additional Bragg peaks}.

Specifically, we find that the elastic component of the scattered light intensity is  given by
\begin{align}
\label{eq:intensity:elastic_Sum}
\frac{I_{\text{e}}(\Delta\vec{k})}{\alpha_{\Delta\vec{k}}B}= &\left(\sqrt{{\sf M}_{\downarrow\downarrow}^{\downarrow \downarrow}} \ev{\op{\rho}_{\bar{\Delta \vec{k}} \downarrow}}
+\sqrt{{\sf M}_{\uparrow\uparrow}^{\uparrow \uparrow}} \ev{\op{\rho}_{\bar{\Delta \vec{k}} \uparrow}}\right)^2\,.
\end{align}
In this expression the dependence on level structure, polarization, and scattering direction are coded in $ {\sf M}^{g_3g_4}_{g_2g_1}$ [Eq.~\eqref{eq:m_tensor}], which contains
 the dipole matrix element [see Eqs.~\eqref{eq:lambda} and~\eqref{eq:dipole_matrix_element}].
$\Delta\vec{k}$ is the change in the momentum of the scattered photons, Eq.~\eqref{eq:Deltak-def},
while $\bar{\Delta \vec{k}}$ is the same momentum projected into the lattice plane. Here the Debye-Waller factor $\alpha_{\Delta\vec{k}}$ depends on the lattice site wave function and determines the overall envelope of the diffraction pattern, see Eqs.~\eqref{eq:B:definition} and~\eqref{eq:alpha_factor}.
The density operator in momentum space for species $g$ is
\begin{equation}
 \op{\rho}_{\vec{q} g} = \sum_{{\vec j}} e^{i \vec{q} \cdot \vec{r}_{\vec{j}}} \: \op{n}_{\vec{j}g}=
\sum_{\vec{k}}\cre{c}{\vec{k}+\vec{q},g}\des{c}{\vec{k},g}  \;,
\end{equation}
where  $\des{c}{\vec{k},g}$ is the annihilation operator at momentum $\vec{k}$  for hyperfine state $g$.

For the AFM state this can be evaluated to be
\begin{equation}
\label{eq:rho-AFM_Sum}
\ev{ \op{\rho}_{\bar{\Delta \vec{k}} g}} = \mathfrak{u}_{\bar{\Delta\vec{k}}} f_g  +
\mathfrak{u}_{\bar{\Delta\vec{k}}+\vec{Q}} \: m  \: \eta(g) \:,
\end{equation}
where $f_g=1/2$ is the atomic filling factor  of species $g$ at half-filling and
$\eta(\uparrow)=1, \; \eta(\downarrow)=-1$  [Eq.~\eqref{eq:etag}].
Here $|\mathfrak{u}_{\bar{\Delta\vec{k}}}|^2$ [Eq.~\eqref{eq:uofk}] generates
the diffraction pattern from the periodic atom density [Eq.~\eqref{eq:UDeltaK}].
The second term of Eq.~\eqref{eq:rho-AFM_Sum}, when substituted into Eq.~\eqref{eq:intensity:elastic_Sum},
 is responsible for these new peaks due to the period doubling in the AFM state.
This effect was first analyzed in Ref.~\cite{PhysRevA.81.013415}. In the absence of true long-range AFM order the staggered magnetization, however, may vary in space. We study the effects of short-range AFM order on the optical response by introducing a phenomenological Ising model that demonstrates a significant suppression of the magnetic Bragg peaks when the correlation length is not much larger than the size of the lattice.

While elastically scattered light gives information about the AFM ordering of the system,
the inelastic component directly probes the quantum atomic correlations of the spin operators
(we follow the notation convention from Ref.~\cite{PhysRevB.39.11663})
\begin{align} \label{eq:spinSzop}
\op{S}^z_{\vec{q}} =&\sum_{\vec{k}} \left(
  \cre{c}{\vec{k}+\vec{q},\uparrow}  \des{c}{\vec{k},\uparrow} \; -\cre{c}{\vec{k}+\vec{q},\downarrow}  \des{c}{\vec{k},\downarrow} \right)\; , %\\
\end{align}
\begin{align}
\label{eq:spinSPlusop}
\op{S}^+_{\vec{q}} =&
\sqrt{2}\sum_{\vec{k}} \cre{c}{\vec{k}+\vec{q}\uparrow}\des{c}{\vec{k}\downarrow} \;, \qquad
\op{S}^-_{\vec{q}} =
\left( \op{S}^+_{-\vec{q}} \right)^{\dagger}
\end{align}
or the total density operator
\begin{equation}
\label{eq:chargeop}
 \op{\rho}_{\vec{q} } = \sum_g \op{\rho}_{\vec{q}  g} =
\sum_{\vec{k},g}\cre{c}{\vec{k}+\vec{q},g}\des{c}{\vec{k},g}  \;.
\end{equation}

We explicitly show how the time-ordered many-body correlation functions obtained in RPA [Eqs.~\eqref{eq:chi:rhorho:RPA}-\eqref{eq:chi:+-:Q:RPA}] are mapped onto the experimental observables
of the optical signal [see, e.g., Eq.~\eqref{eq:chi-to-structurefactorT0}].
We then find that the inelastically scattered light intensity for the two-component $^{40}$K gas is given by Eq.~\eqref{eq:intensity:explicit},
\begin{widetext}
\begin{align}
\frac{I_i(\Delta\vec{k})}{\alpha_{\Delta\vec{k}}B} =&
\frac{1}{4}\left({\sf M}_{\downarrow\downarrow}^{\downarrow \downarrow}
+{\sf M}_{\uparrow\uparrow}^{\uparrow \uparrow}\right)
 \sum\limits_{\vec{q}\neq0}^{\text{RBZ}}
\pmb{u}^\dagger_{\bar{\Delta\vec{k}}-\vec{q}}\:
\notag
\left[  \pmb{\mathcal{S}}^{\rho\rho}(\vec{q})+\pmb{\mathcal{S}}^{zz}(\vec{q})  \right]\pmb{u}_{\bar{\Delta\vec{k}}-\vec{q}}
\\
&
 +\frac{1}{4}\left( {\sf M}_{\uparrow\uparrow}^{\downarrow \downarrow}+{\sf M}_{\downarrow\downarrow}^{\uparrow \uparrow} \right)
 \sum\limits_{\vec{q}\neq0}^{\text{RBZ}}
 \pmb{u}^\dagger_{\bar{\Delta\vec{k}}-\vec{q}}\:
\left[ \pmb{\mathcal{S}}^{\rho\rho}(\vec{q})- \pmb{\mathcal{S}}^{zz}(\vec{q}) \right]\pmb{u}_{\bar{\Delta\vec{k}}-\vec{q}}
 +\frac{1}{2} {\sf M}_{\downarrow \uparrow}^{\downarrow \uparrow} \:\sum\limits_{\vec{q}\neq0}^{\text{RBZ}}
\pmb{u}^\dagger_{\bar{\Delta\vec{k}}-\vec{q}}\: \:\pmb{\mathcal{S}}^{+-}(\vec{q}) \:\pmb{u}_{\bar{\Delta\vec{k}}-\vec{q}} \;,
\label{eq:inelastic-intensity_sum}
\end{align}
\end{widetext}
where the equal time $2\times2$ matrix response function [cf. Eqs.~\eqref{eq:matrix:mathcalS} and \eqref{eq:static-response+-}]
\begin{align}
\pmb{\mathcal{S}}^{ij}  (\vec{q})  =
 \begin{pmatrix}
\ev{ \op{O}^i_{\vec{q}}\op{O}^j_{-\vec{q}} }_c & \ev{ \op{O}^i_{\vec{q}}\op{O}^j_{-\vec{q}-\vec{Q}} }_c \\
\ev{ \op{O}^i_{\vec{q}+\vec{Q}}\op{O}^j_{-\vec{q}} }_c &
\ev{ \op{O}^i_{\vec{q}+\vec{Q}}\op{O}^j_{-\vec{q}-\vec{Q}} }_c
\end{pmatrix} \, .
\end{align}
$\op{O}^i_{\vec{q}}$ can be any of the operators in
Eqs.~\eqref{eq:spinSzop},~\eqref{eq:spinSPlusop}  and~\eqref{eq:chargeop}.
The subscript $c$ in the expectation value denotes a connected correlation function for which $\vec{q}\neq0$. We have also introduced the two-component vector [Eq.~\eqref{eq:uofk:vector}]
$\pmb{u}^\dagger_{\vec{k}} = (\mathfrak{u}_{\vec{k}}^* \: , \: \mathfrak{u}_{\vec{k}+\vec{Q}}^* ) $.
The matrix notation reflects the fact that in the AFM ordered state with period doubling, the original
Brillouin Zone is split up into the  RBZ, and the zone outside the RBZ which is connected
by $\vec{Q}$ to  the RBZ [Fig.~\ref{fig:brillouin_zone}].

In Eq.~\eqref{eq:inelastic-intensity_sum}, the scattering contributions in which spin is conserved are
proportional to density and longitudinal spin correlations. The spin-exchanging transitions
(see Fig.~\ref{fig:sigmatransition}) generate the term depending on the transverse spin correlations.
The two processes exhibit very different angular distribution of the scattered light, as shown in
Eq.~\eqref{eq:inelastic:angular:40K}, and can also be separated in frequency.
Thus, another key result of this paper is that {\it specific quantum correlations can be separated in the detected signal}.

For the detection of inelastically scattered light we  consider two different experimental configurations:
One has the scattered light  collected by a lens in the near-forward direction
with the central diffraction peak  blocked, see Fig.~\ref{fig:experimental_setups}(b), and the other
one has light  collected in
the perpendicular direction, Fig.~\ref{fig:experimental_setups}(c).
The full intensity and spectrum are then evaluated numerically
by integrating over the scattering directions (the momenta) $\Delta\vec{k}$ collected by the lens.

We estimate the number of experimental measurements needed to obtain a given relative
accuracy in the determination of the magnetization or temperature~\cite{ruostekoski:170404}.
This minimum number of measurements is determined by two factors. On one hand, photon shot noise
dictates a minimum difference in the number of photons detected for two close AFM magnetization
values. On the other hand, the total number of photons collected is set by the experimental duration.
This duration is constrained by the  heating rate of the system due to  inelastic scattering of atoms by light. In order to satisfy these two
constraints,  a minimum number of experimental realizations has to be achieved, see
Sec.~\ref{sec:measurement_accuracy} and Figs.~\ref{tau:signal:composite:lensQpeak}-\ref{tau:signal:numap05:lens:perpendicular:s_plusminus_only}.

Armed with these measurement accuracy data, we can propose near-optimal parameters for the experimental configurations mentioned. The best magnetization measurement accuracy
 can be achieved with spin-specific  detection where the photons are collected
near the perpendicular direction or around the direction of the magnetic Bragg peak, while the temperature of the atoms can be measured in the near-forward direction [Fig.~\ref{fig:experimental_setups}(b)].
In the perpendicular direction [Fig.~\ref{fig:experimental_setups}(c) and~\ref{tau:signal:numap05:lens:perpendicular:s_plusminus_only}], it is preferable to detect
only the  light scattered from  spin-exchanging transitions. The light scattered by
spin-conserving or spin-exchanging transitions can be separated owing to the different frequency
and polarization (Sec.~\ref{sec:projected_scattered_light}) of the scattered photons.

We calculate the spectrum of the scattered light in Sec.~\ref{sec:scattered_spectrum}. We obtain an
expression similar to the scattered light intensity, with the static correlations replaced by dynamic ones
$\pmb{\mathcal{S}}^{ij}(\vec{q},\omega)$
[Eqs.~\eqref{eq:spectrum:inelastic} and~\eqref{eq:chi-to-structurefactorT0} and Sec.~\ref{sec:scattered_spectrum}]. The spectrum is calculated for a realistic optical setup where the light is collected using a finite-aperture lens, Fig.~\ref{fig:spectrum:rpa:lenses}.
We find that at moderately large to large $U/J$, the collective excitations
generate a sharp peak in the low energy part of the spectrum, which is separated by an energy gap of order $U$ from a more broad feature generated by single particle excitations. The location and
width of these features can give a quantitative measure of the AFM order.

\section{Optical lattices and the Hubbard model}
\label{sec:optical_lattices}

Optical lattices, generated by a standing wave (periodic) laser potential, provide  ideal tunable systems where almost  every system parameter can be changed independently. In typical experimental situations, the trapping potential for the atoms is a superposition of the lattice and an external, approximately harmonic, trap. For the case of sufficiently weak external potential, the harmonic trap may be ignored. Experimentally, it is also possible to produce entirely homogeneous lattices; a first step towards this has recently been demonstrated for a Bose-Einstein condensate in a uniform trap~\cite{PhysRevLett.110.200406}. Here we will ignore, for simplicity, any modulations of the uniform lattice potential. The effects of the additional harmonic potential, e.g., in the context of the Mott insulator states has already been addressed by several studies~\cite{ISI:000259090800042,U.Schneider12052008}.
For a typical 2D square lattice, the periodic potential then reads (${\vec r} = (x,y,z)$)
\begin{equation}
\label{eq:optical_lattice_potential}
V({\vec r})=s_xE_R \sin^2\left(\frac{\pi x}{a}\right)+s_yE_R \sin^2\left(\frac{\pi y}{a}\right)+\frac12 m \omega_z^2 z^2\,.
\end{equation}
In our case we choose the lattice depth $s=s_x=s_y$, similar to the 2D lattice experiments with a disk-like lattice~\cite{Bloch2DMicroscope}.
Here the frequency of the harmonic confinement in the $z$ direction is denoted by $\omega_z$ and the lattice light recoil energy, $E_{R}$, is defined by
\begin{equation}
\label{eq:E_R}
 E_{R}={(\hbar k_{\text{l}})^2\over 2m} \;
\end{equation}
We define an effective wavenumber for the optical lattice in terms of the lattice spacing $a$ by
\beq
\label{eq:lattice_spacing}
k_{\text{l}} = {\pi\over a}\,.
\eeq
When the effective wavenumber coincides with the laser wavenumber, we have $k_{\text{l}}=2\pi/\lambda_{\rm l}$ and $a=\lambda_{\text{l}}/2$, where $\lambda_{\text{l}}$ denotes the wavelength of the laser beam generating the lattice. In the case of accordion lattices the lattice spacing is manipulated by optical components and can be considerably increased~\cite{Li:08,Williams:08,PhysRevA.82.021604}.

The Hamiltonian that describes the atoms in the optical lattice can be written in terms of second quantization operators.
These operators are related to the field operators $\op{\Psi}_g(\vec{r})$ for the hyperfine state $g$,
 via the expansion in
the complete set of Wannier functions $w_{n,\vec{j}}(\vec{r})=w_n(\vec{r}-\vec{r}_{\vec j})$
 representing a localized state at position $\vec{r}_{\vec j}$ and band $n$~\cite{Ashcroft-Mermin}
\begin{equation}
\label{eq:wannier_function}
\op{\Psi}_g(\vec{r})=\sum_{n,\vec{j}}w_{n,\vec{j}}(\vec{r}) \, \op{c}_{n\vec{j}g} \; ,
\end{equation}
where $\des{c}{n\vec{j}g}$ is the fermionic annihilation operator for hyperfine state $g$ at site $\vec{j}=(j_x,j_y)$. The spatial variation of the density along the lattice is represented by the modulation of the expectation values $\langle \op{c}^\dagger_{n\vec{j}g}\op{c}_{n\vec{j}g}\rangle$. In this study we assumed that the potential in the $xy$ plane can be described by the uniform lattice potential alone and any deviations, e.g., due to harmonic trap may be neglected. The spatial variation of Eq.~\eqref{eq:wannier_function} between different lattice sites is therefore entirely encapsulated in the phase factors of $\op{c}_{n\vec{j}g}$.

We consider only the regime where $U,T$ are much smaller than the bandgap, so that the higher bands
in the initial ground state are not populated. Hence, only the
$n=0$ state in Eq.~\eqref{eq:wannier_function} is included in the analysis of the ground-state properties, and we drop the band index  for now.
(Later, we will consider the effect of light scattering of
atoms into higher bands in Sec.~\ref{subsec:higherband}.)
In order to describe the lattice system we introduce the one-band  Hubbard Hamiltonian~\cite{PhysRevLett.81.3108}
\begin{align}
\label{eq:hubbard-direct-space}
\notag
\mathcal{H}=&
        -J\sum\limits_{
                        \langle\vec{j}_1\vec{j}_2\rangle,g
                        }
                {\left(
                                \cre{c}{\vec{j}_1g}\des{c}{\vec{j}_2g} +\text{H.c.}
                \right)}
        \\&+U\sum\limits_{\vec{j}}{\op{n}_{\vec{j}\uparrow}\op{n}_{\vec{j}\downarrow}}
        -\mu\sum\limits_{\vec{j}g}{\op{n}_{\vec{j}g}} \;.
\end{align}
Here $\mu$ is the chemical potential and the hopping
amplitude $J$ only includes tunneling of atoms between the nearest-neighbor sites as indicated by $\langle\vec{j}_1\vec{j}_2\rangle$ in the summation.
The on-site interaction strength is denoted by $U$. It can be modified by changing the spatial confinement or
via a Feshbach resonance of the $s$-wave scattering length $a_s$~\cite{RevModPhys.82.1225}. In this paper we consider the ground state of the system consisting of atoms trapped in two different hyperfine states, for which we use the ``spin" labels $\ket{\downarrow}$ and $\ket{\uparrow}$, in analogy to electrons in solids.

For deep lattices, the lattice potential  can be approximated locally close to the site minimum as a harmonic
potential  with frequencies
\begin{equation}
\omega_{x}=\omega_y=\frac{2E_R}{\hbar}\sqrt{s}.
\end{equation}
 Then, the ground state Wannier function becomes
\begin{align}
\label{eq:wannier:0}
w(\vec{r})= & \prod\limits_{i=x,y,z}\frac{1}{(\pi l_i^2)^{1/4}}\exp{\left(-\frac{r_i^2}{2l_i^2}\right)}\; , \\
l_i= & (\hbar/m\omega_i)^{1/2} \; ,  i=x,y,z \;,\label{eq:l_i}
\end{align}
where $l_i$ is the oscillator length. With this approximation,
\begin{equation}
\label{eq:onsite_interaction}
U= k_{\rm l}  a_s  \sqrt{\frac{\hbar^3 \omega_x \omega_y \omega_z}{\pi E_R}}
= 2 k_{\rm l} a_s \sqrt{\frac{\hbar E_R s \omega_z }{\pi}}
\; .
\end{equation}
In the limit $s\gg 1$, the hopping matrix element
can be obtained from the 1D Mathieu equation as
\begin{equation}
\label{eq:hopping_amplitude}
J=\frac{4}{\sqrt{\pi}}E_R(s)^{3/4} \exp{\left[ -2 \sqrt{s}\right]} \;.
\end{equation}
Thus  $J$ and  $U$ are  readily changed by tuning the laser intensity
 and  magnetic fields in the experiment.

\section{Approximate solution of the Hubbard model at half-filling}
\label{sec:hubbard_model}

\subsection{Low energy physics of the Hubbard model}
\label{sec:sdw}

In the experimentally relevant large $U/J$ limit, and at half-filling (i.e., on average one atom per site), the ground state of the Hubbard model is a Mott insulator. This is a state where the
energy cost of a doubly occupied site ($\sim U$)  is too large compared to both the hopping
energy $J$ and the temperature $T$ when $k_B T \alt U$.
In the Mott insulator state the on-site \emph{total} atom number fluctuations
are suppressed. Immediately below the onset of the Mott transition there is, however, very little energy cost in mixing the relative populations of
the two spin states and the on-site \emph{relative} atom number may still fluctuate. At even lower temperatures, below
a new characteristic scale determined by the N{\' e}el
temperature, the spins in a  square lattice form into an AFM pattern. Classically, with
equal numbers of $\uparrow$ and $\downarrow$ atoms, this represents a checkerboard
pattern where spin $\uparrow$ and spin $\downarrow$ atoms occupy alternating sites.
In the alternating density pattern, virtual hopping of atoms
between nearest neighbor sites becomes energetically favorable.
Second-order perturbation theory at large $U/J$  then leads to an
effective spin exchange interaction $\sim 4 J^2 /U$ between the atoms in the neighboring sites,
defining the N{\' e}el  temperature scale. At such  low
energies, the system is  described by the (spin only) Heisenberg
model~\cite{RevModPhys.63.1}.

At zero temperature, it is known that the ground state of the 2D  Hubbard model
(or the effective Heisenberg model) has  true long-range AFM order; for a review, see \cite{Fazekas}.
The order parameter for the AFM state is the staggered magnetization, defined as
\begin{equation}
\label{eq:order_parameter}
m = \frac{1}{2} \: e^{i \vec{Q}\cdot \vec{r}_{\vec j}} \; \ev{\op{S^z_{\vec{j}}}},
\end{equation}
where the spin operator in real space is
$\op{S^z_{j}}=(\cre{c}{\vec{j}\uparrow}\des{c}{\vec{j}\uparrow}- \cre{c}{\vec{j}\downarrow}\des{c}{\vec{j}\downarrow})$ [cf. the momentum space version in Eq.~\eqref{eq:spinSzop}].
For the checkerboard pattern in a 2D square lattice,  the ordering momentum is
\begin{equation}
\label{eq:Qvector}
\vec{Q}= (\pi/a,\pi/a).
\end{equation}
This in turn corresponds to  the expectation value of the number operator for
each spin component at half-filling
\begin{equation}
\label{eq:n_mf}
\ev{\op{n}_{\vec{j}g}}=\frac{1}{2} +m \; \eta({g}) \; e^{i\vec{Q}\cdot \vec{r}_{\vec j}}\; ,
\end{equation}
where
\beq \label{eq:etag}
\eta(g)= \left\{ \begin{split} 1 \quad &\text{for} \quad  g=\uparrow \\
-1 \quad &\text{for} \quad g=\downarrow  \end{split}  \right. \;\; .
\eeq

 \begin{figure}
\centering
\includegraphics[width=0.3\textwidth]{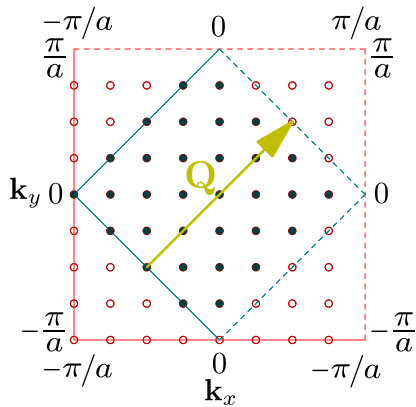}
\caption{Illustration of Brillouin zone indicated by the outer (red) square
and RBZ by the inner (blue) square in a 8$\times$8 lattice. The filled squares represent filled momentum states in
the Fermi sea at half-filling.}
\label{fig:brillouin_zone}
\end{figure}
The AFM state breaks lattice translation symmetry such that the new unit cell in real space
 is doubled in size (to contain both spin components).
 Hence, in momentum space, the Brillouin zone is halved to become the
 RBZ.  In  Fig.~\ref{fig:brillouin_zone}, we show one choice of the RBZ,
 which is bounded by the four lines $k_x \pm k_y = \pm \pi /a$. The
 filled circles denote the momentum states belonging to the RBZ, and the ordering vector $\vec{Q}$
 links a state within the RBZ to one outside (and vice versa). Note that in order to avoid double counting,
 half the states on the bounding lines belong to the RBZ and the other half are outside of the RBZ.

The AFM order can approximately be characterized by a MFT, and results so obtained can be
used to calculate the effect of the ordering on off-resonant light scattering.
However, MFT contains only single particle excitations (of order $U$ at large $U/J$).
Importantly, MFT does not include quantum fluctuations that give rise
 to collective excitations at low energies. Hence MFT fails to capture the  spin waves of the AFM state.
Schrieffer et al.~\cite{PhysRevB.39.11663} and others~\cite{PhysRevB.46.11884,PhysRevB.51.8310,PhysRevB.52.7395} have demonstrated how the RPA can partially incorporate quantum fluctuations and describe the spin wave excitations of the system. Thus, we will use the RPA at zero temperature  to study how collective modes modify the scattered light.

On the other hand, at any non-zero temperatures,  there is no true long-range AFM order in the thermodynamic limit in the  Hubbard
model or the  Heisenberg model~\cite{PhysRevLett.17.1133}. This loss of long-range order
is due to the enhanced  quantum and thermal fluctuations in 2D.
Instead, there is at most quasi-long-range order; for a review, see~\cite{SachdevBook}. For example,
in the Heisenberg model,
the spin correlation function $\langle \op{S^z_{\vec i}} \op{S^z_{\vec j}}  \rangle$ decays with the distance
$r = |\vec{r}_{\vec i} - \vec{r}_{\vec j}|$ as $ \exp (- r / \xi_{\text{AFM}})$, for $r >>  \xi_{\text{AFM}}$.
$\xi_{\text{AFM}}$ is the AFM  correlation length in 2D given by
\begin{equation} \label{HAF-xi}
\xi_{\text{AFM}}(T) \approx c_0 \: a \exp\left( \frac{2 \pi \: b_0 \:J_{H} } { k_B T} \right)   \; ,
\end{equation}
where $c_0 \sim 0.26$, $b_0 \sim 0.2$, $J_H$ is the Heisenberg exchange coupling and $a$ is the lattice spacing~\cite{Chakravarty89,RevModPhys.63.1}.
In contrast, true long-range order is signaled by an additional constant term in the spin correlation
function, which is just $m^2$. In the absence of true long-range order,
the existence of the length scale $\xi_{\text{AFM}}$ leads to the following physical picture:
If the system size $L$ is such that $L \gg \xi_{\text{AFM}}$,  the system is made up of small
domains of size $\sim \xi_{\text{AFM}}$ within which there exists AFM ordering.
The order parameter $m$ in different domains, however, are uncorrelated,
so that there is no net
AFM order overall. Such short-range order can be masked, if the system size is  small
compared to  $\xi_{\text{AFM}}$ and only consists of a single domain (see~\cite{SachdevBook} for the actual  form for the spin correlation function at $r \alt \xi_{\text{AFM}}$).

We can estimate roughly the temperature at which such a finite size effect can become significant in ultracold
atom lattice systems. The crucial ingredient is the strong exponential dependence of  $\xi_{\text{AFM}}$
on $T$ in Eq.~\eqref{HAF-xi}.  The Heisenberg exchange coupling $J_H$
 can be related to the Hubbard model parameters by $J_H \approx 4 J^2 / U$ in the large
 $U/J$ limit.
At $U/J = 6$, we find that $\xi_{\text{AFM}} \sim 1000 a$
 for $k_B T/J \sim 0.1$, or $T/2J \sim 0.05$ where the Fermi temperature is $\sim 2 J/k_B$ at half-filling [$2J$ is half the bandwith of the dispersion relation]. On the other hand,
at $T/2J \sim 0.1$ the correlation length $\xi_{\text{AFM}} \sim 20 a$ is already smaller than current typical optical lattice size of $\sim 30$ sites in each dimension.

\subsection{Mean-field Hamiltonian}
\label{sec:mean_field_hamiltonian}
In order to use the staggered magnetization as the MFT order parameter for the Hubbard Hamiltonian, we write
$\op{n}_{\vec{j}g} = \langle \op{n}_{\vec{j}g} \rangle + (\op{n}_{\vec{j}g}- \langle \op{n}_{\vec{j}g} \rangle)$.
It is then assumed that in the MFT Hamiltonian, terms second order in the fluctuation
$(\op{n}_{\vec{j}g}- \langle \op{n}_{\vec{j}g} \rangle)$ are small. Hence,
the interaction in Eq.~\eqref{eq:hubbard-direct-space}
can be rewritten as: $$U\sum\limits_{\vec{j}}{\op{n}_{\vec{j}\uparrow}\op{n}_{\vec{j}\downarrow}} \approx U\sum\limits_{\vec{j}}
(\op{n}_{\vec{j}\uparrow} \langle \op{n}_{\vec{j}\downarrow} \rangle
+ \op{n}_{\vec{j}\downarrow} \langle \op{n}_{\vec{j}\uparrow} \rangle
-\langle \op{n}_{\vec{j}\uparrow} \rangle  \langle \op{n}_{\vec{j}\downarrow} \rangle).$$
As usual, the real space Hamiltonian can be simplified
by transferring to momentum space via
\begin{equation}
\label{eq:descj:desck}
\des{c}{\vec{j}g}=\frac{1}{N_s}\sum_{\vec{k}}{e^{i \vec{k}\cdot \vec{r}_j}\des{c}{\vec{k}g}}
\end{equation}
The summation over momenta is defined for the whole Brillouin zone: $(k_x,k_y)=\frac{2\pi}{N_s a}(j_x,j_y)$
with $-N_s/2\leq j_{x,y}\leq N_s/2-1$, where we assume $N_s$ sites in each dimension.
If, for simplicity, we assume a translationally invariant system with periodic boundary conditions, we only need to consider coupling between the momentum states $\vec{k}$ and $\vec{k}+\vec{Q}$ in the momentum space representation of the Hamiltonian. Then it is useful to explicitly split up the fermion operator $\des{c}{\vec{k}g}$ defined over
the whole Brillouin zone, to two operators $\des{c}{\vec{k}g}$ and $\des{c}{\vec{k+Q}g}$, where
$\vec{k}$ is now defined only for the RBZ. They are then collected into a two-component Nambu spinor
(in analogy to superconductivity)
\begin{equation}
\vdes{\Psi}{ \vec{k},g}    = \begin{pmatrix}
\des{c}{ \vec{k},g} \\\des{c}{ \vec{k}+\vec{Q},g}
\end{pmatrix} \; , \qquad  \vec{k}    \in \text{RBZ} \; . \label{eq:Nambu}
\end{equation}
Using the definition of the staggered order parameter [Eq.~\eqref{eq:n_mf}] and substituting
Eq.~\eqref{eq:descj:desck} and ~\eqref{eq:Nambu}, the MFT Hamiltonian  $\mathcal{H}$ can be written as a 2$\times$2 matrix
\begin{equation}
 \label{eq:bogoliubov:hamiltonian}
\mathcal{H}= C_m +   \sum\limits_{\vec{k}}^{\text{RBZ}} \sum_g
\vdes{\Psi}{ \vec{k},g}^{\dagger}
\begin{pmatrix}
\epsilon_\vec{k}       & -\Delta_{g}
\\
-\Delta_{g}             & \epsilon_{\vec{k}+\vec{Q}}
\end{pmatrix}
\vdes{\Psi}{ \vec{k},g}  \; ,
\end{equation}
where we have defined the coefficient $C_m\equiv  U N_s^2 (m^2 -1/4)$, single-particle dispersion
\begin{equation}
\epsilon_\vec{k}=-2J(\cos{k_xa}+\cos{k_ya})\,,
\end{equation}
and the order parameter (also called gap parameter)
\begin{equation}
\Delta_{g}=\eta(g)\Delta\:, \qquad  \Delta=mU \;.
\end{equation}
Also, at half-filling we have used  $\mu=U/2$ to simplify the equation.
Note that the Hamiltonian  factors into separate spin sectors due to the choice of ordering in the $z$ direction.

The MFT Hamiltonian \eqref{eq:bogoliubov:hamiltonian} can be diagonalized by a canonical (Bogoliubov) transformation
\begin{equation}\label{eq:Bog-Schr}
\begin{pmatrix}
\des{c}{\vec{k},g} \\\des{c}{\vec{k}+\vec{Q},g}
\end{pmatrix}
=
\begin{pmatrix}
v_{\vec{k},g} & u_{\vec{k},g}
\\
 -\eta(g)u_{\vec{k},g} & \eta(g)v_{\vec{k},g}
\end{pmatrix}
\begin{pmatrix}
\des{c}{1 \vec{k},g} \\\des{c}{2 \vec{k},g}
\end{pmatrix} \; .
\end{equation}
with appropriately chosen $v_{\vec{k},g}$ and $u_{\vec{k},g}$.
The solution is
\begin{align}
v_{\vec{k},g}=&\sqrt{\frac12\left(1-\frac{\epsilon_{\vec{k}}}{E_{\vec{k}g}}\right)},
\quad
u_{\vec{k},g}=\sqrt{\frac12\left(1+\frac{\epsilon_{\vec{k}}}{E_{\vec{k}g}}\right)}\;, \\
E_{\vec{k}g}=&\sqrt{\Delta_g^2+\epsilon_{\vec{k}}^2} = \sqrt{\Delta^2+\epsilon_{\vec{k}}^2}
\equiv E_{\vec{k}}\; .
\end{align}
The last equation holds because $\Delta_g = \eta(g) \Delta$ [see Eq.~\eqref{eq:etag}]. The subscript
of the new operators in Eq.~\eqref{eq:Bog-Schr} refers to the energy bands in RBZ.
 In the new basis, the Hamiltonian reads
\begin{equation} \label{eq:final-MF-H}
\mathcal{H}= C_m + \sum\limits_{\vec{k},g}^{\text{RBZ}}
\begin{pmatrix}
\cre{c}{1 \vec{k},g} &\cre{c}{2 \vec{k},g}
\end{pmatrix}
\begin{pmatrix}
 -E_{\vec{k}g} &      0
\\
0                      &   +E_{\vec{k}g}
\end{pmatrix}
\begin{pmatrix}
\des{c}{1 \vec{k},g} \\\des{c}{2 \vec{k},g}
\end{pmatrix} \,.
\end{equation}

In summary, the MFT ansatz [Eq.~\eqref{eq:order_parameter}] leads to the MFT Hamiltonian
Eq.~\eqref{eq:bogoliubov:hamiltonian} that couples a particle at $\vec{k}$ to a hole at $\vec{k+Q}$.
The Bogoliubov rotation then transforms the original one-band Hubbard model, defined over the full  Brillouin zone,
to a two-band model with energies $\pm E_{\vec{k}g} $.
To accommodate the same number of states, the momenta in the two-band model are defined over the RBZ only.
At temperature $T=0$, the ground state has all the negative energy modes filled at half-filling, i.e., band $1$ is filled.
At finite temperatures, the usual Fermi-Dirac distribution describes the occupation of the two bands.
\begin{figure}
\centering
\includegraphics[width=0.48\textwidth]{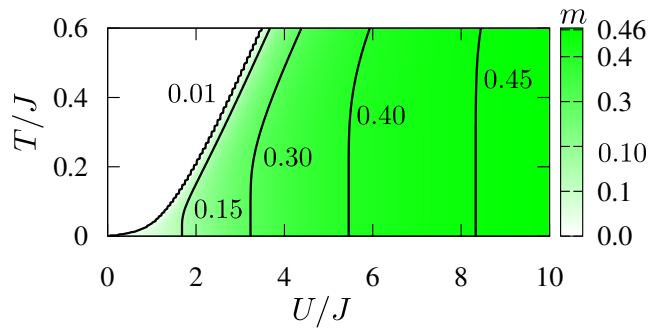}
\caption{Diagram showing the $(U,T)$ dependence of the staggered magnetization, $m$,
obtained by solving Eq.~\eqref{eq:gap_equation} in a 40$\times$40 lattice.
}
\label{fig:phase_diagram_big}
\end{figure}

The MFT  is found by minimizing the total energy with respect to $m$ at a given temperature
using the Hamiltonian~\eqref{eq:final-MF-H}.
The resulting  order parameter equation is
\begin{equation}
\label{eq:gap_equation}
1=\frac{1}{N_s^2}\sum\limits_{\vec{k}'}^{\text{RBZ}}{
    \frac{U  }{E_{\vec{k}'}}
             \tanh{\left(\frac{ E_{\vec{k}'}}{2k_BT}\right)}
                         } \; .
\end{equation}
At half-filling at $T=0$ with  $U\gg J$ the solution saturates towards $m=1/2$.
The MFT critical temperature $T_{C,\text{MFT}}$
can be obtained from solving Eq.~\eqref{eq:gap_equation} with $m=0$.
The value of staggered magnetization $m$ is shown in the $(U,T)$ space in Fig.~\ref{fig:phase_diagram_big}.

\begin{figure}
\centering
\includegraphics[width=0.4\textwidth]{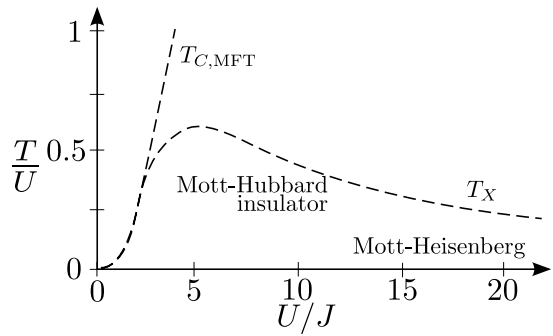}
\caption{Schematic phase diagram of the 2D Hubbard model at half-filling (based on fig. 2 from~\cite{0295-5075-63-5-722}).
$T_{C,\text{MFT}}$ is the critical temperature computed from the gap equation, Eq.~\eqref{eq:gap_equation}.
$T_{X}$ is the  temperature at which the crossover from short-range to (quasi-) long-range order happens.}
\label{fig:phase_diagram_real}
\end{figure}

As mentioned at the beginning of this Section, in 2D, there is strictly no true long-range order
except at $T=0$~\cite{PhysRevLett.17.1133}.
However, one can define instead a cross-over temperature $T_X$ below which  there is
at least short-range order, see Borejsza and
Dupuis~\cite{PhysRevB.69.085119,0295-5075-63-5-722}. In the moderate
to large $U$ regime, $T_X$ is defined as the temperature at which the AFM correlation length
  equals the  lattice spacing $a$: $\xi_\text{AFM}(T_X) = a$.
It turns out that ~\cite{PhysRevB.69.085119} at large $U/J$, $k_B T_X \sim 4 J^2/U$, the
N{\' e}el  temperature scale. On the other hand, at small $U/J$, $T_X \sim T_{C,\text{MFT}}$ where
the critical temperature $T_{C,\text{MFT}}$  is exponentially small in $J/U$.
A detailed analysis of the cross-over phase diagram  can be found in Ref.~\cite{PhysRevB.69.085119}.
Figure~\ref{fig:phase_diagram_real} (based
on Fig.~2 of~\cite{0295-5075-63-5-722}) shows a schematic phase diagram for the Hubbard model in 2D at half-filling.

\subsection{Mean-field susceptibilities at zero temperature}

One key result of this paper is the explicit connection between the physical observables of
inelastic scattered light
intensity and spectrum (coded in static and dynamical structure factors), and susceptibilities
calculated for the  Hubbard model. We first indicate the chain of relations from structure factors
to susceptibilities in Sec.~\ref{sec:retarded-Torder}, and then proceed to give the MFT
susceptibilities in Sec.~\ref{sec:MFChiT0} and the RPA ones in Sec.~\ref{sec:RPA-Chi}.

\subsubsection{Time-ordered and retarded correlation functions} \label{sec:retarded-Torder}

We shall show in Sec.~\ref{sec:quantum_optics} that  the inelastic scattered intensity
[Eq.~\eqref{eq:intensity-inelastic-general}]
and inelastic scattered spectrum  [Eq.~\eqref{eq:spectrum-inelastic-general}]
depends on the static and dynamic {\it structure factors} defined in
Eqs.~\eqref{eq:static_structure_factor} and~\eqref{eq:dynamic_structure_factor}. For the
specific system of two-species atomic gas of $^{40}$K (see Sec.~\ref{sec:40K:properties}),
these general formulae can be reduced [Eqs.~\eqref{eq:intensity:explicit} and~\eqref{eq:spectrum:inelastic}] %
to involve static and dynamic response functions [Eqs.~\eqref{eq:static_response_function},~\eqref{eq:dynamic_response_function} and~\eqref{eq:static-response+-}] %
for the spin operators
$\op{S}^i_{\vec{q}}(t)$ ($i = z, +, -$) [Eqs.~\eqref{eq:spinSzop},~\eqref{eq:spinSPlusop}]
or for the density operator $\op{\rho}_\vec{q}(t)$ [Eq.~\eqref{eq:chargeop}].

On the other hand,  anticipating the calculations of  Sec.~\ref{sec:RPA-Chi},
the RPA method involves a Feynman-Dyson perturbation series that  requires the use
of {\it time-ordered} correlation functions (often called susceptibilities). The Fourier transform in time of time-ordered correlation function is defined as
\begin{equation}
\label{eq:susceptibility}
\chi^{ij}(\vec{q},\vec{q}';\omega)= \int{dt \left[\frac{i}{2N_s^2}\ev{\mathcal{T}\op{O}^i_{\vec{q}}(0)\op{O}^j_{-\vec{q}'}(t)}\right] e^{i\omega t}  } \; ,
\end{equation}
where $\mathcal{T}$ represents the time ordered product, and $\op{O}^i_{\vec{q}}(t)$ can be
$\op{\rho}_\vec{q}(t)$ or $\op{S}^i_{\vec{q}}(t)$.

Fortunately, linear response theory~\cite{fetter2003quantum}
allows to connect time-ordered correlation functions to the
response functions needed for scattered intensity and spectrum, as follows.
First,  at temperature $T=0$, the dynamic response function $\mathcal{S}^{ij}(\vec{q},\vec{q}';\omega)$ [Eq.~\eqref{eq:dynamic_response_function}]
is related to the {\em retarded} susceptibility $\chi^{ij \: R}$ via
\begin{equation} \label{eq:chi-to-structurefactorT0}
\mathcal{S}^{ij}(\vec{q},\vec{q}';\omega)= \frac{-2}{\pi N_s^2}\text{Im}\left[\chi^{ij \: R}(\vec{q},\vec{q}';\omega)\right],
\end{equation}
where the indices $i,j = \rho, z, +,-$. %
Note the factor of $2$ in Eq.~\eqref{eq:chi-to-structurefactorT0} which is there to compensate for the unconventional factor of $2$ in Eq.~\eqref{eq:susceptibility}.
The superscript $^R$ in the susceptibility $\chi^{ij \: R}$ denotes {\it  retarded} correlation functions
that corresponds to the normally ordered physical observables of the scattered light  intensity or  spectrum. Next,
these retarded susceptibilities can be analytically continued from time-ordered susceptibilities, resulting in~\cite{fetter2003quantum}
\begin{align}
\notag
\text{Re}[\chi^{ij}(\vec{q},\vec{q}';\omega)] = & \; \text{Re}[\chi^{ij}{}^R(\vec{q},\vec{q}';\omega)]
\\
\text{Im}[\chi^{ij}(\vec{q},\vec{q}';\omega)] = & \; \text{sgn}(\omega) \; \text{Im}[\chi^{ij}{}^R(\vec{q},\vec{q}';\omega)] \;. \label{eq:Torder2Retard}
\end{align}
 For the MFT (Sec.~\ref{sec:MFChiT0}), and for the  RPA, (Sec.~\ref{sec:RPA-Chi}),
all susceptibilities are time-ordered (and Fourier transformed in time).

For future reference, we also point out that the \emph{static response function} [Eq.~\eqref{eq:static_response_function}]
can be obtained from the dynamic response function by integrating  over $\omega$,
\begin{equation}
\label{eq:static_structure_factor:integrated}
\mathcal{S}^{ij}(\vec{q},\vec{q}')= \: \hbar \int\limits_{-\infty}^{\infty}{d\omega  \; \mathcal{S}^{ij}(\vec{q},\vec{q}';\omega)}\;.
\end{equation}

\subsubsection{Mean-field susceptibilities} \label{sec:MFChiT0}

Because of the RBZ structure of the AFM state, the MFT susceptibilities defined in
Eq.~\eqref{eq:susceptibility} can also be written in a 2$\times$2 matrix form,
\begin{align} \label{eq:susceptibility-Nambu}
\pmb{\chi}^{ij}_{(0)} & (\vec{q}, \omega )  =  \notag\\
%\delta_{\vec{q},\vec{q}'}
& \begin{pmatrix}
\chi^{ij}_{(0)}(\vec{q}, \vec{q}; \omega ) & \chi^{ij}_{(0)}(\vec{q}, \vec{q}+\vec{Q}; \omega ) \\
\chi^{ij}_{(0)}(\vec{q}+\vec{Q}, \vec{q}; \omega ) & \chi^{ij}_{(0)}(\vec{q}+\vec{Q}, \vec{q}+\vec{Q}; \omega )
\end{pmatrix} \, .
\end{align}
The subscript $_{(0)}$ in all the susceptibilities signifies that these correlation functions are calculated within MFT.
A similar 2$\times$2 structure can also be written for the RPA susceptibilities.
We shall use a boldface $\pmb{\chi}$ to denote the susceptibility  matrix.
Here, and for the rest of the Section, unless it is explicitly indicated that this is not the case, we use the notation that the momentum
$\vec{q}$ belongs to the RBZ only.
Schrieffer {\it et al.}~\cite{PhysRevB.39.11663} and others~\cite{PhysRevB.46.11884}
have calculated the zero temperature susceptibilities using MFT. Here  we will present the results;
 see Appendix \ref{App:RPA-Feynman} for an outline of the derivation.

For the 2D Hubbard model in a square lattice at half-filling in the AFM ground state, the MFT density susceptibility is
\begin{equation}
 \pmb{\chi}^{\rho\rho}_{(0)}(\vec{q}, \omega)=
 \begin{pmatrix}
\chi^{\rho\rho}_{(0)}(\vec{q},\vec{q};  \omega ) & 0 \\
0 & \chi^{\rho\rho}_{(0)}(\vec{q}+\vec{Q},\vec{q}+\vec{Q};   \omega )
\end{pmatrix} \, , \label{eq:chi:rhorho:0:delta}
\end{equation}
\begin{widetext}
\begin{align}
 \label{eq:chi:rhorho:0}
\chi^{\rho\rho}_{(0)}(\vec{q},\vec{q}; \omega) =& -\frac{1}{2N_s^2}\sum_{\vec{k}}^{\text{RBZ}}
\left(1-\frac{\epsilon_{\vec{k}}\epsilon_{\vec{k}+\vec{q}}+\Delta^2}{E_{\vec{k}}E_{\vec{k}+\vec{q}}}\right)
\left(        \frac{1}{\hbar \omega-E_{\vec{k}}-E_{\vec{k}+\vec{q}}+i\delta}
        +\frac{1}{-\hbar \omega-E_{\vec{k}}-E_{\vec{k}+\vec{q}}+i\delta}
        \right) \; .
 \end{align}
\end{widetext}
Here in Eq.~\eqref{eq:chi:rhorho:0} the momentum $\vec{q}$ belongs to the full BZ for both sides of the equation.
We have explicitly shown the convergence factor $+ i \delta$ ($\delta>0$) appropriate for time-ordered
susceptibilities. For the calculations in this paper we set $\delta$ a small
but finite value which determines the frequency resolution in calculated spectra.
Since the total density does not distinguish between the spin components, there can be no
component in the susceptibility matrix which transfers momentum $\vec{Q}$ between atoms,
hence the diagonal nature of the matrix in Eq.~\eqref{eq:chi:rhorho:0:delta}. The form of the
susceptibility in Eq.~\eqref{eq:chi:rhorho:0} is similar in structure to susceptibilities
for BCS superfluidity.

 It turns out that at the MFT level, the longitudinal spin susceptibility
 is equal to the density susceptibility
\begin{align}
\pmb{\chi}^{zz}_{(0)}({\vec{q}}, \omega) = & \; \pmb{\chi}^{\rho\rho}_{(0)}({\vec{q}}, \omega) \; .\label{eq:chi:zz:0}
\end{align} This however, is no longer true when  RPA is used to compute the susceptibility, cf.
Eqs.~\eqref{eq:chi:rhorho:RPA} and~\eqref{eq:chi:zz:RPA}.

The transverse spin susceptibility, on the other hand, is sensitive to the coupling of atoms
that differ in momentum by $\vec{Q}$. This then leads to  two distinct components
in the transverse spin susceptibility matrix with nonzero off-diagonal components
\begin{equation}
\pmb{\chi}^{+-}_{(0)} (\vec{q}, \omega ) =
\begin{pmatrix}
\chi^{+-}_{(0)}(\vec{q},\vec{q}; \omega ) & \chi^{+-}_{(0)}(\vec{q}, \vec{q}+\vec{Q};  \omega ) \\
\chi^{+-}_{(0)}(\vec{q}+\vec{Q}, \vec{q};\omega ) & \chi^{+-}_{(0)}(\vec{q}+\vec{Q},\vec{q}+\vec{Q}; \omega )
\end{pmatrix}.
\label{eq:chi:+-:rbz}
\end{equation}
We find
\begin{widetext}
\begin{align}
\label{eq:chi:+-:0}
%\notag
\chi^{+-}_{(0)}(\vec{q},\vec{q}; \omega) =& -\frac{1}{2N_s^2}\sum_{\vec{k}}^{\text{RBZ}}
\left(1-\frac{\epsilon_{\vec{k}}\epsilon_{\vec{k}+\vec{q}}-\Delta^2}{E_{\vec{k}}E_{\vec{k}+\vec{q}}}\right)
%\times\\\notag&\times
  \left(        \frac{1}{\hbar \omega-E_{\vec{k}}-E_{\vec{k}+\vec{q}}+i\delta}
%\right.\\&\left.
     +\frac{1}{-\hbar\omega-E_{\vec{k}}-E_{\vec{k}+\vec{q}}+i\delta}
        \right),\\
\chi^{+-}_{(0)}(\vec{q},\vec{q}+\vec{Q}; \omega) =& -\frac{1}{2N_s^2}\sum_{\vec{k}}^{\text{RBZ}}
\frac{\Delta(E_{\vec{k}}+E_{\vec{k}+\vec{q}})}{E_{\vec{k}}E_{\vec{k}+\vec{q}}}
%\times\\\notag&\times
       \left(        \frac{1}{\hbar \omega-E_{\vec{k}}-E_{\vec{k}+\vec{q}}+i\delta}
%\right.\\&\left.
        -\frac{1}{-\hbar \omega-E_{\vec{k}}-E_{\vec{k}+\vec{q}}+i\delta}
        \right) \; . \label{eq:chi:+-:Q}
\end{align}
\end{widetext}
Similarly to Eq.~\eqref{eq:chi:rhorho:0} the momentum label $\vec{q}$ in Eqs.~\eqref{eq:chi:+-:0} and~\eqref{eq:chi:+-:Q} is valid for the full BZ
on both sides of the equation and $\chi^{+-}_{(0)}(\vec{q},\vec{q}+\vec{Q}; \omega)=\chi^{+-}_{(0)}(\vec{q}+\vec{Q},\vec{q}; \omega)$.
Note that the transverse spin susceptibilities are different to the longitudinal one. This is
due to  isotropy in spin space being broken in our MFT: we have assumed in
Eq.~\eqref{eq:order_parameter} that the AFM
ordering occurs in  a specific spin  direction (along $z$ axis).

We also generalize these MFT susceptibilities to finite temperatures in Appendix \ref{app:mfa_susceptibilities}. At finite
temperatures, there are more available scattering processes as the lower effective band (band 1)
is no longer fully filled, leading to several additional terms in the expressions for the susceptibilities.

\subsection{RPA susceptibilities}\label{sec:RPA-Chi}

The MFT results of the previous subsection capture only the ground state order and single particle
excitations. The latter have energies  of order $U$ at  $U/J \gg 1$, and hence only describe high energy excitations.
However, the low energy physics is that for spin, coming
from the effective Heisenberg exchange interaction at large $U/J$. In particular, there are low energy quantum
fluctuations that lead to gapless collective excitations, corresponding to the spin waves of the Heisenberg antiferromagnet.
The simplest approximate theory that can capture these collective modes is the
RPA~\cite{PhysRevB.39.11663}, which we will therefore employ here.

The basic idea behind the RPA is that certain terms in the perturbation  expansion
in $U$  in the Dyson equation  can be summed to infinite order as a  geometric series (see Appendix \ref{App:RPA-sus}). Formally, at least for
the non-magnetically ordered state,  RPA  can be justified as a series expansion in the small parameter
$k_F a_s$ where $k_F$ is the Fermi wavevector, and $a_s$ is the scattering length related to $U$~\cite{negele1998quantum}.
However, it is interesting  that the RPA for the AFM ordered state also captures the collective modes at large $U/J$.
These collective modes  show up as
bosonic {\it gapless} modes in  the transverse spin susceptibility $\chi^{+-}_{\text{RPA}}$.

In Appendix \ref{App:RPA-sus}, we outline the derivation of the RPA susceptibilities, here we only state the relevant results.
For the density correlation,
\begin{align}
\label{eq:chi:rhorho:RPA}
 \pmb{\chi}^{\rho\rho}_{\text{RPA}}(\vec{q}, \omega)=
  \pmb{\chi}^{\rho\rho}_{(0)}(\vec{q}, \omega) \left[
\pmb{1} + U \pmb{\chi}^{\rho\rho}_{(0)} ({\vec{q}}, \omega ) \right]^{-1}
\end{align}
and for the longitudinal spin correlation,
\begin{align}
\label{eq:chi:zz:RPA}
 \pmb{\chi}^{zz}_{\text{RPA}}(\vec{q}, \omega)=
  \pmb{\chi}^{zz}_{(0)}(\vec{q}, \omega) \left[
\pmb{1} - U \pmb{\chi}^{zz}_{(0)} ({\vec{q}}, \omega ) \right]^{-1}
\end{align}
Note the only difference between these two is the sign of the term proportional to $U$ in the denominator.
Both have the form recognizable from summing a geometric series; indeed they have the same form
 as the more familiar  RPA result for the non-ordered interacting Fermi gas.

\begin{figure}
\centering
\includegraphics[width=0.49\textwidth]{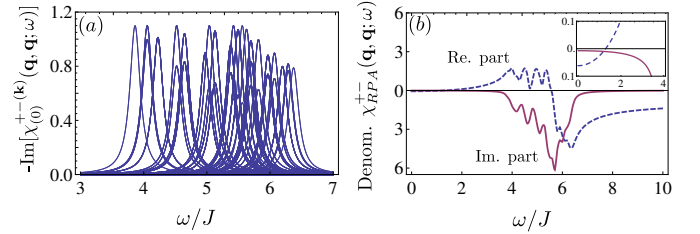}
\caption{Transverse spin susceptibility  for $U=5J$ and $\vec{q}=(\pi/a,0)$
computed with a finite convergence factor $\delta=0.1J$ in a 20$\times$20 lattice.
(a) We show for the MFT case the individual momentum terms $\chi_{(0)}^{+-(\mathbf{k})}(\mathbf{q},\mathbf{q};\omega)$ of Eq.~\eqref{eq:chi:+-:0}
$\chi^{+-}_{(0)}(\mathbf{q},\mathbf{q};\omega)=\sum_{\vec{k}}{\chi_{(0)}^{+-(\mathbf{k})}(\mathbf{q},\mathbf{q};\omega)}$.
Each term contributes with a Lorentzian shaped peak at $\omega=E_{\vec{k}}-E_{\vec{k}+\vec{q}}$.
(b) Imaginary (solid) and real (dashed) parts of the denominator in the RPA case
$\chi^{+-}_{\text{RPA}}(\vec{q},\vec{q};\omega)$ [Eq.~\eqref{eq:chi:+-:0:RPA}].
The inset in (b) zooms in close to the origin and shows the zero-crossing in the real part.
}
\label{fig:RPA_poles20}
\end{figure}

\begin{figure}
\centering
\includegraphics[width=0.49\textwidth]{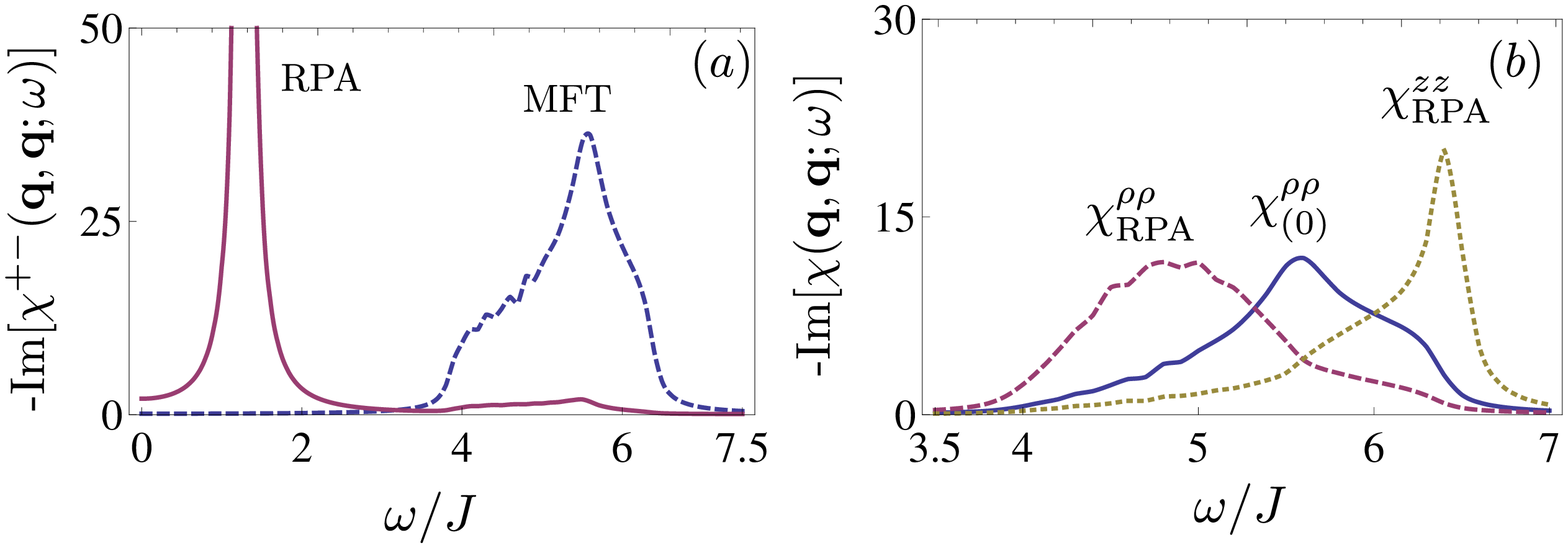}
\caption{Imaginary part of the MFT and RPA density, longitudinal spin and transverse spin susceptibilities for
a lattice size of 40$\times$40. Other parameters are as in Fig.~\ref{fig:RPA_poles20}.
(a) Imaginary part of the transverse spin susceptibility
for the MFT case $\chi^{+-}_{(0)}(\vec{q},\vec{q};\omega)$
of Eq.~\eqref{eq:chi:+-:0} (dashed) and for the RPA case $\chi^{+-}_\text{RPA}(\vec{q},\vec{q};\omega)$
of Eq.~\eqref{eq:chi:+-:0:RPA} (solid).
(b) RPA density susceptibility $\chi^{\rho\rho}_\text{RPA}(\vec{q},\vec{q};\omega)$
of Eq.~\eqref{eq:chi:rhorho:RPA} (dashed),
RPA longitudinal spin susceptibility $\chi^{\rho\rho}_\text{RPA}(\vec{q},\vec{q};\omega)$
of Eq.~\eqref{eq:chi:zz:RPA} (dotted).
Note that the in the MFT case $\chi^{\rho\rho}_{(0)}(\vec{q},\vec{q};\omega)$ (solid) coincides
with the MFT $\chi^{zz}_{(0)}(\vec{q},\vec{q};\omega)$, see
Eqs.~\eqref{eq:chi:rhorho:0} and~\eqref{eq:chi:zz:0}. However,
the  RPA renormalizes these two susceptibilities differently.
}
 \label{fig:RPA_chi20}
\end{figure}

The transverse spin susceptibility is more complicated: the AFM state doubles the unit cell and
halves the Brillouin zone, coupling a spin $g$ particle at $\vec{k}$ to a spin $g$ hole at $\vec{k+Q}$
[see Eq.~\eqref{eq:bogoliubov:hamiltonian}]. This gives rise to the non-diagonal matrix form of
Eq.~\eqref{eq:chi:+-:rbz} for the MFT
susceptibility, which then feeds into the RPA one. The RBZ structure can be accommodated in a 2$\times$2 matrix  version of
Dyson's equation to give (see Appendix \ref{App:RPA-sus})
\begin{align}
\label{eq:chi:+-:0:RPA:matrix}
\pmb{\chi}^{+-}_{\text{RPA}} (\vec{q}, \omega ) =
\pmb{\chi}^{+-}_{(0)} (\vec{q}, \omega ) \left[
\pmb{1} - U \pmb{\chi}^{+-}_{(0)} (\vec{q}, \omega ) \right]^{-1} \;.
\end{align}
Or in the explicit form given by~\cite{PhysRevB.46.11884},
\begin{widetext}
\begin{align}
\label{eq:chi:+-:0:RPA}
\chi^{+-}_{\text{RPA}}(\vec{q},\vec{q};\omega) &=
                \frac{
                \chi^{+-}_{(0)}(\vec{q},\vec{q};\omega)[1-U\chi^{+-}_{(0)}(\vec{q}+\vec{Q},\vec{q}+\vec{Q};\omega)]+U[\chi^{+-}_{(0)}(\vec{q},\vec{q}+\vec{Q};\omega)]^2
                }{
                [1-U\chi^{+-}_{(0)}(\vec{q},\vec{q};\omega)][1-U\chi^{+-}_{(0)}(\vec{q}+\vec{Q},\vec{q}+\vec{Q};\omega)]-U^2[\chi^{+-}_{(0)}(\vec{q},\vec{q}+\vec{Q};\omega)]^2}
\\ \label{eq:chi:+-:Q:RPA}
\chi^{+-}_{\text{RPA}}(\vec{q},\vec{q}+\vec{Q};\omega) &=
                \frac{
                \chi^{+-}_{(0)}(\vec{q},\vec{q}+\vec{Q};\omega)
                }{
                [1-U\chi^{+-}_{(0)}(\vec{q},\vec{q};\omega)][1-U\chi^{+-}_{(0)}(\vec{q}+\vec{Q},\vec{q}+\vec{Q};\omega)]-U^2[\chi^{+-}_{(0)}(\vec{q},\vec{q}+\vec{Q};\omega)]^2} \; .
\end{align}
\end{widetext}
Here in Eqs.~\eqref{eq:chi:+-:0:RPA} and~\eqref{eq:chi:+-:Q:RPA}, as in Eqs.~\eqref{eq:chi:rhorho:0},~\eqref{eq:chi:+-:0} and~\eqref{eq:chi:+-:Q}, the momentum $\vec{q}$ belongs to the full BZ on both sides of the equation and $\chi^{+-}_{\text{RPA}}(\vec{q},\vec{q}+\vec{Q};\omega)=\chi^{+-}_{\text{RPA}}(\vec{q}+\vec{Q},\vec{q};\omega)$.
The appearance of new poles in $\chi^{+-}_{\text{RPA}} $ of Eq.~\eqref{eq:chi:+-:0:RPA:matrix} is
displayed in
Fig.~\ref{fig:RPA_poles20}. Figure~\ref{fig:RPA_poles20}(a) shows the individual poles originating from the MFT transverse susceptibility $\chi^{+-}_{(0)}$
[Eq.~\eqref{eq:chi:+-:0}]. These poles start from $\hbar\omega \geq 2 \Delta$, and  the width of the
peaks is determined by $\delta$ [see discussion after Eq.~\eqref{eq:chi:rhorho:0:delta}].
Figure~\ref{fig:RPA_poles20}(b) shows the real and imaginary parts of the
denominator of Eq.~\eqref{eq:chi:+-:0:RPA}.
When the real part of the denominator crosses zero, a new pole emerges.
We can estimate the energy of this pole as follows.  The Heisenberg model has  the spin wave dispersion
given by~\cite{RevModPhys.63.1}
\begin{equation}
\label{eq:omega:heisenberg}
\hbar \omega_{\vec{q}} =2J_{\text{H}}\sqrt{1-\gamma_\vec{q}^2} \: ,
\end{equation}
with $\gamma_\vec{q}=(\cos{q_x a}+\cos{q_y a})/2$. The effective coupling constant $J_H$
 is related to the Hubbard model parameters via $J_{\text{H}}=4J^2/U$
in the large $U/J$ limit.
Then, in Fig.~\ref{fig:RPA_poles20} with $\vec{q} = (\pi/a, 0)$ and $U= 5 J$,
the pole occurs at energy %
$\hbar \omega_{\vec{q}} \sim 8 J^2 /U \sim 1.6 J$. It can be seen that this is an overestimate, because
we are not strictly in the $U\gg J$ limit.
The imaginary part of both MFT and RPA transverse spin susceptibilities are compared in Fig.~\ref{fig:RPA_chi20}(a)
to illustrate how the RPA renormalizes the single particle excitations and induces collective modes.
The RPA transverse susceptibility notably differs from the MFT result owing to the collective excitations.

In Fig.~\ref{fig:RPA_chi20}(b), we show the corresponding RPA renormalization of the MFT
density and longitudinal spin susceptibilities. The different signs in the denominators in
Eqs.~\eqref{eq:chi:rhorho:RPA} vs. \eqref{eq:chi:zz:RPA} indicate that the RPA corrections have
opposite effects on the MFT density and longitudinal spin susceptibilities, as shown in Fig.~\ref{fig:RPA_chi20}(b).

\subsubsection{RPA correction to the AFM order parameter} \label{sec:RPA-m}

One can also incorporate quantum fluctuations and the effect of collective excitations in the calculation of
the AFM order parameter $m$ by including the RPA corrections to the MFT result of solving the MFT
Eq.~\eqref{eq:gap_equation}.
Indeed, quantum fluctuations are known to strongly suppress $m$ (by about $40 \%$)
\cite{RevModPhys.63.1} in the Heisenberg model. For the half-filled Hubbard model,
Schrieffer et al.~\cite{PhysRevB.39.11663} have computed the RPA corrections to $m$ numerically
(see their Fig.~\ref{fig:magnetic_bragg_peak}). We will use their RPA-corrected data for computing  the elastic scattering intensity
in Secs.~\ref{subsec:K40:scattered_intensity:elastic} and~\ref{sec:measurement_accuracy}.

\section{Optical diagnostics} \label{sec:quantum_optics}

\subsection{Scattered intensity}\label{sec:intensity}

In the previous Section we showed how the time-ordered and normally-ordered correlation functions can be calculated for the AFM ordered fermionic atoms in an optical lattice.
Here we show how these quantities are related to measurements on the light scattered from the atoms. In optical lattice systems the atoms can be confined by highly anisotropic trapping potentials in which the atom dynamics is restricted to 1D or 2D. This makes the atomic samples particularly suitable for imaging. For the incident light tuned off from the atomic transition resonance frequency, the sample is then optically thin and the quantum statistical correlations of the atoms can be mapped onto fluctuations of the scattered light~\cite{PhysRevLett.91.150404,PhysRevA.77.013603,PhysRevLett.98.100402,PhysRevA.76.053618,ruostekoski:170404,PhysRevA.80.043404,PhysRevA.81.013404,PhysRevA.82.033434, PhysRevA.84.033637,PhysRevA.81.063618,PhysRevA.84.053608,PhysRevA.83.051604,PhysRevA.86.023607,PhysRevA.84.043825}.
Measurements on the scattered light therefore convey information about the many-body state of the atoms and can be employed in the diagnostics of the correlated phases of the ultracold atoms in the lattice. Moreover, it was shown in Ref.~\cite{ruostekoski:170404} that by collecting the scattered light into the forward direction by a lens with the diffraction maxima blocked, an experimentally feasible thermometer for fermionic atoms can be realized.
The sensitivity of the thermometer was analyzed by comparing the shot noise of the scattered light with fluctuations in the far-field diffraction pattern that arise from thermal correlations of the atoms. In this paper we study quantum correlations in AFM-ordered strongly-correlated phase of fermionic atoms in an optical lattice. In this Section we introduce the formalism describing the relationship between optical signal (the intensity and spectrum of the scattered light) and the atomic correlations.

\begin{figure}
\centering
\includegraphics[scale=0.5]{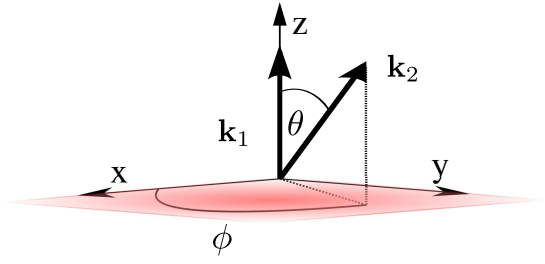}
\caption{Schematic illustration of the light scattering setup. The atoms are confined in the optical lattice close to the $xy$ plane. The incident light field with the wavevector $\vec{k}_1$ propagates perpendicular to lattice in the positive $z$ direction. The wavevector of the scattered light is denoted by $\vec{k}_2$ with the scattering direction determined by the coordinates $\theta$ and $\phi$ }
\label{fig:setup}
\end{figure}
We assume that the atoms in the lattice are illuminated by light that can be approximated by a monochromatic plane wave with the frequency $\Omega_{\rm in}$, propagating perpendicular to the lattice (in the positive $z$ direction). The setup is illustrated in Fig.~\ref{fig:setup}. We write the positive frequency electric field component $\mathbf{E}^+_{\text{in}}$ as
\beq
\mathbf{E}^+_{\text{in}} (\vec{r},t)=
\frac{1}{2}\xi \,\hat{\vec{e}}_{\text{in}}e^{i\vec{k}_1\cdot \vec{r}-i \Omega_{\rm in} t}\,,
\eeq
where $\hat{\vec{e}}_{\text{in}}$ and $\vec{k}_1=k \hat{\vec{e}}_z$ ($k=\Omega_{\rm in}/c$) denote the polarization and wavevector, respectively, of the incoming light.
In an optically thin sample, the dynamics of the electronically excited atomic state may be adiabatically eliminated and the scattered field amplitude $\mathbf{E}_{\rm sc}^{+}(\vec{r},t)$ is proportional to the transition amplitude of atoms between the initial and final hyperfine electronic ground states $g$ and $g'$~\cite{PhysRevA.52.3033},
\begin{equation}
\label{eq:E+:sc}
\mathbf{E}_{\rm sc}^{+}(\vec{r},t) = C \sum\limits_{g'g}\mathbf{\Lambda}_{g'g}\int d^3r'e^{-\mathrm{i}\Delta\vec{k}\cdot\vec{r'}}\hat{\Psi}_{g'}^{\dagger}(\vec{r'},t)\hat{\Psi}_{g}(\vec{r'},t)\,.
\end{equation}
Here the scattered field at $\vec{r}$ is evaluated in the far radiation zone, so that $|\vec{r}-\vec{r'}|\simeq r-\hat{\bf n}\cdot\vec{r'}$ with
\beq
\hat{\bf n}={(\vec{r}-\vec{r'})\over |\vec{r}-\vec{r'}|}\,,
\eeq
and the origin is located inside the atomic sample. The integration is over all the radiating atomic dipole sources at the positions $\vec{r'}$. The field is scattered in the direction $\vec{k}_2$ and  the change of the wavevector of light upon scattering is, see Fig.~\ref{fig:setup},
\begin{eqnarray} \label{eq:Deltak-def}
\Delta \vec{k} &=& \vec{k}_2-\vec{k}_1=k(\hat{\bf n}-\hat{\bf e}_z) \nonumber \\
&=&  k \left(\sin \theta \sin \phi,  \sin \theta \cos \phi, \cos \theta -1 \right) \; .
\end{eqnarray}
In Eq.~\eqref{eq:E+:sc} the effects of the level structure are incorporated in $\mathbf{\Lambda}_{g'g}$ defined by
\begin{equation}
\label{eq:lambda}
\mathbf{\Lambda}_{g'g} ={1\over \mathfrak{D}^2} \sum\limits_{e}\hat{\vec{n}}\times(\hat{\vec{n}}\times  \vec{d}^*_{g'e})( \hat{\vec{e}}_{\rm in}\cdot    \vec{d}_{eg})\,.
\end{equation}
Here the atomic transition dipole matrix elements between the ground state $g$ and the excited state $e$ are
\beq
\label{eq:dipole_matrix_element}
\mathbf{d}_{eg}=(\mathbf{d}_{ge})^{*} = \langle e|\mathbf{d}|g\rangle=\mathfrak{D}\sum_{\sigma} \langle e| \sigma g\rangle\hat{\mathbf{e}}_{\sigma}^{*}\,,
\eeq
where $\langle e|\sigma g\rangle $ denotes the corresponding Clebsch-Gordan coefficient. The summation runs over the circular polarization vectors $\hat{\mathbf{e}}_{\sigma}$ ($\sigma=-1,0,1$),
\begin{equation}
\label{eq:circular_polarization_vectors}
\hat{\mathbf{e}}_{+}=-\frac{1}{\sqrt{2}}\left(\hat{\mathbf{e}}_{x}+ i\hat{\mathbf{e}}_{y}\right),\; \,
\hat{\mathbf{e}}_{0}=\hat{\mathbf{e}}_{z},\; \,
\hat{\mathbf{e}}_{-}=\frac{1}{\sqrt{2}}\left(\hat{\mathbf{e}}_{x}- i\hat{\mathbf{e}}_{y}\right),\; \,
\end{equation}
and  $\mathfrak{D}$ is the reduced dipole matrix element.
The latter is related to the Weisskopf-Wigner radiative resonance linewidth $\gamma$ by
\beq
\gamma={\mathfrak{D}^2 k^3\over 6\pi \hbar \epsilon_0}\,,
\eeq
where $\epsilon_0$ is the vacuum permittivity.
In Eq.~\eqref{eq:E+:sc} we defined the prefactor $C$ by
\beq
\label{eq:detuning}
C={3\xi e^{ikr} \gamma\over 4\delta kr}\,,\quad\delta\equiv \Omega_{\rm in}-\omega_0
\eeq
Here $\delta$ denotes the detuning of the incident light frequency $\Omega_{\rm in}$ from the atomic resonance frequency $\omega_0$.

Typically the wavenumber for the probe light is not the same as the effective wavenumber of the optical lattice potential. In order to suppress the spontaneous emission due to the lattice light lasers, the lattice lasers are considerably more off resonant. The probe and the lattice lasers may also be tuned to different transitions and the lattice spacing can be modulated by optical components, resulting, e.g., in the accordion lattices where the lattice spacing may vary for a given lattice light laser frequency~\cite{Li:08,Williams:08,PhysRevA.82.021604}. We investigate the effect of different ratios $\kappa$ between the probe light wavenumber $k$ and the effective lattice light wavenumber $k_{\rm l}$ [Eq.~\eqref{eq:lattice_spacing}], so that
\beq
\label{eq:kappa}
\kappa = {k\over k_{\rm l}} = {k a\over \pi}\,.
\eeq

The intensity of the scattered light is given by
\begin{equation}
I =2\epsilon_0 c\ev{\vec{E}_{\rm sc}^-(\vec{r},t) \vec{E}_{\rm sc}^+(\vec{r},t)}
\end{equation}
where $c$ denotes the speed of light in vacuum.
By substituting the field amplitudes from Eq.~\eqref{eq:E+:sc} in the expression for the light intensity, we obtain the intensity in terms of the atomic correlation functions. The atoms are initially assumed to occupy the ground state in the lowest energy band. In the tight-binding regime the atomic field operators in the correlation function may be expanded in the series of the Wannier functions [Eq.~\eqref{eq:wannier_function}] where $\op{c}_{\vec{j} g}$ denotes the annihilation operator for the atoms in the electronic ground state $g$ and the lattice site $\vec{j}=(j_x,j_y)$.  We obtain for the scattered light intensity
\begin{equation}
\label{eq:intensity}
\frac{I}{B}=\alpha_{\Delta\vec{k}}
\sum\limits_{\substack{g_1,g_2\\g_3,g_4 \\\vec{i},\vec{j}}}{
     {\sf M}^{g_3g_4}_{g_2g_1}
     e^{i\Delta \vec{k}\cdot(\vec{r}_i-\vec{r}_j)}
            \ev{\op{c}^\dagger_{\vec{i} g_4 }\op{c}_{\vec{i} g_3}\op{c}^\dagger_{\vec{j} g_2}\op{c}_{\vec{j} g_1}}}\,.
\end{equation}
Here we have defined
\beq
\label{eq:B:definition}
B\equiv I_{\rm in}\left({3\gamma\over 2\delta kr}\right)^2, \quad I_{\rm in} = {1\over 2} \epsilon_0 c \xi^2\,,
\eeq
where $I_{\rm in}$ denotes the intensity of the incoming light, and
\begin{equation}
\label{eq:m_tensor}
{\sf M}_{g_2g_1}^{g_3g_4} = \mathbf{\Lambda}_{g_3g_4}^* \mathbf{\Lambda}_{g_2g_1}\,.
\end{equation}
In deriving Eq.~\eqref{eq:intensity} we have assumed that the spatial overlap between different sites is negligible. If the lattice potential is approximately independent of the different ground state levels $g$, the Debye-Waller factor $\alpha_{\Delta\vec{k}}$ exhibits a simple format of the Fourier transform of the lattice site density that can be evaluated by means of the Wannier functions [Eq.~\eqref{eq:wannier:0}],
\begin{align}
\label{eq:alpha_factor}
\alpha_{\Delta\vec{k}}= & \left| \int d^3 r\,e^{-i\Delta \vec{k}\cdot \vec{r}}| w_0(\vec{r})|^2\right|^2\nonumber\\ = &\prod_{i=x,y,z}{\exp{\left[-{(\Delta k_i)^2 l_i^2\over 2}\right]}},
\end{align}
where the oscillator length $l_i$ is defined by Eq.~\eqref{eq:l_i}.

In the expression~\eqref{eq:intensity} for the scattered light intensity, the spatial variation of atomic correlations is encapsulated in the operators $\op{c}_{\vec{i} g}$. The result is general and also includes the cases where the translational invariance of the lattice is broken owing to finite-size effects. In this work, we neglect any additional potential superposed with the lattice that would lead to a nonuniform density distribution, so that $\langle \op{c}^\dagger_{\vec{j}g}\op{c}_{\vec{j}g}\rangle$ is here constant. The spatial profile in Eq.~\eqref{eq:intensity} is therefore solely determined by the phase factors of $\op{c}_{\vec{i} g}$'s. The simple relationship \eqref{eq:intensity} between the scattered light intensity and the atomic correlation functions is a consequence of the weak off-resonant coupling of light. For near-resonant light the coupling is strong even in a 2D lattice and results in excitations of collective polarization modes~\cite{Jenkinsoptlattice}.

We will consider the scattering processes of the atoms to higher energy bands in Sec.~\ref{subsec:higherband}. Although such processes result in the photon frequencies that are shifted by the energy difference between the bands and can be filtered from the signal, they can contribute to the heating rate of the atoms in the lattice.

In the specific analysis of the optical signatures of the atomic correlations it is beneficial to separate in the scattered intensity the contributions from the {\em elastic} and {\em inelastic} scattering events. In the following study we define the elastic scattering processes as those in which the atom scatters back to its original momentum state. We evaluate Eq.~\eqref{eq:intensity} in terms of the elastically and inelastically scattered light intensities $I_\text{e}(\Delta \vec{k})$ and $I_\text{i}(\Delta \vec{k})$, respectively,
\begin{align}
\label{eq:intensity-elastic+inelastic}
I(\Delta \vec{k}) =& I_\text{e}(\Delta \vec{k}) + I_\text{i}(\Delta \vec{k}) \\
I_\text{e}(\Delta \vec{k}) =&   B\alpha_{\Delta\vec{k}}
\left| \sum\limits_{g}{
      \mathbf{\Lambda}^*_{gg}\:  \langle\op{\rho}_{\bar{\Delta \vec{k}} g}\rangle
      }\right|^2
\notag\\=&   B\alpha_{\Delta\vec{k}}
\sum\limits_{g_1 g_3}{
      {\sf M}^{g_3g_3}_{g_1g_1}\: \langle \op{\rho}_{\bar{\Delta \vec{k}} g_3} \rangle \: \langle\op{\rho}_{-\bar{\Delta \vec{k}} g_1}\rangle
      }\,, \label{eq:intensity-elastic-general}
\\
I_\text{i}(\Delta \vec{k})  =& B\alpha_{\Delta\vec{k}}
\sum\limits_{\substack{g_1,g_2\\g_3,g_4 }}{
      {\sf M}^{g_3g_4}_{g_2g_1}} S^{g_3g_4}_{g_2g_1} (\bar{\Delta\vec{k}})
      \, . \label{eq:intensity-inelastic-general}
\end{align}
Here $\bar{\Delta \vec{k}}$ denotes the change of wave vector of light  on the
$xy$ plane and the density operator $\op{\rho}_{\vec{k} g}$ for the spin state $g$ is given by Eq.~\eqref{eq:chargeop}. We have defined
\begin{equation}
\label{eq:uofk}
\mathfrak{u}_{\vec{k}} \equiv \sum\limits_{\vec{j}}{e^{-i\vec{k}\cdot\vec{r}_{j}}}\,,
\end{equation}
and the static structure factor
\begin{align}
S^{g_3g_4}_{g_2g_1} (\bar{\Delta\vec{k}}) & \equiv {1\over N_s^4} \sum\limits_{\vec{q},\vec{q}'\neq0}
          \mathfrak{u}^*_{\bar{\Delta\vec{k}}-\vec{q}}\mathfrak{u}_{\bar{\Delta\vec{k}}-\vec{q}'}\notag\\
           &\times \sum\limits_{\vec{k},\vec{k}'}{ \ev{\op{c}^\dagger_{\vec{k}+\vec{q} g_4}\op{c}_{\vec{k} g_3}\op{c}^\dagger_{\vec{k}'-\vec{q}' g_2}\op{c}_{\vec{k}' g_1}}}_c\,.
\label{eq:static_structure_factor}
\end{align}
All the $\vec{q} = \vec{q}' = 0$ terms in the previous equation are included in the elastic part. This corresponds to incorporating only the connected Feynman diagrams in the correlation function of the static structure factor (indicated by the subscript $c$) and the  disconnected ones in the relevant expansion are precisely those that go into the elastic part, see Ch. 13.4 of~\cite{bruus2004many}.

The finite-size effects of the lattice contribute in Eq.~\eqref{eq:static_structure_factor} via the $\mathfrak{u}_{\vec{k}}$ factors.
In the limit of a large lattice, we may approximate $S^{g_3g_4}_{g_2g_1} $ by translationally invariant atomic mode functions, so that the summation over the sites approaches a delta function
$
\mathfrak{u}_{\vec{k}} \rightarrow N_s^2 \delta_{\vec{k},0}
$ and
\begin{equation}
\label{eq:static_structure_factor_second}
S^{g_3g_4}_{g_2g_1} (\bar{\Delta\vec{k}})\simeq
           \sum\limits_{\vec{k},\vec{k}'}{ \ev{\op{c}^\dagger_{\vec{k}+\bar{\Delta\vec{k}} g_4}\op{c}_{\vec{k} g_3}\op{c}^\dagger_{\vec{k}'-\bar{\Delta\vec{k}} g_2}\op{c}_{\vec{k}' g_1}}}_c\,.
\end{equation}
In a typical 40$\times$40 lattice we are investigating, taking the continuum limit changes the integrated inelastically scattered light intensity by less than 2\%.

In order to calculate the intensity of the scattered light [Eq.~\eqref{eq:intensity-inelastic-general}] and the corresponding static structure factor [Eq.~\eqref{eq:static_structure_factor}] it is useful to define a static response function as
\begin{align}
\label{eq:static_response_function}
\mathcal{S}^{g_3g_4}_{g_2g_1}(\vec{q},\vec{q}') \equiv &\frac{1}{N_s^4} \sum\limits_{\vec{k},\vec{k}'}{ \ev{\op{c}^\dagger_{\vec{k}+\vec{q} g_4}\op{c}_{\vec{k} g_3}\op{c}^\dagger_{\vec{k}'-\vec{q}' g_2}\op{c}_{\vec{k}' g_1}}}_c\,.
\end{align}
In Sec.~\ref{sec:MFChiT0} and \ref{sec:RPA-Chi} we showed how both the RPA correlations based on the Feynman-Dyson perturbation series as well as the MFT correlations for the AFM state in the large lattice limit can then be efficiently calculated in a compact form by evaluating the diagonal components $\mathcal{S}^{g_3g_4}_{g_2g_1}(\vec{q},\vec{q})$. In this paper we approximate
the inelastically scattered light intensity by these diagonal expressions, while still including finite-size contribution from the diffraction pattern via $|\mathfrak{u}_{\bar{\Delta\vec{k}}-\vec{q}}|^2$. The corresponding intensity expression reads
\begin{align}
\frac{I_\text{i}(\Delta \vec{k})}{\alpha_{\Delta\vec{k}}B}\simeq
 \sum_{\vec{q}\neq0}  |\mathfrak{u}_{\bar{\Delta\vec{k}}-\vec{q}}|^2    &
  %\left[
\sum\limits_{\substack{g_1,g_2\\g_3,g_4 }}  {\sf M}^{g_3g_4}_{g_2g_1} \, {\cal S}^{g_3g_4}_{g_2g_1}(\vec{q},\vec{q})
%\right]
\label{eq:intensity-inelastic-general:mathcalS}
\end{align}
In Sec.~\ref{sec:40K:intensity} we will show how $\mathcal{S}^{g_3g_4}_{g_2g_1}(\vec{q},\vec{q})$ can be related to the density as well as longitudinal and transverse spin correlation functions [Eq.~\eqref{eq:static-response+-}].

The general Eqs.~\eqref{eq:intensity-elastic+inelastic}-\eqref{eq:intensity-inelastic-general} give the scattered light intensity for an arbitrary lattice system. The scattered light carries information about the atomic correlation functions. The lattice structure generates the diffraction pattern and the overall envelope of the pattern is produced by the Debye-Waller factors that depend on the profile of the atomic wave functions on individual sites.
The dependence of the scattered light on the polarization, atomic level structure, and the scattering direction is incorporated in ${\sf M}^{g_3g_4}_{g_2g_1}$.

The elastic scattering produces a diffraction pattern from a non-fluctuating atom density in the lattice where the Wannier site wave functions play the role of the diffraction slit profile~\cite{ruostekoski:170404}. The inelastic scattering processes are those in which an atom scatters from one quasimomentum state to another different state. The inelastic scattering is sensitive to the fluctuations of the atoms and reflects the underlying statistical correlations between the atoms. It produces scattered light into angles outside the diffraction orders, generating fluctuating shifts in the diffraction pattern that result from the atom-lattice system absorbing recoil kicks from the scattered photons~\cite{ruostekoski:170404}.

The role of the elastically scattered light intensity is easiest to analyze in the case of a uniformly filled lattice (when the translation symmetry of the lattice is not broken by any of the atomic species). In that case the density operator expectation value reads
\begin{equation} \label{eq:rho_k-simple}
\ev{ \op{\rho}_{\bar{\Delta \vec{k}} g}} = \mathfrak{u}_{\bar{\Delta\vec{k}}} f_g \:,
\end{equation}
where $f_g$ is the atomic filling factor of species $g$ (the total number of atoms of species
$g$ divided by the total number of sites, $N_s^2$). $\bar{\Delta\vec{k}}$ is defined after Eq.~\eqref{eq:intensity-inelastic-general}. We then find that the elastically scattered light intensity,
\begin{equation}
I_\text{e}(\Delta \vec{k}) = B\alpha_{\Delta\vec{k}} |\mathfrak{u}_{\bar{\Delta\vec{k}}}|^2
\left| \sum\limits_{g}{
      \mathbf{\Lambda}_{gg} f_g
      }\right|^2\, ,
\label{eq:intensity:elastic:finite_size_scaling}
\end{equation}
is determined by the Bragg diffraction pattern of the lattice $|\mathfrak{u}_{\bar{\Delta\vec{k}}}|^2$, weighted by the contributions from the atomic level structure via $\mathbf{\Lambda}_{gg}$, and modulated by the Debye-Waller factor $\alpha_{\Delta\vec{k}} $. In the specific case of a 2D square lattice
we obtain the familiar diffraction pattern of a 2D square array of $N_s\times N_s$ diffracting apertures
\begin{equation}
\label{eq:UDeltaK}
|\mathfrak{u}_{\bar{\Delta \vec{k}}}|^2 =\prod\limits_{\alpha=x,y}{
        \frac{\sin^2\left(\frac{ N_s\bar{\Delta \vec{k}}_\alpha a}{2}\right)}{\sin^2\left(\frac{ \bar{\Delta \vec{k}}_\alpha a}{2}\right)}
}\,.
\end{equation}
For uniformly filled lattice the elastic part contains no information about the atomic correlations in the system. Since the atom statistics is mapped onto the inelastically scattered light intensity according to Eq.~\eqref{eq:intensity-inelastic-general}, it is beneficial to block the elastically scattered light before the measurement~\cite{ruostekoski:170404}. We will explain this procedure in detail in Sec.~\ref{subsec:K40:scattered_intensity:elastic}.
For the AFM state studied in Sec.~\ref{sec:mean_field_hamiltonian},
lattice translation symmetry is broken, resulting in the Brillouin Zone being halved in size.
We shall show in Sec.~\ref{subsec:K40:scattered_intensity:elastic} that in this case,
there are new diffraction peaks that correspond to this new lattice periodicity, and is therefore
a key signature of the AFM order. This effect was first analyzed in Ref.~\cite{PhysRevA.81.013415}.

\subsection{Scattered spectrum}
\label{subsec:spectrum}

In the previous Section we established the relation between the scattered light intensity with the equal time
atomic correlations [Eqs.~\eqref{eq:intensity-elastic+inelastic}-\eqref{eq:intensity-inelastic-general}]. We now analyze the spectrum of the scattered light and show how it conveys information about
the excitation spectrum of the atoms in the lattice. The scattered light spectrum may
be obtained as a Fourier transform of the two-time correlation function of the scattered electric field~\cite{PhysRevA.52.3033}
\beq
\mathbb{S}(\Delta\vec{k},\omega)= A \int dt\, e^{i\omega t} \ev{{\bf E}^-(\vec{r},0) {\bf E}^+(\vec{r},t)}\,,
\eeq
where $A$ denotes the normalization factor. We can then write the scattered spectrum in terms of the two-time correlation functions of the atoms in the optical lattice. The spectrum can be separated
into an elastic and inelastic component and we obtain
\begin{align}
\label{eq:spectrum-elastic+inelastic}
\mathbb{S}(\Delta\vec{k},\omega) =& \mathbb{S}_\text{e}(\Delta\vec{k},\omega) +
\mathbb{S}_\text{i}(\Delta\vec{k},\omega) \; , \\
\mathbb{S}_\text{e}(\Delta\vec{k},\omega) =&   A'\alpha_{\Delta\vec{k}}\delta(\omega)
\left|
\sum\limits_{g}{
      \mathbf{\Lambda}_{gg}\:  \langle\op{\rho}_{-\bar{\Delta \vec{k}} g}\rangle
      }
\right|^2
\,,\label{eq:spectrum-elastic-general}  \\
\mathbb{S}_\text{i}(\Delta\vec{k},\omega) =&
A'
\alpha_{\Delta\vec{k}}
\sum\limits_{\substack{g_1,g_2\\g_3,g_4 }}
     {\sf M}^{g_3g_4}_{g_2g_1}
           S^{g_3g_4}_{g_2g_1}(\Delta\vec{k},\omega)
    .\,\label{eq:spectrum-inelastic-general}
\end{align}
where $A'\equiv AB/(2\epsilon_0 c)$. In the last equation we have introduced the dynamic structure factor, which is analogous to the static case of Eq.~\eqref{eq:static_structure_factor}
\begin{widetext}
\begin{equation}
S^{g_3g_4}_{g_2g_1}(\Delta\vec{k},\omega)
             =\frac{1}{N_s^4}
             \sum\limits_{\vec{q},\vec{q}'\neq 0}
             \mathfrak{u}^*_{\bar{\Delta\vec{k}}-\vec{q}}\mathfrak{u}_{\bar{\Delta\vec{k}}-\vec{q}'}
             \sum\limits_{\vec{k},\vec{k}'} \int dt\, e^{i\omega t}
             \ev{\op{c}^\dagger_{\vec{k}+\vec{q} g_4}(0)\op{c}_{\vec{k} g_3}(0)\op{c}^\dagger_{\vec{k}'-\vec{q}' g_2}(t)\op{c}_{\vec{k}' g_1}(t)}_c\,.
\label{eq:dynamic_structure_factor}
\end{equation}
The elastic component corresponds to a peak at $\omega = 0$. The subscript $c$ indicates the connected diagrams for which $\omega\neq0$, revealing the excitations of the system [see Sec.~\ref{sec:scattered_spectrum}]. Analogously to Eq.~\eqref{eq:static_response_function} we define the dynamical response functions as
\begin{align}
\mathcal{S}^{g_3g_4}_{g_2g_1}(\vec{q},\vec{q}';\omega)
             = &\frac{1}{N_s^4}
             \sum\limits_{\vec{k},\vec{k}'} \int dt\, e^{i\omega t}
             \ev{\op{c}^\dagger_{\vec{k}+\vec{q} g_4}(0)\op{c}_{\vec{k} g_3}(0)\op{c}^\dagger_{\vec{k}'-\vec{q}'g_2}(t)\op{c}_{\vec{k}' g_1}(t)}_c\,.
\label{eq:dynamic_response_function}
\end{align}
\end{widetext}
We approximate Eq.~\eqref{eq:spectrum-inelastic-general} in a similar fashion as Eq.~\eqref{eq:intensity-inelastic-general:mathcalS}
\begin{align}
\frac{\mathbb{S}_\text{i}(\Delta\vec{k},\omega)}{\alpha_{\Delta\vec{k}}A'}\simeq
        \sum\limits_{\vec{q}\neq 0}   \left|\mathfrak{u}^*_{\bar{\Delta\vec{k}}-\vec{q}}\right|^2
      \sum\limits_{\substack{g_1,g_2\\g_3,g_4 }}  {\sf M}^{g_3g_4}_{g_2g_1} \mathcal{S}^{g_3g_4}_{g_2g_1}(\vec{q},\vec{q};\omega)
    .\,\label{eq:spectrum-inelastic-general:mathcalS}
\end{align}

\subsection{Atom losses to higher bands} \label{subsec:higherband}

In Sec.~\ref{sec:intensity} we calculated the optical intensity signal for probing the atoms in an optical lattice. This consists of the elastic scattering processes in which the final state of the atoms is the same as the initial state as well as the inelastic scattering processes within the lowest energy band where the quasimomentum state of the atoms changes. The atoms that initially occupy the lowest energy band may also undergo scattering to higher bands. Owing to the energy splitting between the adjacent bands, which is of the order of $2 s^{1/2} E_R$ [$E_R$ was defined in Eq.~\eqref{eq:E_R}], the photons that scatter to higher bands are frequency-shifted from the optical signal and could be filtered out. This is because the maximum recoil kick absorbed by the atom within the lowest band on the $xy$ plane is $k$ [the recoil component on the lattice plane satisfies $|\Delta\vec{q}|=k\sin(\theta)$, see Eq.~\eqref{eq:Deltak-def}], corresponding to the energy shift of $\kappa^2 E_R$ [Eqs.~\eqref{eq:E_R} and~\eqref{eq:kappa}], which is less than the energy difference between the bands~\footnote{analogous argument applies to the scattering in the $z$ direction}. The scattering to higher bands, however, provides a loss mechanism which we will estimate when calculating the measurement accuracy of AFM correlations of the atoms.

Here we extend the analysis of the loss rates of Ref.~\cite{PhysRevA.84.033637} to our multi-level formalism. In the evaluation of the scattered intensity the following correlation functions involving states in higher energy bands yield nonvanishing contributions
\begin{equation}
\sum\limits_{\vec{j},m\neq0} {\cal G}^{0\vec{j}_4,m\vec{j}}_{m\vec{j},0\vec{j}_1}
            \ev{\op{c}^\dagger_{0\vec{j}_4 g_4}\op{c}_{m \vec{j} g_3}\op{c}^\dagger_{m\vec{j} g_2}\op{c}_{0\vec{j}_1 g_1}}\,,
            \label{loss1}
\end{equation}
where $\op{c}_{m \vec{j} g}$ denotes the annihilation operator for the atoms in the band $m$, site $\vec{j}$, and ground state $g$, Eq.~\eqref{eq:wannier_function}. The nonvanishing contribution from the empty excited energy band $m\neq0$  results from  $\ev{\op{c}_{m \vec{j} g_3}\op{c}^\dagger_{m\vec{j} g_2}}$ by the creation of an atom at $(m,\vec{j})$ followed by the annihilation of an atom at $(m,\vec{j})$. We concentrate on the case that the interactions do not mix the spin states, so that only the term $g_3=g_4$ is nonvanishing. The coefficients ${\cal G}$ are defined in terms of the Wannier function integrals
\begin{widetext}
\beq
{\cal G}^{0\vec{j}_4,m\vec{j}}_{m\vec{j},0\vec{j}_1}=\int d^3r d^3r' w^*_{0\vec{j}_4}(\vec{r}) w_{m\vec{j}}(\vec{r}) w^*_{m\vec{j}}(\vec{r'}) w_{0\vec{j}_1}(\vec{r'}) e^{i \Delta\vec{k}(\vec{r}-\vec{r'})}\,.
\label{loss2}
\eeq
\end{widetext}
The mode functions in each site form a complete basis and we have
\beq
\sum_{\vec{j},m\neq0} w_{m\vec{j}}(\vec{r}) w^*_{m\vec{j}}(\vec{r'})=\delta(\vec{r}-\vec{r'})- \sum_{\vec{j}} w_{0\vec{j}}(\vec{r}) w^*_{0\vec{j}}(\vec{r'})\,.
\eeq
This can be used to simplify Eqs.~\eqref{loss1} and~\eqref{loss2}. We find that the contribution to the scattered intensity of this process reads
\begin{equation}
\label{eq:intensity:losses}
I_{\text{hb}}=B N_s^2\left[1-\alpha_{\Delta\vec{k}}\right]
\sum\limits_{g,g'}{
      {\sf M}^{g 'g}_{g' g}
      }\,,
\end{equation}
where $I_{\text{hb}}$ refers to the scattered light intensity resulting from the scattering events where the atoms end up in the higher energy bands. Here $N=N_s^2$ is the total number of atoms in the lattice.
This result remains valid for both fermions and bosons as long as the assumption of unpopulated higher bands is valid.

\section{Optical signatures of magnetic ordering in scattered intensity}
\label{sec:optical_signatures}
\subsection{Two-species atomic gas of $^{40}$K}
\label{sec:40K:properties}
\begin{figure}
\centering
        \includegraphics[width=0.35\textwidth]{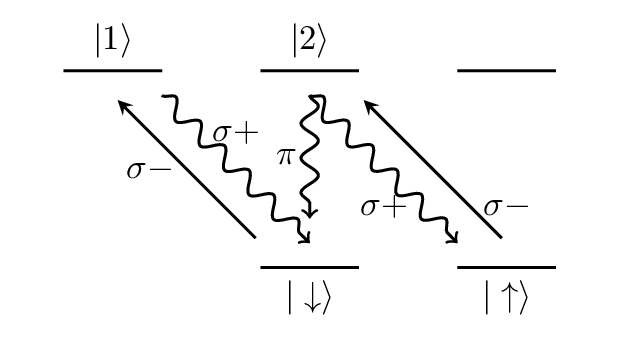}
\caption{
Schematic illustration of the atomic level structure. The atoms are illuminated by an incident light with the $\sigma^-$ polarization, exciting atoms $\ket{\uparrow}{\rightarrow} \ket{2}$ and $\ket{\downarrow}{\rightarrow} \ket{1}$.
The state $\ket{1}$ decays to $\ket{\downarrow}$, while the state $\ket{2}$ can decay either to $\ket{\uparrow}$ or to $\ket{\downarrow}$.}
\label{fig:sigmatransition}
\end{figure}
In the previous Section we presented the general expressions for the dependence of the scattered light
on the atomic correlation functions in an optical lattice system. Next we analyze the 2D square lattice
system of  two-species fermionic atoms introduced in Sec.~\ref{sec:optical_lattices}, with  equal
population of $N_s^2/2$ atoms of  both species.
As a specific example we consider $^{40}$K, which has been used in an experimental realization of fermionic
Mott insulator states in lattices~\cite{ISI:000259090800042,U.Schneider12052008,Greif14062013} and recently AFM ordering~\cite{Greif14062013}. We consider the two electronic ground states
\begin{align}
\label{eq:states:g_levels}
\notag
&\ket{\downarrow}=\ket{4S_{1/2},F_g=9/2,m_F=-9/2},
\\
&\ket{\uparrow}=\ket{4S_{1/2},F_g=9/2,m_F=-7/2}\,.
\end{align}
The incident field is assumed to be $\sigma^-$ polarized, so that the two ground states are coupled to the electronically excited states
\begin{align}
\label{eq:states:e_levels}
\notag
&\ket{1}=\ket{4P_{3/2},F_e=11/2,m_F=-11/2},
\\
&\ket{2}=\ket{4P_{3/2},F_e=11/2,m_F=-9/2}\,.
\end{align}
The level scheme and the transitions are illustrated in Fig.~\ref{fig:sigmatransition}. The atoms in $\ket{\downarrow}$ undergo a cycling transition in which case they are only excited to $\ket{1}$, decaying back to the original state $\ket{\downarrow}$. The atoms in $\ket{\uparrow}$ are excited to $\ket{2}$ from where they can decay either back to $\ket{\uparrow}$ or to $\ket{\downarrow}$. The latter represents a spin-exchanging transition. We will find that the transitions in which the spin is changed convey information about transverse spin correlation functions, while those associated with the scattering processes in which the spin is conserved are proportional to density and longitudinal spin correlation functions. Specifically, the different transitions can be identified with the different RPA susceptibilities in Eqs.~\eqref{eq:chi:rhorho:RPA},~\eqref{eq:chi:zz:RPA} and~\eqref{eq:chi:+-:0:RPA:matrix}, respectively.

In our system we vary the lattice size between $N_s=16$ and $40$ sites along each direction. We consider two lattice heights, $7.8E_R$ and $25E_R$. The trap frequency perpendicular to the lattice is chosen as $\omega_z=10E_R/\hbar$.
For $^{40}$K we take experimentally realistic~\cite{ISI:000259090800042,U.Schneider12052008,Greif14062013} values for the incoming light $\lambda=766.5$nm and $I_{\text{in}}=5 W/m^2$, and assume it to be detuned from the atomic resonance $\delta=20\gamma$ [Eq.~\eqref{eq:detuning}]. This yields in Eq.~\eqref{eq:B:definition} $Br^2\approx 1615 \text{ photons}/s$. We vary the ratio between the lattice spacing and the wavelength of the incident light by changing $\kappa$ [Eq.~\eqref{eq:kappa}] and take $\kappa=0.66\text{, }1.05\text{ and }1.5$. All these correspond to subwavelength lattice spacing, but the additional magnetic peak due to period doubling may only be observed for $\kappa=1.5>\sqrt{2}$.

\subsection{Scattered intensity}
\label{sec:40K:intensity}
In Sec.~\ref{sec:intensity} the total scattered intensity was separated into the elastic component
$I_\text{e}(\Delta \vec{k})$ and an intraband inelastic component $I_\text{i}(\Delta \vec{k})$, see
Eqs.~\eqref{eq:intensity-elastic-general} and~\eqref{eq:intensity-inelastic-general}.
Furthermore, in Sec.~\ref{subsec:higherband} we have taken
into account the inelastic scattering of atoms to higher bands in Eq.~\eqref{eq:intensity:losses},
$I_{\text{hb}}(\Delta \vec{k})$, that is not assumed to contribute to the detected signal but affects the heating rate of the atoms. The total scattered intensity is thus
\begin{equation}
\label{eq:intensity:split}
I =I_{\text{e}} +I_{\text{i}} +I_{\text{hb}} \,.
\end{equation}
We now analyze each of these components for the specific case of a two-species $^{40}$K. Applying the level structure of Fig.~\ref{fig:sigmatransition} we find the explicit expressions for all the scattered elastic and inter-band intensity components
\begin{align}
\label{eq:intensity:elastic1}
\frac{I_{\text{e}}(\Delta\vec{k})}{\alpha_{\Delta\vec{k}}B}= &\left(\sqrt{{\sf M}_{\downarrow\downarrow}^{\downarrow \downarrow}} \ev{\op{\rho}_{\bar{\Delta \vec{k}} \downarrow}}
+\sqrt{{\sf M}_{\uparrow\uparrow}^{\uparrow \uparrow}} \ev{\op{\rho}_{\bar{\Delta \vec{k}} \uparrow}}\right)^2,\\
\label{eq:intensity:losses:explicit1}
I_\text{hb}(\Delta \vec{k}) =& B N_s^2\left(1-\alpha_{\Delta\vec{k}}\right)
\left(
      {\sf M}^{\uparrow \uparrow}_{\uparrow \uparrow}+{\sf M}^{\downarrow \downarrow}_{\downarrow \downarrow}+{\sf M}^{\downarrow \uparrow}_{\downarrow \uparrow}
      \right).%,\\
\end{align}
Here the Debye-Waller factor $\alpha_{\Delta\vec{k}}$ is given in
Eq.~\eqref{eq:alpha_factor} and the coefficient $B$ in Eq.~\eqref{eq:B:definition}.
In the elastically scattered intensity, Eq.~\eqref{eq:intensity:elastic1}, the Fourier transform of the density operator is defined in Eq.~\eqref{eq:rho-AFM}.
The components of ${\sf M}^{g_3g_4}_{g_2g_1}$ in
Eq.~\eqref{eq:m_tensor} read [see Eqs.~\eqref{eq:states:g_levels}
and~\eqref{eq:states:e_levels}]
\begin{align}
\label{eq:inelastic:angular:40K}
\notag
{\sf M}^{\downarrow\downarrow}_{\downarrow\downarrow}=&\frac{1}{4}\left(3+\cos{2\theta}\right)
,&
{\sf M}^{\uparrow\uparrow}_{\downarrow\downarrow}=&
{\sf M}^{\downarrow\downarrow}_{\uparrow\uparrow}=\frac{9}{44}\left(3+\cos{2\theta}\right)
,
\\
{\sf M}^{\uparrow\uparrow}_{\uparrow\uparrow}=&\frac{81}{484}\left(3+\cos{2\theta}\right),&
{\sf M}^{\downarrow\uparrow}_{\downarrow\uparrow}=&\frac{18}{121}\sin^2{\theta}\,.
\end{align}

We now consider the scattered inelastic intraband intensity component $I_{\text{i}}$.
Due to the broken translation symmetry in the AFM state, the RPA susceptibilities of
Eqs.~\eqref{eq:chi:rhorho:RPA},~\eqref{eq:chi:zz:RPA} and~\eqref{eq:chi:+-:0:RPA:matrix}
have a matrix structure
[cf. the MFT susceptibility matrix Eq.~\eqref{eq:susceptibility-Nambu}]. To exhibit clearly this RBZ structure, the
inelastic intensity component of Eq.~\eqref{eq:intensity-inelastic-general:mathcalS} can be written by generalizing the static response functions
 of Eq.~\eqref{eq:static_response_function}  to  a 2$\times$2  matrix with a RBZ momentum structure analogous to that of Eq.~\eqref{eq:susceptibility-Nambu} as
\begin{equation}
\pmb{ \mathcal{S}}^{g_3g_4}_{g_2g_1}(\vec{q})=
\begin{pmatrix}
\label{eq:matrix:mathcalS}
\mathcal{S}^{g_3g_4}_{g_2g_1}(\vec{q},\vec{q})&\mathcal{S}^{g_3g_4}_{g_2g_1}(\vec{q},\vec{q}+\vec{Q})
\\
\mathcal{S}^{g_3g_4}_{g_2g_1}(\vec{q}+\vec{Q},\vec{q})&\mathcal{S}^{g_3g_4}_{g_2g_1}(\vec{q}+\vec{Q},\vec{q}+\vec{Q})
\end{pmatrix}.
\end{equation}
The static response functions that appear in each element of the previous matrix are defined in of Eq.~\eqref{eq:static_response_function}. %
To relate to the RPA density and spin susceptibilities
[Eqs.~\eqref{eq:chi:rhorho:RPA},~\eqref{eq:chi:zz:RPA}, and~\eqref{eq:chi:+-:0:RPA:matrix}]
we first note that using Eqs.~\eqref{eq:chargeop}, ~\eqref{eq:spinSPlusop} and ~\eqref{eq:spinSzop}, we can define
the matrix of static response functions for the density, longitudinal and transverse spin operators
\begin{align}
\pmb{\mathcal{S}}^{\rho\rho}(\vec{q}) = & \: \sum_{g, g' }\pmb{\mathcal{S}}^{g'g'}_{gg}(\vec{q})  \nonumber \\
\pmb{\mathcal{S}}^{zz}(\vec{q})   = & \:\sum_{g, g' } \eta(g) \: \eta(g') \:
\pmb{\mathcal{S}}^{g' g' }_{g\,g}(\vec{q})    \nonumber \\
\pmb{\mathcal{S}}^{+-}(\vec{q})  = & \:2\: \pmb{\mathcal{S}}^{\downarrow \uparrow}_{\downarrow \uparrow}(\vec{q}) \;, \label{eq:static-response+-}
\end{align}
where  $\eta(g)$ is defined in Eq.~\eqref{eq:etag}.
These static response functions can in turn be related to the time-ordered correlation functions
(susceptibilities) of
Eqs.~\eqref{eq:chi:rhorho:RPA},~\eqref{eq:chi:zz:RPA} and~\eqref{eq:chi:+-:0:RPA:matrix}
of Sec.~\ref{sec:hubbard_model},
using the relationship in Eqs.~\eqref{eq:Torder2Retard} and ~\eqref{eq:chi-to-structurefactorT0}
together with frequency integration from Eq.~\eqref{eq:static_structure_factor:integrated}.

The diffraction factors $\mathfrak{u}_{\vec{q}}$ defined in Eq.~\eqref{eq:uofk}
can also be accommodated into the RBZ structure by defining
\begin{equation}
\label{eq:uofk:vector}
\pmb{u}_{\vec{k}} = \begin{pmatrix}
\mathfrak{u}_{ \vec{k}} \\ \mathfrak{u}_{ \vec{k}+\vec{Q}}
\end{pmatrix} \; , \qquad  \vec{k}    \in \text{RBZ}.
\end{equation}
Hence, the inelastic component of the intensity of
 light scattering off atoms in the AFM state of Eq.~\eqref{eq:intensity-inelastic-general:mathcalS} can be written as
\begin{widetext}

\begin{align}
\frac{I_i(\Delta\vec{k})}{\alpha_{\Delta\vec{k}}B} =&
\; \frac{1}{4} \left({\sf M}_{\downarrow\downarrow}^{\downarrow \downarrow}
+{\sf M}_{\uparrow\uparrow}^{\uparrow \uparrow}\right)
 \sum\limits_{\vec{q}\neq0}^{\text{RBZ}}
\pmb{u}^\dagger_{\bar{\Delta\vec{k}}-\vec{q}}\:
\notag
\left[  \pmb{\mathcal{S}}^{\rho\rho}(\vec{q})+\pmb{\mathcal{S}}^{zz}(\vec{q})  \right]\pmb{u}_{\bar{\Delta\vec{k}}-\vec{q}}
\\
&
 + \frac{1}{4} \left( {\sf M}_{\uparrow\uparrow}^{\downarrow \downarrow}+{\sf M}_{\downarrow\downarrow}^{\uparrow \uparrow} \right)
 \sum\limits_{\vec{q}\neq0}^{\text{RBZ}}
 \pmb{u}^\dagger_{\bar{\Delta\vec{k}}-\vec{q}}\:
\left[ \pmb{\mathcal{S}}^{\rho\rho}(\vec{q})- \pmb{\mathcal{S}}^{zz}(\vec{q}) \right]\pmb{u}_{\bar{\Delta\vec{k}}-\vec{q}}
 + \frac{1}{2}{\sf M}_{\downarrow \uparrow}^{\downarrow \uparrow} \:\sum\limits_{\vec{q}\neq0}^{\text{RBZ}}
\pmb{u}^\dagger_{\bar{\Delta\vec{k}}-\vec{q}}\: \:\pmb{\mathcal{S}}^{+-}(\vec{q}) \:\pmb{u}_{\bar{\Delta\vec{k}}-\vec{q}} \; .
\label{eq:intensity:explicit}
\end{align}

\end{widetext}
In the following we will analyze the different scattering contributions. The spectrum and the excitations
will be studied in Sec.~\ref{sec:scattered_spectrum}.

\subsubsection{Elastic scattering}
\label{subsec:K40:scattered_intensity:elastic}
This Section is devoted to studying the elastic component of the scattered light.
We show that the emergence of AFM ordering in the system is directly observable in the
elastically scattered light intensity as this results in \emph{magnetic Bragg peaks} in the scattered light signal.
For the two-species system studied here (Fig.~\ref{fig:sigmatransition}) the total scattered light intensity can be computed
from Eqs.~\eqref{eq:intensity:elastic1} and \eqref{eq:inelastic:angular:40K}. This results in
\begin{equation}
\label{eq:intensity:elastic}
\frac{I_{\text{e}}(\Delta\vec{k})}{\alpha_{\Delta\vec{k}}B}=
\frac{1}{4}\left(3+\cos{2\theta}\right)\left(
\ev{\op{\rho}_{\bar{\Delta \vec{k}} \downarrow}}+\frac{9}{11}\ev{\op{\rho}_{\bar{\Delta \vec{k}} \uparrow}}
\right)^2 \; .
\end{equation}
Note that the  two spin terms contribute unequally because
 the dipole matrix elements are different for each hyperfine state, see Eq.~\eqref{eq:inelastic:angular:40K}.

When both atomic species fill the lattice with a uniform density, the Fourier transforms of the density terms in Eq.~\eqref{eq:intensity:elastic} are given by Eq.~\eqref{eq:rho_k-simple}. In the AFM state the total atom density is still uniform, but this is no longer true for the densities of the individual spin components. Each spin component favors the occupations of alternating sites, indicating broken lattice translation symmetry and period doubling [see Eq.~\eqref{eq:n_mf}]. Consequently, the Fourier transform has a new term. It can be written as
\begin{align}
\label{eq:rho-AFM:general}
\ev{ \op{\rho}_{\bar{\Delta \vec{k}} g}} =&
 \sum_{{\vec j}} e^{i \bar{\Delta \vec{k}} \cdot \vec{r}_{\vec{j}}}  \op{n}_{\vec{j}g}=
  \sum_{{\vec j}} e^{i \bar{\Delta \vec{k}} \cdot \vec{r}_{\vec{j}}}{  \left(f_g +m  \eta({g})  e^{i\vec{Q}\cdot \vec{r}_{\vec j}} \right)}
\\
=& \mathfrak{u}_{\bar{\Delta\vec{k}}} \: f_g  +
\mathfrak{u}_{\bar{\Delta\vec{k}}+\vec{Q}} \: m  \: \eta(g) \:,
\label{eq:rho-AFM}
\end{align}
where $f_g=1/2$ is the atomic filling factor  of species $g$ at half-filling and
$\eta(g)$ is defined in Eq.~\eqref{eq:etag}. Comparing to Eq.~\eqref{eq:rho_k-simple}, the
new term is proportional to the AFM order parameter $m$, Eq.~\eqref{eq:order_parameter}, and is centered at the
ordering vector $\vec{Q}=(\pi/a,\pi/a$). The order parameter $m$ can be obtained by solving the implicit MFT [Eq.~\eqref{eq:gap_equation}].
The MFT phase diagram of Fig.~\ref{fig:phase_diagram_big} shows its dependence on $T$ and $U$.
However, as discussed in Sec.~\ref{sec:RPA-Chi},
quantum fluctuations around the MFT solution have significant effects not only on susceptibilities, but
also on the AFM order parameter $m$: the ordering can decrease by up to $\sim 40 \%$ at large $U$. Thus,
we shall use the RPA-corrected $m$ values as computed by Schrieffer et al.~\cite{PhysRevB.39.11663}.

\begin{figure}
\centering
\includegraphics[width=0.4\textwidth]{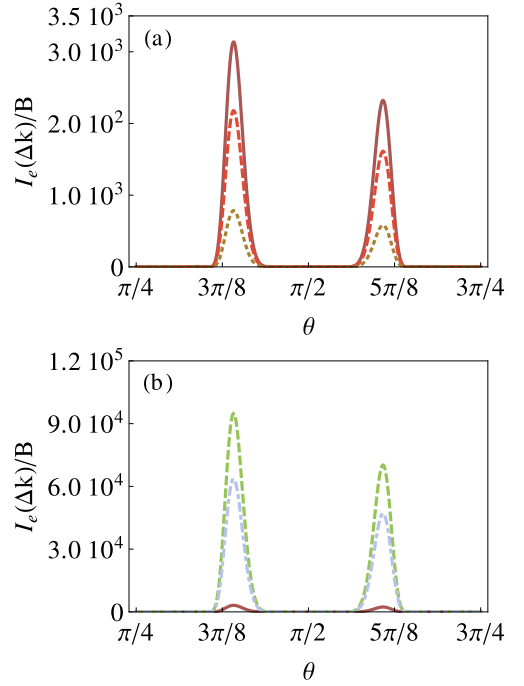}
\caption{
Angular distribution of the elastic component of the
scattered light intensity along the direction $\phi=\pi/4$.
The calculations use the AFM order parameter $m$ computed with RPA corrections.
Here the number of sites is 40$\times$40 and the lattice depth $s=25$.
(a) $I_{\text{e}}(\Delta\vec{k})$ for different values of $m$, when both species are detected.
Different curves represent
$U=7.3J$ and $m_{\text{RPA}}=0.3$ (solid),
$U=3.9J$ and $m_{\text{RPA}}=0.25$ (dashed),
$U= 2.0J$ and $m_{\text{RPA}}=0.15$ (short dashed).
In (b) we compare the results for the total density with the single species detection
($m_\text{RPA}=0.3$). Curves from top to bottom:
$I^{\downarrow}_{\text{e}}(\Delta\vec{k})$ (dashed),
$I^{\uparrow}_{\text{e}}(\Delta\vec{k})$ (dash-dotted),
and $I_{\text{e}}(\Delta\vec{k})$ (solid).
The magnetic Bragg peak is observable since $\kappa=1.5>\sqrt{2}$.
Note that the highest peak in (a) corresponds
to the smallest one in (b).
}
\label{fig:magnetic_bragg_peak}
\end{figure}
Substituting Eq.~\eqref{eq:rho-AFM} into Eq.~\eqref{eq:intensity:elastic},
we see that in addition to the usual diffraction
term  $\mathfrak{u}^*_{\bar{\Delta \vec{k}}}\mathfrak{u}_{\bar{\Delta \vec{k}}}$ centered at $\Delta \vec{k} = 0$, there is a new magnetic  Bragg peak centered at $\bar{\Delta\vec{k}} = \vec{Q}$, proportional
to $m^2$ (see Fig.~\ref{fig:magnetic_bragg_peak}).
This new peak can be detected by collecting scattered light around
$\bar{\Delta\vec{k}}\sim \vec{Q}$, which, according to the definition of $\Delta\vec{k}$
in Eq.~\eqref{eq:Deltak-def},  corresponds to $\theta_B=\arcsin{\frac{\sqrt{2}}{\kappa}}$
and $\phi_B=\pi/4$ (see Sec.~\ref{sec:optical_componets}).
The position of the magnetic peak depends on the
ratio $\kappa$ [Eq.\eqref{eq:kappa}] between the probe light wavevector and the effective wavevector of the lattice light, see Eqs.~\eqref{eq:Deltak-def}. Hence it will only be observable  if $ \kappa \geq \sqrt{2}$. Magnetic Bragg peaks were first studied in optical lattices in Ref.~\cite{PhysRevA.81.013415} and experimentally observed for an artificially prepared density pattern in Ref.~\cite{PhysRevLett.106.215301}.

The dependence of the magnetic Bragg peak on the staggered magnetization $m$ is illustrated in
Fig.~\ref{fig:magnetic_bragg_peak}(a) that shows the elastically scattered intensity from a lattice populated by both atomic species for different values of $m$ [Eq.~\eqref{eq:intensity:elastic}]. The different values of $m$ may correspond, e.g., to different values of temperature or the on-site interaction strength $U$.

On the other hand, if for example only the $\downarrow$ species is imaged, according to
Eq.~\eqref{eq:intensity:elastic}, the elastic part of the intensity
 becomes
\begin{align}
\label{eq:intensity:elastic:explicit:downdown:general}
\frac{I^{\downarrow}_{\text{e}}(\Delta\vec{k})}{B\alpha_{\Delta\vec{k}}}=&
\frac{\left(3+\cos{2\theta}\right)}{4}
\ev{\op{\rho}_{\bar{\Delta \vec{k}} \downarrow}}^2
\\
=& \frac{\left(3+\cos{2\theta}\right)}{4} \left(
\frac{\mathfrak{u}_{\bar{\Delta\vec{k}}}}{2} -
\mathfrak{u}_{\bar{\Delta\vec{k}}+\vec{Q}} m \right)^2 \,.
\label{eq:intensity:elastic:explicit:downdown}
\end{align}
The resulting magnetic Bragg peak is now strongly enhanced, see Fig.~\ref{fig:magnetic_bragg_peak}(b).
We show two different cases;
the case when both species are present in the system [Eq.~\eqref{eq:intensity:elastic}]
and the case when imaging is done with only a single species present in the lattice [Eq.~\eqref{eq:intensity:elastic:explicit:downdown}].
In the first case, without spin-specific detection,
the total signal is very weak because of destructive interference between the scattered light from the two species. In fact, the magnetic Bragg peak is observable only because the dipole transition matrix elements
between the two species are not equal and, according to Eqs.~\eqref{eq:intensity:elastic} and~\eqref{eq:rho-AFM}, is weaker than the single species response by the factor of $(2/11)^2$.
If the transition matrix elements were the same, the spin-independent imaging would only probe the total density without revealing the AFM order.
%%%%%%

\subsubsection{Elastic scattering in the presence of short-range correlations}

So far we have considered long-range AFM order where the staggered magnetization $m$ is constant throughout the lattice. In Sec.~\ref{sec:sdw}, we discussed how there is no genuine long range AFM order in the Hubbard model (or its strong coupling limit, the Heisenberg model), at any finite temperature in the thermodynamic limit. It is beyond the scope of
this paper to calculate the finite temperature short-range effects, but we can simulate short-range ordering effects in a phenomenological manner, as follows:
At finite temperatures, the spins are correlated up to the AFM correlation length $\xi_\text{AFM}$ which leads to domains of size $\sim \xi_\text{AFM}$.
We introduce the spatial variation of the AFM order parameter by letting $m_{\bf i}$ to
depend on the site index ${\bf i}$, with the amplitude fixed at the $T=0$ magnetization value $m_\text{RPA}$.
For simplicity, we assume that $m_{\bf i}$'s in different domains are not oriented in random directions, but that there exists a preferred axis, generated, e.g., by a small imbalance in the Fermi levels of the two species. We therefore introduce a
sign $s_{\bf i} = \pm 1 $ (Ising variable) for $m_{\bf i}$'s  that fluctuates from site to site
\begin{figure}
\centering
\includegraphics[width=0.5\textwidth]{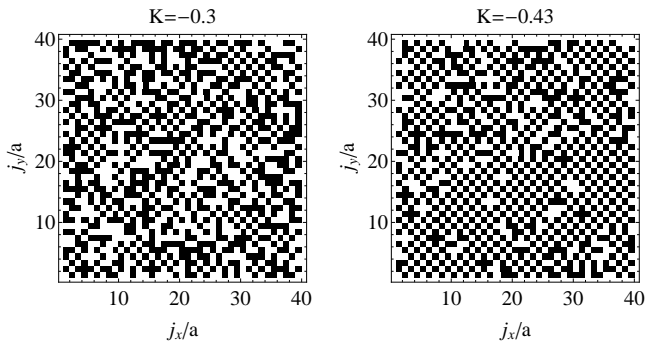}
\caption{Typical configurations for the Ising model at two different temperatures. We show the cases far above ($K=-0.3$ on left) and slightly above ($K=-0.43$ on right) the transition temperature. The corresponding correlation lengths are $\xi_\text{Ising}(-0.3)\approx 3a$ and $\xi_\text{Ising}(-0.43)\approx 40a$.}
\label{fig:ising:configurations}
\end{figure}
\begin{equation}
m_{\bf i} = m_\text{RPA} s_{\bf i} \; .
\end{equation}
The configuration of $s_\mathbf{i}$ is modelled using the nearest-neighbor  Ising Hamiltonian
\begin{equation}
\label{eq:IsingModel}
\frac{\mathcal{H}_\text{I}}{k_BT}=-\frac{J_\text{Ising}}{k_BT}\sum\limits_{\langle\vec{i},\vec{j}\rangle}{s_\vec{i} s_\vec{j}}
=- K\sum\limits_{\langle\vec{i},\vec{j}\rangle}{s_\vec{i}s_\vec{j}}\,.
\end{equation}
$J_\text{Ising}$ is the coupling strength, and $K=J_\text{Ising}/k_BT$.
For $J_\text{Ising}<0$, the ground state of the system at $T=0$ is an AFM Ne\'el state.
In contrast to the Heisenberg model, the Ising model has a nonzero critical temperature.
The transition temperature for a finite system with periodic boundary conditions can be estimated as~\footnote{We present results for a lattice size $40\times40$, and finite-size effects shift the transition point from $K_c(N_s=\infty)$ to the finite-size critical point $K_c(N_s)$. The finite-size critical temperature can be estimated from the the infinite system using the scaling arguments of~\cite{PhysRevLett.19.169}, along with the numerical fit to the finite-size scaling hypothesis from~\cite{PhysRevB.13.2997}. The finite-size shifted value of the critical temperature can be estimated as $K_c(N_s)\approx\frac{N_s K_c(\infty)}{N_s-\alpha}$. For a system with periodic boundary conditions $\alpha\approx -0.36$~\cite{PhysRevB.13.2997}.}
\begin{equation}
\label{eq:shifted:Kc:40}
K_c(N_s=40)\approx -0.437\,.
\end{equation}
We approximate the the correlation length $\xi_\text{Ising}(K)$ for a finite-size system ($N_s=40$) and for $T>T_c(40)$ [$K_c(40)<K$] by~\cite{pathria2011statistical}
\begin{equation}
\label{eq:xi_ising:above}
\xi_\text{Ising}(K)\approx\frac{1}{4[K_c(40)-K]}\,.
\end{equation}
Using the Wolff algorithm \cite{PhysRevLett.62.361}, we can numerically generate a specific configuration for the Ising variable $s_{\bf i}  $,
with a  correlation length determined by the temperature in the Ising model.
To ensure that the number of up and down spins is equal, we impose a constrain
$
|m_{\text{F}}|= |\sum_{\vec{i}}{s_\vec{i}} |<0.01
$ on the
ferromagnetic Ising order parameter.
Figure~\ref{fig:ising:configurations} shows two of the generated configurations at two different temperatures.
A specific configuration of $m_{\bf i}  $ is then used in Eq.~\eqref{eq:rho-AFM:general} to calculate the single-species
($\downarrow$ atom only) elastic part of the scattered intensity of Eq.~\eqref{eq:intensity:elastic:explicit:downdown:general} as
\begin{align}
%\notag
I^\downarrow_\text{e}(\Delta \vec{k}) =&  B\alpha_{\Delta\vec{k}}
      \mathsf{M}^{\downarrow\downarrow}_{\downarrow\downarrow} m_\text{RPA}^2 \left|\sum_{{\vec j}} e^{i \bar{\Delta \vec{k}}\cdot \vec{r}_{\vec{j}}} s_\vec{j}\right|^2
      \, \label{eq:ising:intensity:elastic:downdown}.
\end{align}
Note that this expression does not contain the diffraction peak in the forward direction, as we are only interested in the magnetic Bragg peaks
whose contribution is significant around the perpendicular direction.
\begin{figure}
\centering
\includegraphics[width=0.45\textwidth]{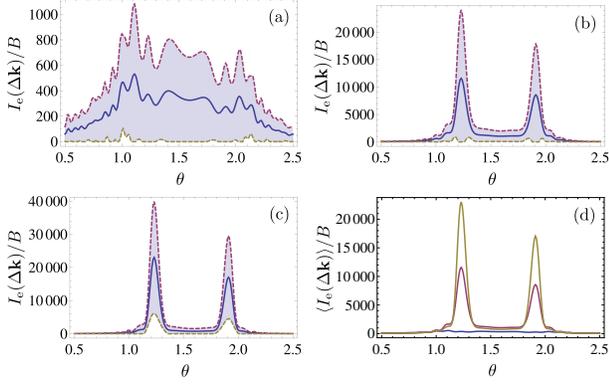}
\caption{Angular distribution of the elastic component of the scattered light intensity in the presence of short-range order [Eq.~\eqref{eq:ising:intensity:elastic:downdown}]
 along the $\phi=\pi/4$ direction. The lattice depth is taken to be $s=25$,
$\kappa=1.5$ [Eq.~\eqref{eq:kappa}],  and $m_{\text{RPA}}\approx0.3$.
We only show the magnetic Bragg peaks for the case that only the down species is imaged.
In (a)-(c) the solid lines represent the ensemble averages of the intensity over 100 stochastic realizations at different temperatures: (a) $K=-0.3$; (b) $K=-0.42$; (c) $K=-0.43$. The corresponding fluctuations (sample standard deviation) are
given by the dashed lines. The different average intensities of (a)-(c) are shown together in (d).
Typical individual lattice configurations are shown in Fig.~\ref{fig:ising:configurations}.}
\label{fig:ising:scattered_light}
\end{figure}
The numerically calculated scattered light intensity at different temperatures is shown in Fig.~\ref{fig:ising:scattered_light}(a)-(c).
The average value of the order parameter $m$ is smaller than $0.02$ in all the cases.
The average scattered light intensity grows at temperatures close to the transition [Fig.~\ref{fig:ising:scattered_light}(d)], and the magnetic Bragg peaks emerge as the correlation length increases. The large fluctuations between different realizations in the simplified model  (Fig.~\ref{fig:ising:scattered_light}) are likely to significantly overestimate the fluctuations of the true AFM state.

\subsubsection{Inelastic intraband scattering}
\label{subsec:K40:scattered_intensity:intraband}
\begin{figure}
  \centering
\includegraphics[width=0.425\textwidth]{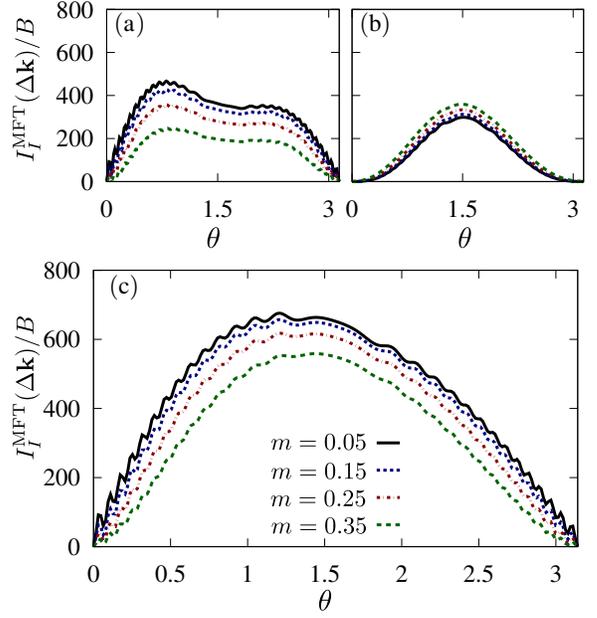}
  \caption{
Angular distribution of the inelastically
scattered light intensity
for different values of the interaction strength $U$ at $T=0$
along the direction $\phi=\pi/4$.
The calculations are based on MFT. Here the number of sites is 40$\times$40.
The ratio between the wavenumber
of the  probe light to the effective
wavenumber of the optical lattice light
$\kappa=1.05$ [Eq.~\eqref{eq:kappa}].
The lattice height is $s=25$.
Intensity contributions from
(a) the density and longitudinal spin components;
(b) transverse spin component;
(c) total scattered light intensity.
The scattered intensity decreases with increasing magnetization
because of the changes in the density and longitudinal spin susceptibility.}
  \label{fig:mfa:Ns40_dU_Angular:Tzero}
\end{figure}
As was demonstrated in the previous Section, the elastic part of the scattered light intensity conveys information
about the atom density in the lattice and generates the diffraction pattern of the atomic lattice structure. If the
detected signal cannot distinguish the two spin components, the light provides almost no information about the AFM
order. On the other hand, when the contributions of the two spin components can be separated in the scattered
light, the emerging AFM order and the period doubling can be identified as additional Bragg peaks. In addition
to the elastic signal, one may also study the inelastically scattered light intensity
[Eq.~\eqref{eq:intensity-inelastic-general}]. The inelastic scattering processes are proportional to static structure
factors $S^{g_3 g_4}_{g_2 g_1}(\Delta\vec{k})$ [Eq.~\eqref{eq:static_structure_factor}] that represent scattering events in which an atom is excited from a
quasimomentum state $\vec{q}$ and scatters to a different quasimomentum state $\vec{q}'$.
Atoms absorb recoil kicks from the scattered photons. The recoil events depend on the statistical correlations between the
atoms, generating fluctuating shifts in the diffraction pattern and significant scattering outside the diffraction
orders. In the process, the atomic correlations are mapped onto the properties of the emitted light. Inelastically
scattered light in a single-component fermionic gas in a lattice can reveal thermal
correlations~\cite{ruostekoski:170404} and has in a two-component case previously been proposed as a
detection method for topological order of the atoms~\cite{PhysRevLett.91.150404}.

The inelastically scattered light intensity for the two-component $^{40}$K gas is given by Eq.~\eqref{eq:intensity:explicit}. The scattering contributions in which the spin is conserved are proportional to density and longitudinal spin susceptibilities. The spin-exchanging transitions (see Fig.~\ref{fig:sigmatransition}) generate the term depending on the transverse spin susceptibility. The two processes exhibit very different angular distribution of the scattered light as shown in Eq.~\eqref{eq:inelastic:angular:40K}. In the spin-conserving processes the emitted photons are generated by the $\sigma^+$ transition in which case the intensity in the forward direction is twice the intensity in the perpendicular direction. The spin-exchanging process, on the other hand, produces scattered photons via the $\pi$ transition which is oriented parallel to the propagation direction of the incident field. Therefore the scattering reaches its maximum in the perpendicular direction ($\theta=\pi/2$) and entirely vanishes in the forward direction.

We have calculated the angular distribution of the inelastically scattered light for different values of the on-site interaction strength $U$, and hence the staggered magnetization $m$ of the AFM ordering. In Fig.~\ref{fig:mfa:Ns40_dU_Angular:Tzero} we show the results based on MFT at $T=0$. In this case the intensity is obtained using the MFT static structure factor
[Eqs.~\eqref{eq:corr:inelastic} and \eqref{eq:g_function}] in Eq.~\eqref{eq:intensity-inelastic-general}.
The corresponding MFT susceptibilities required in the calculation are provided by Eqs.~\eqref{eq:chi:rhorho:0}-\eqref{eq:chi:+-:Q},~\eqref{eq:chi-to-structurefactorT0},~\eqref{eq:Torder2Retard}, and~\eqref{eq:static_structure_factor:integrated}. The different angular distribution of the different scattering contributions is clearly visible in Fig.~\ref{fig:mfa:Ns40_dU_Angular:Tzero}. We find that in MFT the density and longitudinal spin susceptibilities are more sensitive to the variation of $U$ than the transverse susceptibility.
\begin{figure}
\centering
\includegraphics[width=0.45\textwidth]{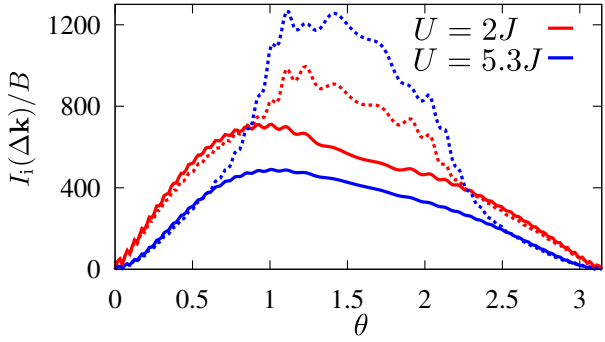}
\caption{
Comparison of the angular distribution of the inelastic
scattered light intensity based on MFT (solid) and RPA (dashed)
at $T=0$
along the direction $\phi=\pi/4$, with $\kappa=1.5$.
The rest of the parameters are as in Fig.~\ref{fig:mfa:Ns40_dU_Angular:Tzero}.
The MFT and the RPA results notably differ, owing to the collective modes that significantly modify
the transverse spin component.
The magnitude of the order parameter [Eq.~\eqref{eq:order_parameter}]
is $m=0.19$ ($m_\text{RPA}\simeq 0.15$) for $U=2J$ and $m=0.4$ ($m_\text{RPA}\simeq 0.28$) for $U=5.3J$.}
\label{fig:intensity:mfa_vs_rpa:Ns40_dM_Angular}
\end{figure}
\begin{figure}
\centering
\includegraphics[width=0.45\textwidth]{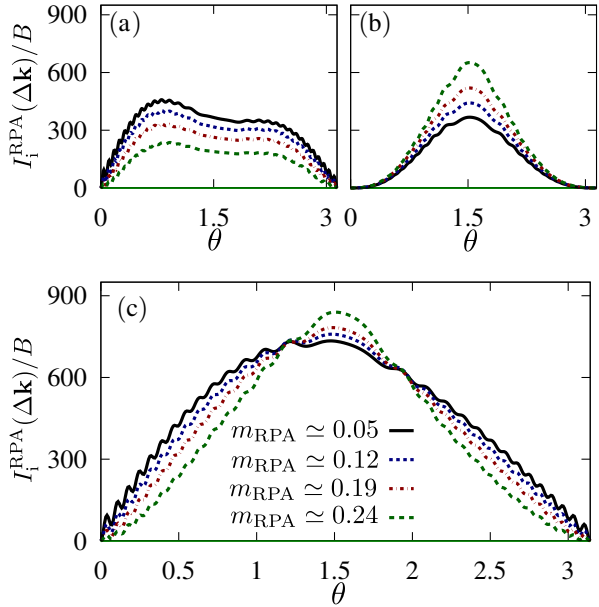}
\caption{
Angular distribution of the inelastically
scattered light intensity
for different values of the interaction strength $U$ at $T=0$
along the direction $\phi=\pi/4$.
The calculations are based on RPA.
We use the same parameters as in Fig.~\ref{fig:mfa:Ns40_dU_Angular:Tzero}.
Intensity contributions from
(a) the density and longitudinal spin components;
(b) transverse spin component;
(c) total scattered light intensity.
The scattered intensity increases with increasing magnetization near the perpendicular direction
$\theta \sim \pi/2$ because, due to the collective modes,
the transverse spin component dominates the scattered light.
}
\label{fig:intensity:rpa:Ns40_dM_Angular}
\end{figure}

The intensity calculations based on MFT fail to capture the effects of collective excitations.
We showed in Sec.~\ref{sec:RPA-Chi} how the collective modes emerge in RPA. (See Figs.~\ref{fig:RPA_poles20} and~\ref{fig:RPA_chi20} for comparisons of the RPA and MFT.) The difference between the two approaches in the scattered intensity distribution is illustrated in Fig.~\ref{fig:intensity:mfa_vs_rpa:Ns40_dM_Angular}. The low-energy collective modes are notable in the transverse spin correlations, corresponding to the spin-flip transitions, but in the case of the spin-conserving scattering processes the two approaches yield almost identical intensity distributions.
The scattered light intensity distributions based on the calculation of the atomic correlations within RPA at $T=0$ for different $U$ are shown in Fig.~\ref{fig:intensity:rpa:Ns40_dM_Angular}. The variation of the signal as a function of $U$ and the AFM order are most notable in the perpendicular direction,
compared with the MFT case of Fig.~\ref{fig:mfa:Ns40_dU_Angular:Tzero}.

Finally, to produce an example how the scattered signal also depends on the temperature of the system, we show the calculated light intensity distributions for different $T$ based on the MFT
Eq.~\eqref{eq:intensity-inelastic-general} [the finite temperature factors are presented in
Eq.~\eqref{eq:corr:inelastic} in Appendix~\ref{app:mfa_susceptibilities}]. Although the collective excitations in this example are ignored, the comparison between the MFT and RPA $T=0$ results suggests
that the scattered intensity is less sensitive to low-energy collective excitations in the near-forward direction where the the spin-conserving transitions are dominant.
A more accurate description of the temperature dependence would require a finite-temperature version of RPA which is beyond the scope of the present study.

\begin{figure}
  \centering
\includegraphics[width=0.45\textwidth]{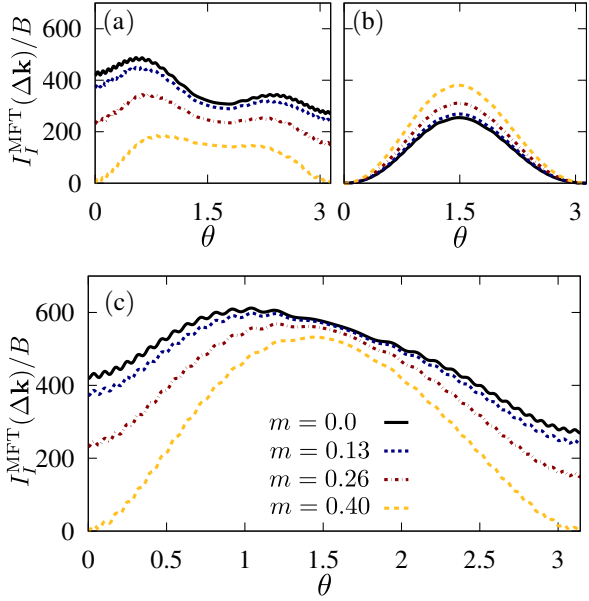}
  \caption{Angular distribution of the inelastically
scattered light intensity at different temperatures along the direction  $\phi=\pi/4$.
The calculations are based on MFT. Here the on-site interaction is fixed at
 $U=5.3J$. This means that lower $T$ corresponds to higher values of $m$.
We use the same parameters as in Fig.~\ref{fig:mfa:Ns40_dU_Angular:Tzero}.
The intensity contributions from
(a) the density and longitudinal spin components;
(b) transverse spin component;
(c) total scattered light intensity.
An increase in temperature enhances
scattering in the near-forward direction.}
  \label{fig:mfa:Ns40_dM_Angular}
\end{figure}

We find in Fig.~\ref{fig:mfa:Ns40_dM_Angular}(a) a significant dependence of the light intensity on the
temperature $T$ of the atoms in the near-forward direction that is analogous to the temperature sensitivity of a
single-component noninteracting fermionic gas~\cite{ruostekoski:170404}. (Note that in this figure,
lower $T$ is represented by increased staggered magnetization $m$, for a given fixed value of $U=5.3J$.) The suppression of small-angle
scattering at low $T$ can be understood in terms of the Fermi blocking: the scattering events in which an atom
would recoil to an already occupied state are forbidden and in MFT the small-momentum recoil events can take
atoms out of the Fermi sea only near the Fermi surface. Owing to the sensitivity of the signal to temperature,
optical diagnostics could be used as a thermometer of the atoms also for an interacting two-component case.
We study the detection accuracy of this method in Sec.~\ref{sec:measurement_accuracy}.

\subsubsection{Inelastic losses to higher bands}
\label{subsec:K40:scattered_intensity:higher_bands}

\begin{figure}
  \centering
  \includegraphics[width=0.47\textwidth]{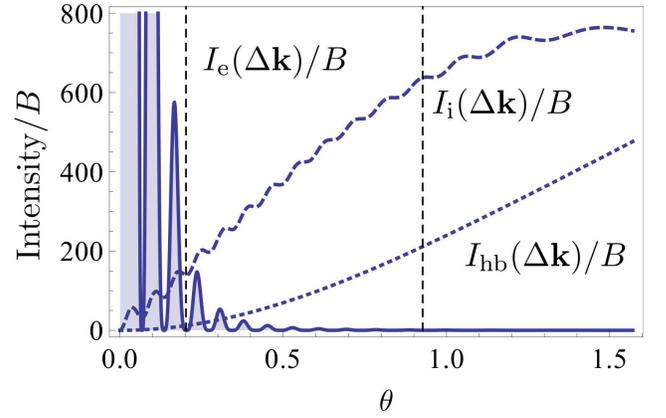}
  \caption{
Comparison of the angular distribution of the
different components of the
scattered light intensity
at $T=0$ along the direction $\phi=\pi/4$.
Here the on-site interaction
strength $U=1.76J$ and the order parameter $m=0.16$ ($m_\text{RPA}\simeq 0.13$).
We use the same parameters as in Fig.~\ref{fig:mfa:Ns40_dU_Angular:Tzero}.
MFT elastic component (solid),
RPA inelastic intraband component (dashed) and
inelastic interband component (dotted).
We schematically represent the position of a block and a NA=0.8 lens by vertical lines.
}
  \label{fig:intensity_angular_Vs_elastic}
\end{figure}

The detected light intensity consists of the elastic and inelastic intraband components that were calculated for $^{40}$K in Secs.~\ref{subsec:K40:scattered_intensity:elastic} and~\ref{subsec:K40:scattered_intensity:intraband}. In addition, the atoms can scatter to higher bands as demonstrated in Sec.~\ref{subsec:higherband}. The interband scattering can be separated from the detected signal (owing to the different frequency of the photons), but still contributes to the heating rate of the atoms. For the level structure of $^{40}$K we can write the intensity of the scattered light corresponding to the interband transitions, Eq.~\eqref{eq:intensity:losses}, as
\begin{equation}
\label{eq:intensity:losses:explicit}
I_\text{hb}(\Delta \vec{k}) = B N_s^2\left(1-\alpha_{\Delta\vec{k}}\right)
\frac{192+ 10 \cos 2\theta}{121} \:.
\end{equation}
We show the angular distribution in Fig.~\ref{fig:intensity_angular_Vs_elastic}. The inelastic interband scattering is proportional to $1-\alpha_{\Delta \vec{k}}$, where  $\alpha_{\Delta \vec{k}}$ denotes the Debye-Waller factor [Eq.~\eqref{eq:alpha_factor}]. On the other hand, the intraband inelastic scattering is proportional to $\alpha_{\Delta \vec{k}}$. %
Consequently, in deep lattices atoms are more strongly confined and the loss rate to higher bands can be suppressed, while the inelastic intraband scattering, that provides information on correlations, can be enhanced.

\section{Detection of scattered light}
\label{sec:detection}

\subsection{Optical components for light detection}
\label{sec:optical_componets}
As discussed in Sec.~\ref{sec:optical_signatures}, the elastically and inelastically scattered light from the atoms can provide different information on the AFM order parameter $m$,
the temperature of the system, and collective and single particle excitations. Here in this subsection, we
discuss in sequence how one might optimize the experimental configurations for i) detection of the
extra intensity peak due to AFM ordering and the measurement of the AFM order parameter $m$,
ii) temperature measurement, and iii) measuring the effects of atomic correlations on inelastically scattered light.

\begin{figure}[h,t,b]
\centering
\includegraphics[width=0.5\textwidth]{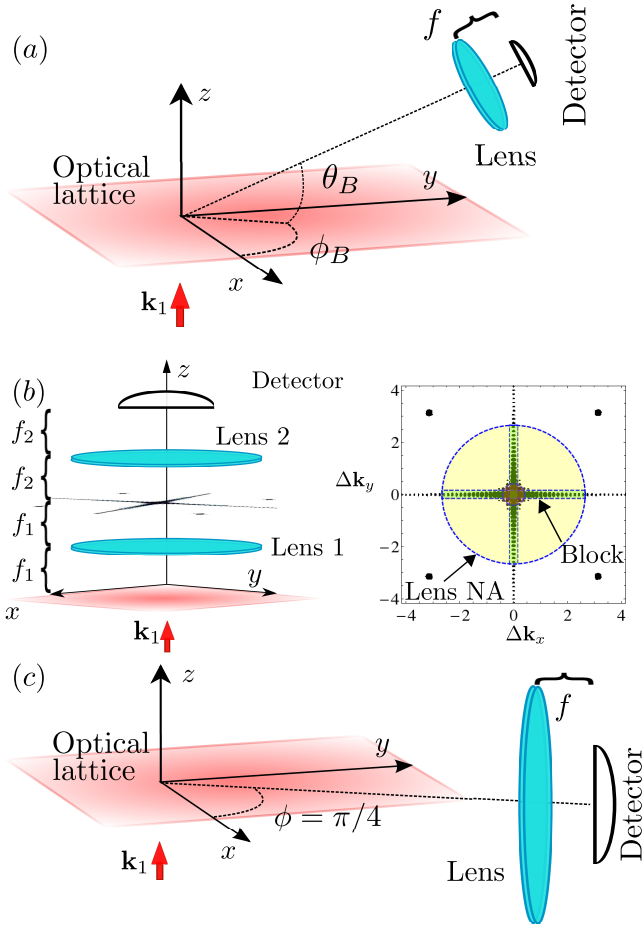}
\caption{Schematic illustration of the experimental configurations to detect AFM ordering of fermionic atoms in
an optical lattice. The atoms are confined in the 2D optical lattice close to the $z=0$ plane and the incident light
propagates towards the positive $z$ direction. In (a) the elastically scattered light corresponding to the emerging
additional Bragg peak generated by the AFM ordering is collected by a small lens. In (b) the setup is closely
related to that of Ref.~\cite{ruostekoski:170404}. The two lenses have focal lengths $f_1$ and $f_2$. The light
scattered in the near-forward direction is first collected by  lens 1. In the focal plane the scattered light is
selectively stopped by a block in order to suppress the intensity of the elastically scattered light at the detector.
The shape and the size of the block can be optimized for different measurements  of collective excitations or of
temperature. In (c) the scattered light  is collected near the perpendicular scattering direction of $\theta=\pi/2$,
 to measure transverse spin correlations.
}
\label{fig:experimental_setups}
\end{figure}

The elastically scattered light intensity in the presence
of the AFM ordering contains extra peaks that result from the period doubling of the atom density
when  atoms in only one of the two spin states are measured [Eqs.~\eqref{eq:intensity:elastic} and~
\eqref{eq:rho-AFM}]. We consider detection of this emerging ${\bf Q}$ peak by a small optical lens when
the lens is placed at the appropriate angle so as to maximize the contribution of the peak [Fig.~\ref{fig:experimental_setups}(a)]. This then can be used to to measure
directly the AFM order parameter $m$.

In order to detect atomic correlations or to measure temperature of the system from inelastically scattered light, we consider two experimental
configurations: i) a lens is placed in the near-forward direction, as depicted in
Fig.~\ref{fig:experimental_setups}(b), and ii)
a lens is centered close to the perpendicular direction, see Fig.~\ref{fig:experimental_setups}(c).
The total light intensity collected by the lens (L) with a given NA can be obtained by integrating the scattered intensity, Eqs.~\eqref{eq:intensity:elastic1}-\eqref{eq:intensity:explicit},
over the solid angle ($d\Omega=\sin\theta d\theta d\phi$) determined by the corresponding scattering angles,
\begin{equation}
\label{eq:I:L:alpha}
     \mathfrak{I}^{\text{L}}_{\alpha}(m)= \int_{\text{L}}{d\Omega\, I_{\alpha}(\Delta \vec{k},m) } \; ,
\end{equation}
where $\Delta \vec{k}$ denotes the change of momentum upon scattering [Eq.~\eqref{eq:Deltak-def}] and the index  $\alpha=\text{e},\text{i},\text{hb}$ refers to the elastic, inelastic intraband or inelastic interband (higher band) components, respectively Eqs.~\eqref{eq:intensity:elastic1},~\eqref{eq:intensity:explicit} and~\eqref{eq:intensity:losses:explicit1}.%

We have already shown in Sec.~\ref{subsec:K40:scattered_intensity:intraband} that an increase in temperature leads to enhanced
inelastic scattering in the near-forward direction, see Fig.~\ref{fig:mfa:Ns40_dM_Angular}. However, a lens placed near
the forward direction will also capture the much stronger elastic scattering signal, see, e.g., Fig.~\ref{fig:intensity_angular_Vs_elastic}.
Now, the elastically scattered light generates a diffraction pattern [Eq.~\eqref{eq:UDeltaK}] and, for the
subwavelength lattice spacing we consider, only the zeroth order Bragg peak is collected by the lens. Thus, in order to maximize the proportion of inelastically scattered light in the
measurement, we can block the high-intensity regions of the elastically scattered light by placing an appropriately
designed block on the focal plane of the lens, see Fig.~\ref{fig:experimental_setups}(b).
This setup is similar to the one proposed in
Ref.~\cite{ruostekoski:170404} to measure the temperature of an ideal single-species fermionic gas in a lattice.
In that case the inelastically scattered light varied as a function of the temperature in the near-forward direction resulting from the enhancement of the Pauli blocking effect at low temperatures. In the two-species interacting case within the MFT we observe the same behavior as shown in Fig.~\ref{fig:mfa:Ns40_dM_Angular}.
We correspondingly achieve the optimized detection efficiency by selecting a narrow cross-shaped block that covers the Bragg peak and high-intensity regions along the principal axis of the lattice, analogously to Ref.~\cite{ruostekoski:170404}, as shown in Figs.~\ref{tau:signal:composite:lens:forward:crossblock1peak:mft}(a) and (b).

We now consider how to optimize the detection of AFM correlations in the inelastically scattered light.
At  zero temperature, the effect of the AFM ordering on the angular distribution of the scattered light  calculated
with RPA  is displayed in Fig.~\ref{fig:intensity:rpa:Ns40_dM_Angular}. The staggered magnetization  most
noticeably changes the signal  away from the forward direction. Optimal shape and size of the block can be
estimated from the angular distribution of scattered light that shows both the elastic and inelastic contributions
(see Fig.~\ref{fig:intensity_angular_Vs_elastic}). Since near-forward scattering provides only little information on magnetization in this case, it is beneficial to consider a block that combines a narrow cross block with a circular block located at the center. The light can then be collected with a lens of large NA. For a lattice of 40$\times$40 sites the block sizes used are listed in Table~\ref{table:block_parameters}. In Fig.~\ref{fig:experimental_setups}(b) we show an example circular block with the angular size $\Theta_\text{Circ-Block}\approx 0.20$ rad.
\begin{table}
\caption{
Parameters of the block used to estimate the measurement accuracy estimates
shown in Fig.~\ref{tau:signal:composite:lens:forward:crossblock1peak_circular2peaks}.
}
\label{table:block_parameters}
\begin{ruledtabular}
\begin{tabular}{ccc}
$\kappa$ & Cross width (rad)& Circular radius(rad)
\\\hline
$0.66$ &$0.08$ & $0.22$
\\
$1.05$ & $0.05$ & $0.14$
\\
$1.5$ & $0.03$  & $0.10$
\end{tabular}
\end{ruledtabular}
\end{table}

As has been discussed in Sec.~\ref{sec:optical_signatures}, spin-exchanging scattering processes dominate the inelastic signal near the perpendicular direction [Fig.~\ref{fig:experimental_setups}(c)].
For $\kappa<\sqrt{2}$ the elastic scattering contribution to the collected signal is negligible (Fig.~\ref{fig:intensity_angular_Vs_elastic}) and no block is needed.
On the other hand, for $\kappa\geq \sqrt{2}$ the elastic component near the perpendicular  direction
strongly depends on the AFM order
parameter $m$ [see Fig. \ref{fig:magnetic_bragg_peak}, and Sec.~\ref{subsec:K40:scattered_intensity:elastic}, in particular Eqs.~\eqref{eq:intensity:elastic} and~\eqref{eq:intensity:elastic:explicit:downdown}].
As we will show in next subsection, this  enhances  the sensitivity of the signal to changes in $m$.

\subsection{Measurement accuracy}
\label{sec:measurement_accuracy}

We will analyze the accuracy of the optical measurements of the AFM correlations in the lattice when the light is collected by a lens. We will follow the procedure introduced in Ref.~\cite{ruostekoski:170404}
where the optical detection accuracy of temperature in a single-species fermionic atomic gas in an optical lattice was calculated.
In an inelastic scattering event an atom scatters to a different quasimomentum state owing to the photon recoil kick. Inelastic scattering leads to heating of the atomic gas and perturbs the many-body state of the atoms. In order to limit the effect of heating, the number of inelastic scattering events in each experimental realization of the lattice system should be limited to a small fraction of the total number of atoms in the lattice. We set the maximum number of allowed inelastic scattering events to be $W$, so that $W/N_s^2$ is sufficiently small. In the example analysis we take $W/N_s^2=0.1$.
We specify the fraction of inelastically scattered photons that are collected by the lens and constitute the measured signal for a given magnetization $m$ by $\eta(m)$,
\begin{equation}
\eta(m)=
\frac{
    \mathfrak{I}^{\text{L}}_{\text{i}}(m)
  }{
       \mathfrak{I}^{\text{tot}}_{\text{i}}(m)
  +
  \mathfrak{I}^{\text{tot}}_{\text{hb}}
    }\,.   \label{eq:eta-m}
\end{equation}
Here $ \mathfrak{I}^{\text{tot}}_{\text{i}}(m)$ and $\mathfrak{I}^{\text{tot}}_{\text{hb}}$ denote the total rate of inelastic intraband and interband scattering events, respectively.  We assume that the scattered light corresponding to interband transitions is filtered out of the signal so the corresponding rate is excluded from the numerator. If, for simplicity, we assume a 100\% photon detector efficiency, the number of detected inelastically scattered photons in each experimental realization of the lattice system is given by $N^{\rm i}_c(m) = \eta(m) W$. If the lattice system is prepared and the experiment is repeated $\tau$ times, we find that the total number of detected photons is
\begin{align}
\tau N_c(m) &=\tau [N^{\rm i}_c(m) + N^{\rm e}_c(m)], \\ N^{\rm e}_c(m) &= {\mathfrak{I}^{\text{L}}_{\text{e}}(m) \over \mathfrak{I}^{\text{L}}_{\text{i}}(m)} N^{\rm i}_c(m)
\end{align}
where $N^{\rm e}_c(m)$ denotes the total number of detected elastically scattered photons in a single experimental realization of the lattice system and $\mathfrak{I}^{\text{L}}_{\text{e}}(m)$ is the scattering rate of elastically scattered photons that are collected by the lens. In optical diagnostics of AFM ordering one would need to distinguish in the scattered light signal two fermionic states in the lattice that exhibit different magnetic orderings $m_1$ and $m_2$. After $\tau$ experimental realizations the difference in the total number of detected photons between the two ordered states is $\tau [N_c(m_2) -N_c(m_1)]$. The minimum requirement for the two states to be distinguishable is that this difference is at least equal to the photon shot-noise $\sqrt{\tau N_c(m_2)}$, so that $\tau [N_c(m_2) -N_c(m_1)]\agt \sqrt{\tau N_c(m_2)}$. Thus the minimum number of experimental realizations $\tau_{\rm min}$ required to distinguish between the optical signal from two magnetization states $m_1$ and $m_2$ approximately satisfies
\beq
\label{eq:tau:min}
\tau_{\rm min}\simeq \frac{N_c(m_2)}{\left[N_c(m_2)-N_c(m_1)\right]^2}\,.
\eeq

\begin{figure}
  \centering
  \includegraphics[width=0.45\textwidth]{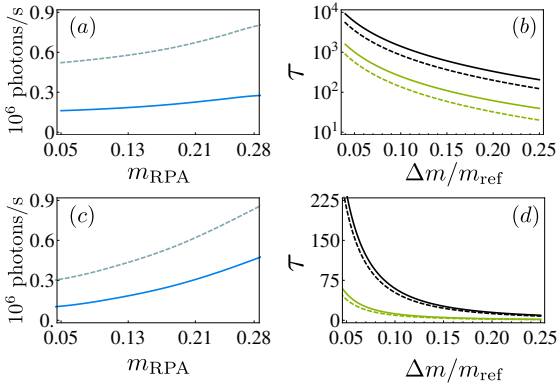}
  \caption{
Measurement accuracy at $\kappa = 1.5$ when light along the direction of the emerging magnetic Bragg peak  is detected. The calculations are based on RPA susceptibilities at $T=0$ with an RPA corrected order parameter $m_{\text{RPA}}$ for the elastic component.
We show the collected photon rates vs. the RPA corrected AFM order parameter $m_\text{RPA}$ (left column) and
the number of experimental realizations $\tau$ to achieve a relative accuracy $\Delta m/m_{\text{ref}}$ (right column).
In all the plots (a)-(d), dashed curves correspond to the configuration of a lens  with NA$=0.4$
pointing in the perpendicular direction [Fig.~\ref{fig:experimental_setups}(c)], and
solid curves are for a small lens with NA$=0.2$ pointing in the direction of the emerging magnetic Bragg
peak [Fig.~\ref{fig:experimental_setups}(a)]. Note that the lens in the perpendicular direction also
collects the signal from two magnetic Bragg peaks.
(a), (b) show the case when both spin species are detected.   (c), (d) show the case when
only the $\ket{\downarrow}$  atoms are detected. Far fewer experimental realizations are needed for
a given accuracy when only one species is detected.
In (b) and (d) we show the $\tau$ values for two different reference states:
$m_\text{ref}\simeq 0.12$ in black (top two curves) and $m_\text{ref}\simeq0.19$ in green (bottom two curves).
}
  \label{tau:signal:composite:lensQpeak}
\end{figure}

\begin{figure}
\centering
\includegraphics[width=0.5\textwidth]{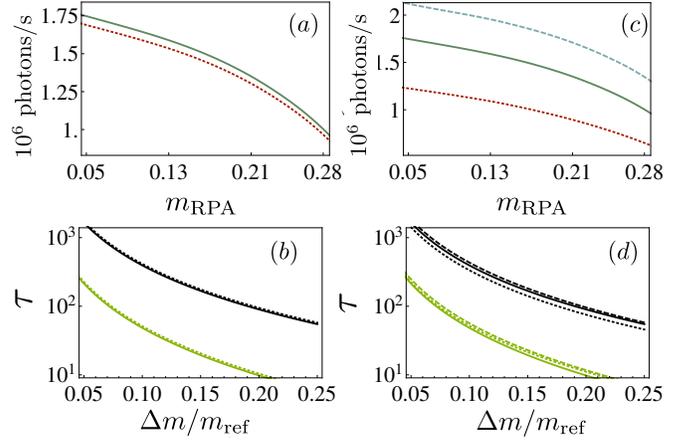}
\caption{
$T=0$ collected photon rates vs. the RPA corrected AFM order parameter $m_\text{RPA}$ with a lens of NA$=0.8$ in the forward direction (top row) and the corresponding estimated number
of experimental realizations to achieve a relative accuracy $\Delta m/m_{\text{ref}}$ (bottom row).
The calculations are computed with RPA susceptibilities at $T=0$ and with an RPA corrected order parameter
$m_{\text{RPA}}$ for the elastic component.
In both (a) and (b), we show a fixed value of $\kappa=1.05$ and compare two different lattice depths
$s=7.8$ (dotted) and $s=25$ (solid). In (b), the top two curves (in black) are for $m_\text{ref}\simeq 0.12$ and
the bottom two curves (in green) are for $m_\text{ref}\simeq 0.19$.
The top two curves are essentially on top of  one another, similarly for the bottom two curves.
 In both (c) and (d) we show a fixed lattice depth $s=25$ and compare three different values of the
 parameter $\kappa=0.66$ (dotted line), $\kappa=1.05$ (solid) and $\kappa=1.5$ (dashed).
 Note that varying $\kappa$ changes the width of the elastic diffraction peak.  Thus, to block out the main elastic diffraction peaks, a  different  block
width is required for  each $\kappa$ value, see Table~\ref{table:block_parameters}.
In (d), the top three curves (in black) are for $m_\text{ref}\simeq 0.12$ and the bottom three curves (in green) are
for $m_\text{ref}\simeq 0.19$. The top three curves are close together, and even more so for the bottom three curves.
}
\label{tau:signal:composite:lens:forward:crossblock1peak_circular2peaks}
\end{figure}

\begin{figure}
\centering
\includegraphics[width=0.5\textwidth]{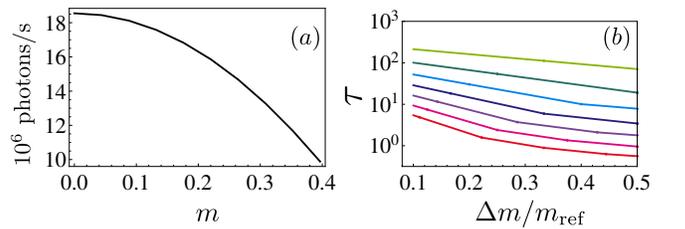}
\caption{
Finite temperature plots of (a) collected photon rates vs. the MFT AFM order parameter $m$ with a lens of NA$=0.8$ in the forward direction,
 and (b) the corresponding estimated number
of experimental realizations to achieve a relative accuracy $\Delta m/m_{\text{ref}}$.
Results are obtained with MFT at finite $T$ with fixed $U=5.3J$, and with $s=25$ and $\kappa=1.05$.
The cross block width is $0.048$ rad.
At fixed $U$,  increasing $T$ leads to decreasing $m$, and
in (b), we show from top to bottom (corresponding to lowering $T$), $m_{\text{ref}}=0.13,0.18,0.22,0.26,0.31,0.35,0.4$.
In (b)  the data points are joined with straight lines.
}
\label{tau:signal:composite:lens:forward:crossblock1peak:mft}
\end{figure}

\begin{figure}
  \centering
\includegraphics[width=0.5\textwidth]{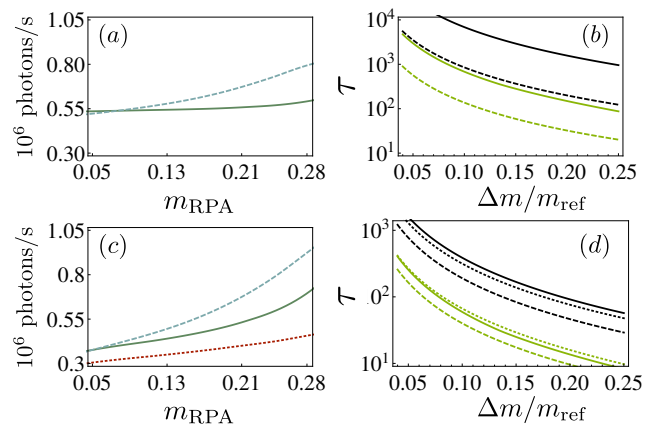}
  \caption{
Collected photon rates vs. the AFM order parameter $m$ with a lens pointing in the perpendicular direction (left column) and the
corresponding estimated number of experimental realizations $\tau$ to achieve a relative accuracy $\Delta m/m_{\text{ref}}$
(right column). The calculations are computed with RPA susceptibilities at $T=0$ and with an RPA corrected order parameter
$m_{\text{RPA}}$ for the elastic component.
We compare the case when [(a) and (b)] all the density and spin components are collected by a lens with NA$=0.4$, to
the case when  [(c) and (d)] only the transverse spin component of the scattered light is collected by a lens with NA$=0.5$.
In both (a) and (b), solid lines are for $\kappa=1.05$  and dashed lines for $\kappa=1.5$.
In (b), the top two curves (in black) are for $m_\text{ref}\simeq 0.12$ and bottom two curves (in green) are for $m_\text{ref}\simeq 0.19$.
Note that the bottom black dashed curve  almost overlaps the top green solid curve.
In both (c) and (d), dotted lines are for  $\kappa=0.66$, solid lines for  $\kappa=1.05$  and dashed lines for $\kappa=1.5$.
In (d), the top three curves (in black) are for $m_\text{ref}\simeq 0.12$ and bottom three curves (in green) are for
$m_\text{ref}\simeq 0.19$.
}
 \label{tau:signal:numap05:lens:perpendicular:s_plusminus_only}
\end{figure}

In the rest of this Section, we present results for the rate of detected photons as a function of the staggered magnetization $m$
and the number of experimental realizations $\tau$ required to determine changes in the AFM order parameter $m$. The changes are calculated with respect to a reference value, $m_{\rm ref}$, and for
a given relative accuracy $\Delta m/m_{\rm ref}$ [Eq.~\eqref{eq:tau:min}].
At $T=0$, the inelastic scattering is calculated using the RPA susceptibilities.
In Figs.~\ref{tau:signal:composite:lensQpeak},~\ref{tau:signal:composite:lens:forward:crossblock1peak_circular2peaks} and~\ref{tau:signal:numap05:lens:perpendicular:s_plusminus_only} $m_\text{ref}$ is the RPA corrected order parameter (Sec.~\ref{sec:RPA-m}) and also the elastic component of the scattered
light [Eqs.~\eqref{eq:intensity:elastic} and ~\eqref{eq:intensity:elastic:explicit:downdown}] has been computed using the
RPA corrected order parameter $m_\text{RPA}$ (Sec.~\ref{sec:RPA-m}).
At finite temperature we use only the MFT results.
For the temperature dependent MFT results of Fig.~\ref{tau:signal:composite:lens:forward:crossblock1peak:mft} the $m_\text{ref}$ is the MFT order parameter
obtained from the solution of Eq.~\eqref{eq:order_parameter}.
Except the scaling analysis of $\tau$ with lattice size, all the results in this Section are for a lattice of size 40$\times$40.

We first study the AFM order parameter measurement accuracy when a lens  (NA$=0.2$) is used to collect the light along the direction of the emerging magnetic Bragg peak
[Fig.~\ref{fig:experimental_setups}(a)]. We also study the configuration in which a lens (NA$=0.4$) is placed perpendicular to the propagation direction of the incident probe laser [Fig.~\ref{fig:experimental_setups}(c)]. In both cases, the ratio between the wavenumber of probe light and the effective lattice light $\kappa =1.5>\sqrt{2}$ [Eq.~\eqref{eq:kappa}]. This allows the NA$=0.4$ lens to collect
the signal also from two magnetic Bragg peaks [Eq.~\eqref{eq:rho-AFM}].
In Sec.~\ref{subsec:K40:scattered_intensity:elastic},
we have shown that the elastic signal is strongly enhanced when only the $\downarrow$
atoms  are detected. Figure~\ref{tau:signal:composite:lensQpeak}
compares this case [Fig.~\ref{tau:signal:composite:lensQpeak}(c),(d)] to
the case when both species are detected [Fig.~\ref{tau:signal:composite:lensQpeak}(a),(b)]. In both lens configurations
the number of experimental realizations $\tau$ needed to achieve a given accuracy
drops dramatically when only $\downarrow$ atoms are detected.
 Some specific values of $\tau$ for single species detection are presented in
Table~\ref{table:lensQpeak_results}. These values are (much) lower than  all other experimental
configurations to be presented later [see Table~\ref{table:optimal_block_results:ds}]. We conclude that when available,
\emph{the single-species detection scheme provides the most accurate determination of the AFM order.}
Experimentally, the single-species imaging can be realized by transferring the other spin component to a different hyperfine state~\cite{PhysRevLett.106.215301}.
\begin{table}
\caption{
Specific values of the estimated number of experimental realizations, $\tau(m_\text{ref})$
 for single species detection [Fig.~\ref{tau:signal:composite:lensQpeak}(d)] with $\kappa=1.5$.
The two lenses with
NA$=0.2$ and NA$=0.4$ are pointing in the direction of the magnetic Bragg peak [Fig.~\ref{fig:experimental_setups}(a)] and in the direction perpendicular to the incident field[Fig.~\ref{fig:experimental_setups}(c)], respectively.
 These values are for a relative accuracy of $10\%$.
Here $m_\text{ref}$ is the RPA corrected order parameter (Sec.~\ref{sec:RPA-m}).
}
\label{table:lensQpeak_results}
\begin{ruledtabular}
\begin{tabular}{cccc}
NA &  $\tau(0.08)$&  $\tau(0.12)$&$\tau(0.19)$
\\\hline
$0.2$ &$210 $ & $60 $ & $10 $
\\
$0.4$ & $200 $ &$50$ & $ 10$
\end{tabular}
\end{ruledtabular}
\end{table}

In Fig.~\ref{tau:signal:composite:lens:forward:crossblock1peak_circular2peaks} we show the effect of lattice depth $s$ [Eq.~\eqref{eq:optical_lattice_potential}] and $\kappa$ on the detection accuracy of $m$.
The scattered light is collected by a lens of NA=0.8 in the forward direction, corresponding to the experimental arrangement of Fig.~\ref{fig:experimental_setups}(b). We find that a deeper lattice generally enhances the scattering rate, but the effect of $s$ on $\tau$ in the studied cases is negligible [Fig.~\ref{tau:signal:composite:lens:forward:crossblock1peak_circular2peaks}(a),(b)]. Some example values are shown in Table~\ref{table:optimal_block_results:ds}. The number of required experimental realizations for a 40$\times$40 lattice drops rapidly when the desired accuracy is reduced and the staggered magnetization is increased.
\begin{table}
\caption{
Specific values of the estimated number of experimental realizations, $\tau(m_\text{ref})$, for two
lattice depths with $\kappa=1.05$
presented in Fig.~\ref{tau:signal:composite:lens:forward:crossblock1peak_circular2peaks}(b).
The parameters of the block are given in Table~\ref{table:block_parameters}.
Here $m_\text{ref}$ is the RPA corrected order parameter (Sec.~\ref{sec:RPA-m}).
}
\label{table:optimal_block_results:ds}
\begin{ruledtabular}
\begin{tabular}{ccccc}
 & \multicolumn{2}{c}{$\Delta m/m_\text{ref}=10\%$}& \multicolumn{2}{c}{$\Delta m/m_\text{ref}=20\%$}
\\
$s$ &  $\tau(0.12)$& $\tau(0.19)$&  $\tau(0.12)$& $\tau(0.19)$
\\\hline
$7.8$ & $410$ & $50$ & $90$ & $10$
\\
$25$  & $390$ & $50$& $90$ & $10$
\end{tabular}
\end{ruledtabular}
\end{table}

Similarly, increasing $\kappa$ enhances the rate of detected photons
[Figs.~\ref{tau:signal:composite:lens:forward:crossblock1peak_circular2peaks}(c),(d)].
Generally however, the detection accuracy is lower for larger values of $\kappa$. This is so
 because for larger $\kappa$, there are more inelastic scattering events
particularly near the ordering wavevector. Such inelastic signal contributes to inelastic losses and  heating, but is not captured by the forward direction lens considered here.

As explained previously (Sec.~\ref{subsec:K40:scattered_intensity:intraband}), we only provide a
qualitative analysis of finite temperature effects using MFT, without taking into account the collective excitations included in
RPA. Using a narrow cross-shaped block of width $0.048$ rad, the light can be collected near the
forward direction where the temperature strongly affects the scattering rate
(Fig.~\ref{fig:mfa:Ns40_dM_Angular}). Lower $T$ corresponds to stronger magnetization values and  fewer collected photons, as shown in
Fig.~\ref{tau:signal:composite:lens:forward:crossblock1peak:mft}(a), and the temperature changes in $m$ can be accurately detected [Fig.~\ref{tau:signal:composite:lens:forward:crossblock1peak:mft}(b)]. For example,  $m_{\text{ref}}=0.18$ can be measured with the accuracy of $10\%$ with $100$ realizations.

Finally in Fig.~\ref{tau:signal:numap05:lens:perpendicular:s_plusminus_only}, we show the RPA results for the accuracy in $m_\text{RPA}$ when
the scattered light is collected perpendicular to the propagation direction of the incident laser [Fig.~\ref{fig:experimental_setups}(c)].
The light scattered from the spin-conserving and spin-exchanging transitions may be separated owing to the different frequency of the scattered photons (or the polarization, see the next Section). If the transitions are not separated, the measurement accuracy  in the perpendicular direction is significantly lower than, e.g., for forward direction measurements.
There is a notable improvement in the detection accuracy in the perpendicular direction when only the spin-exchanging scattering processes (representing the transverse spin correlations) are selected (for the angular distribution of the scattered light for the different components, see Fig.~\ref{fig:intensity:rpa:Ns40_dM_Angular}). For instance, for $\kappa=1.5$, $10\%$ accuracy for $m_{\text{ref}}\simeq 0.19$ can now be achieved after $50$ realizations for NA$=0.4$. By increasing the size of the lens to NA$=0.5$ this can be further improved to $40$ realizations.
In general, for the perpendicular direction the large
$\kappa=1.5$ case gives the highest number of scattered photons because of the strong enhancement of spin-exchanging scattering processes [cf. Fig.~\ref{fig:intensity:rpa:Ns40_dM_Angular:combined}(a) vs Fig.~\ref{fig:intensity:rpa:Ns40_dM_Angular}(c)]. Also the $\tau$ values are substantially lower for the $\kappa=1.5$ case [Fig.~\ref{tau:signal:numap05:lens:perpendicular:s_plusminus_only}(b),(d)].

Our example calculations are for  a 40$\times$40 lattice. Smaller values of $\tau$ can be
obtained for larger lattices. The number of required experimental realizations of the lattice system $\tau$ is approximately inversely proportional to the number of sites $\tau\propto N_s^{-2}$. We have
simulated lattices sizes between $N_s=16$ and $N_s=40$, and our  results confirm this scaling to be
qualitatively accurate for reasonably large lattice systems $N_s\agt 25$
for both forward and perpendicular directions measurements.
For smaller systems the choice of the block size
and shape can result in larger variations owing to the dependence of the width of the
diffraction peak on the lattice size.

\subsection{Distinguishability of transitions by light polarization}
\label{sec:projected_scattered_light}
\begin{figure}
\centering
\includegraphics[width=0.45\textwidth]{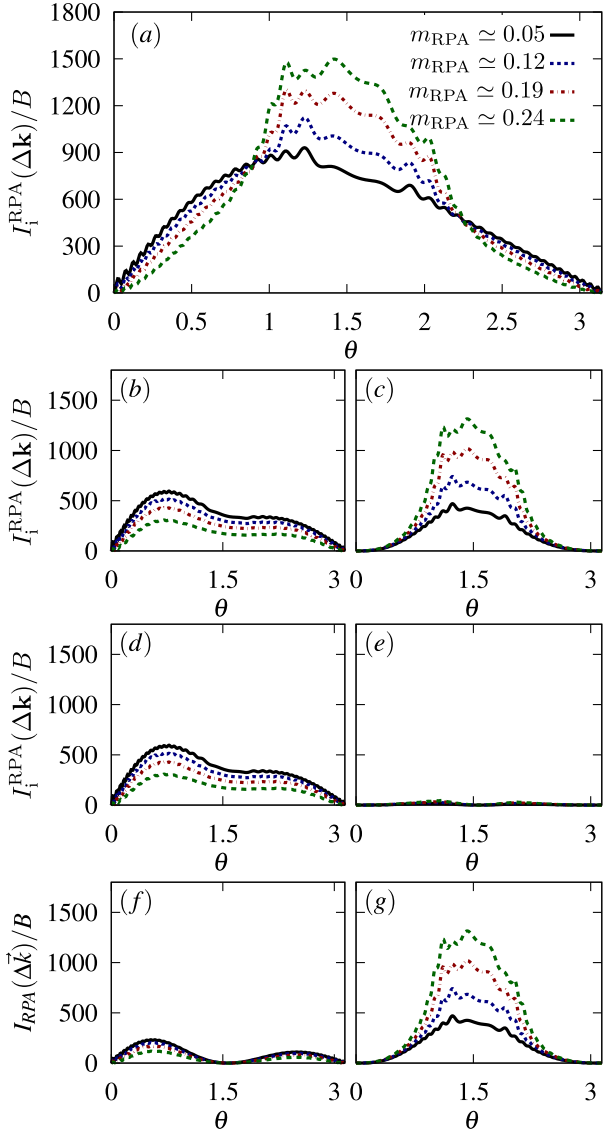}
\caption{
Angular distribution of the inelastically scattered light intensity
along the direction $\phi=\pi/4$ when the scattered light is projected along specific polarization directions. We show different values of the interaction strength $U$ at $T=0$.
The calculations are based on RPA,
$\kappa=1.5$ and the other parameters are as in Fig.~\ref{fig:mfa:Ns40_dU_Angular:Tzero}.
(a) The total scattered light intensity; (b) the total scattered intensity from the spin-conserving transition; (c) the total scattered intensity from the spin-exchanging transition;
In (d) and (e) the light has been projected along the direction of the scattered light
from the spin-conserving transition. (d) projected component of the scattered intensity from the spin-conserving transition which in this case equals to (b); (e) projected component of the scattered intensity from the spin-exchanging transition.
In (f) and (g) the light has been projected along the direction of the scattered light
from the spin-exchanging transition. (f) projected component of the scattered intensity from the spin-conserving transition; (g) projected component of the scattered intensity from the spin-exchanging transition  which in this case equals to (c).
}
\label{fig:intensity:rpa:Ns40_dM_Angular:combined}
\end{figure}
In general the scattered light corresponding to different transitions are separated in frequency space and could be identified by filtering the relevant frequencies. If the sufficient frequency resolution is not achievable the transitions could still be partially distinguished by the polarization of the scattered light. The polarization of the scattered light from the spin-conserving and spin-exchanging transitions depend on the scattering direction and generally they are not orthogonal.
For any given scattering direction we may project, e.g., the light scattered from the spin-exchanging transition to the direction of the polarization of the scattered light from the spin-conserving transitions. This provides the optimum value how much for the given scattering direction the light from the spin-conserving transition can be distinguished from the total scattered intensity. In order to analyze this we modify the polarization vector of Eq.~\eqref{eq:lambda} $\mathbf{\Lambda}_{g'g}$ by
\begin{equation}
\label{eq:lambda:projected}
\mathbf{\Lambda}_{g'g}\to\mathbf{\Lambda}_{g'g}|_{\hat{\epsilon}_{g_1g_2}}= \mathbf{\Lambda}_{g'g}\cdot\mathbf{\hat{\epsilon}}_{g_1g_2}\mathbf{\hat{\epsilon}}_{g_1g_2}\,,
\end{equation}
where the polarization of the scattered is given by
\begin{equation}
\hat{\epsilon}_{g'g}=\frac{\mathbf{\Lambda}_{g'g}}{\left|\mathbf{\Lambda}_{g'g}\right|}\,.
\label{eq:scattered_field_polarization}
\end{equation}
In a studied example case of Fig.~\ref{fig:intensity:rpa:Ns40_dM_Angular:combined} we show the angular dependence of the scattered light intensity and compare this with the projected intensities along the polarization of the scattered light from either of the two transitions. The relative part of the signal from the density and longitudinal spin correlations is enhanced in Figs.~\ref{fig:intensity:rpa:Ns40_dM_Angular:combined}(d) and~(e), while the contribution of the transverse spin correlations is particularly strong in Figs.~\ref{fig:intensity:rpa:Ns40_dM_Angular:combined}(f) and~(g) in the direction around $\theta\simeq \pi/2$.

\section{Diagnostics of excitations from scattered spectrum}
\label{sec:scattered_spectrum}

In this Section we calculate the spectrum of scattered light and show how intraband inelastic scattering can reveal the single-particle and collective excitations in an AFM ordered lattice system.
In Sec.~\ref{subsec:spectrum}, the scattered spectrum is split up into an elastic [Eq.~\eqref{eq:spectrum-elastic-general}] and an inelastic part
[Eq.~\eqref{eq:spectrum-inelastic-general}].  The elastically scattered light has no nontrivial spectral structure, consisting of only the zero frequency part, by definition.
Inelastic losses due to scattering to higher bands
occur at high frequency (on the order of the band gap) and can therefore be filtered out and
ignored, as this part does not contain information about the state probed.

The inelastic spectrum of Eq.~\eqref{eq:spectrum-inelastic-general},
or the more specific form
of Eq.~\eqref{eq:spectrum-inelastic-general:mathcalS}, contains
useful information about the excitation spectrum of the system being probed.
 In Sec.~\ref{sec:RPA-Chi}, we show how the MFT
contains only single particle excitations in the density, longitudinal and transverse spin susceptibilities.
The RPA partially takes into account quantum fluctuations around the AFM ordered state
 and can  capture the collective excitations (spin waves) emerging in the transverse spin susceptibility.
 RPA also renormalizes the single particle excitations.
(See Figs.~\ref{fig:RPA_poles20} and~\ref{fig:RPA_chi20}  for  comparisons of the RPA and MFT.)
Via linear response theory (see Sec.~\ref{sec:retarded-Torder}), the various RPA susceptibilities can be related to
 the dynamic structure factor
Eq.~\eqref{eq:dynamic_structure_factor}, or the dynamic response function Eq.~\eqref{eq:dynamic_response_function}. The spectrum of inelastically scattered light reads
\begin{widetext}
\begin{align}
\notag
\frac{\mathbb{S}_{\text{i}}(\Delta\vec{k},\omega)}{\alpha_{\Delta\vec{k}}B}=&
\frac{1}{4}\left({\sf M}_{\downarrow\downarrow}^{\downarrow \downarrow}+{\sf M}_{\uparrow\uparrow}^{\uparrow \uparrow}\right)
 \sum\limits_{\vec{q}\neq0}^{\text{RBZ}}
\pmb{u}^\dagger_{\bar{\Delta\vec{k}}-\vec{q}}
\left[
 \pmb{\mathcal{S}}^{\rho\rho}(\vec{q},\omega)+
\pmb{\mathcal{S}}^{zz}(\vec{q},\omega)
\right]\pmb{u}_{\bar{\Delta\vec{k}}-\vec{q}}
\\%\notag
&+\frac{1}{4}\left({\sf M}_{\downarrow\downarrow}^{\uparrow \uparrow}+{\sf M}_{\uparrow\uparrow}^{\downarrow \downarrow}\right)
 \sum\limits_{\vec{q}\neq0}^{\text{RBZ}}
\pmb{u}^\dagger_{\bar{\Delta\vec{k}}-\vec{q}}
\left[
 \pmb{\mathcal{S}}^{\rho\rho}(\vec{q},\omega)-
\pmb{\mathcal{S}}^{zz}(\vec{q},\omega)
\right]\pmb{u}_{\bar{\Delta\vec{k}}-\vec{q}}
%\\
%&
+
\frac{1}{2}{\sf M}_{\downarrow \uparrow}^{\downarrow \uparrow}
 \sum\limits_{\vec{q}\neq0}^{\text{RBZ}}
\pmb{u}^\dagger_{\bar{\Delta\vec{k}}-\vec{q}}
            \pmb{\mathcal{S}}^{+-}(\vec{q},\omega)\;\pmb{u}^\dagger_{\bar{\Delta\vec{k}}-\vec{q}}\;.
\label{eq:spectrum:inelastic}
\end{align}
\end{widetext}
This spectrum has been derived using the same procedures as the analogous expression for the intensity
in Sec.~\ref{sec:40K:intensity},
generalizing  the definitions of Eq.~\eqref{eq:static-response+-} to the frequency-dependent case here.
We can translate the  dynamical response function of  Eq.~\eqref{eq:dynamic_response_function}
to the density, longitudinal and transverse spin dynamical response functions
$\pmb{\mathcal{S}}^{ij}(\vec{q},\omega)$ ($i,j = \rho, z, + , -$) using a frequency dependent version of Eq.~\eqref{eq:static-response+-}. In turn, these dynamical response functions are related
via Eq.~\eqref{eq:chi-to-structurefactorT0}  to the various RPA spin and
density susceptibilities Eqs.~\eqref{eq:chi:rhorho:RPA},~\eqref{eq:chi:zz:RPA}
and~\eqref{eq:chi:+-:0:RPA:matrix} of Sec.~\ref{sec:hubbard_model},
which are then used to compute the scattered spectrum.

We first compare in Sec.~\ref{subsubsec:compare-spectrum-lens} the physical quantities of interest,
the angle-resolved spectrum [Eq.~\eqref{eq:spectrum:inelastic}] and susceptibilities
[Eqs.~\eqref{eq:chi:rhorho:RPA},~\eqref{eq:chi:zz:RPA} and~\eqref{eq:chi:+-:0:RPA:matrix}],
with the angle-integrated spectra [Eq.~\eqref{eq:integratedspectrum}] that corresponds to a measurement of the spectrum by a lens over a range of scattering angles.
We then analyze  in Sec.~\ref{subsubsec:singleparticle} the features corresponding to single particle excitations and  in Sec.~\ref{subsubsec:collective},  the collective mode peak.

\subsubsection{Comparison between angle-resolved spectrum and angle-integrated spectrum}
\label{subsubsec:compare-spectrum-lens}

\begin{figure}
  \centering
\includegraphics[width=0.48\textwidth]{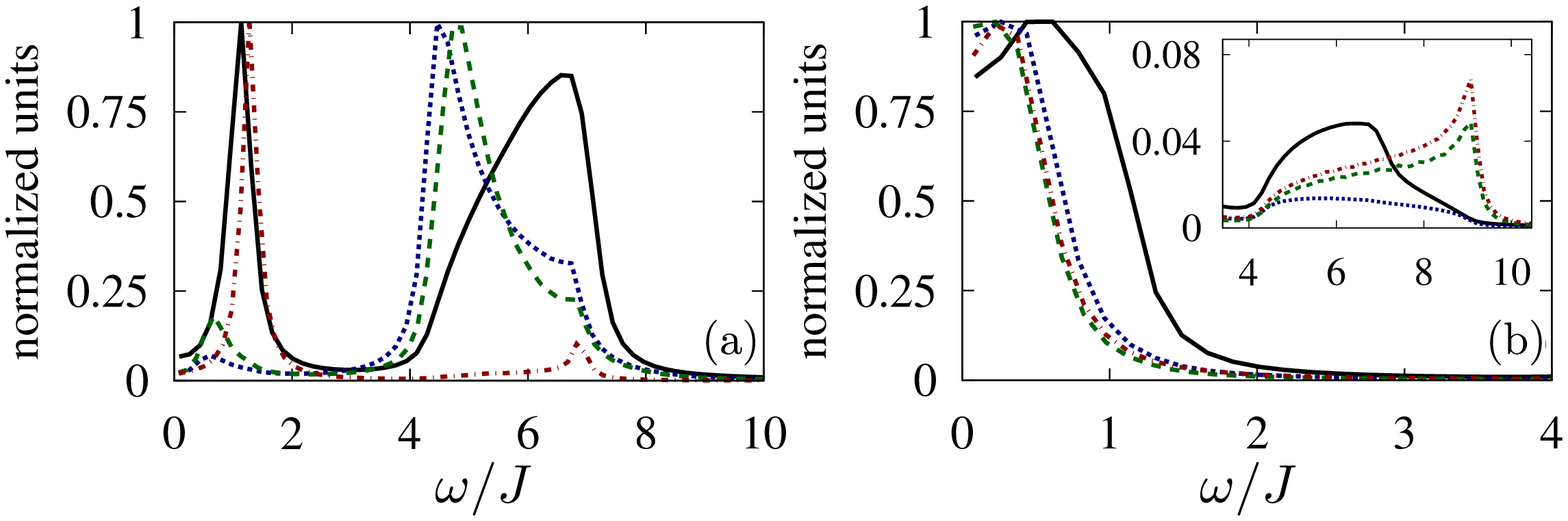}
\caption{
Comparison between the RPA susceptibilities, angle-resolved spectrum, and the spectrum collected by a lens over a range of scattering angles. The ratio between the wavenumber of the  probe light to the effective
wavenumber of the optical lattice light is $\kappa=1.5$ [Eq.~\eqref{eq:kappa}] and $U=5J$.
Each line shown has been normalized to its maximum value for comparison.
We show $\mathfrak{s}^\text{L}_{0.5}(\omega)$ [solid (black)], $\mathfrak{s}^\text{L}_{0.1}(\omega)$ [dotted (blue)], $\mathbb{S}(\Delta \vec{k},\omega)$ [dashed (green)] and $\chi(\Delta \vec{k},\Delta\vec{k};\omega)$ [dotdashed (red)].
(a) shows the configuration with a lens in the forward direction [Fig.~\ref{fig:experimental_setups}(b)].
The angle-resolved quantities, the spectrum $\mathbb{S}(\Delta \vec{k},\omega)$ and total susceptibility  $\chi(\Delta \vec{k},\Delta\vec{k};\omega)$
are computed at a wavevector close to the axis of the lens ($\theta \sim 0, \: \phi \sim 0$).
(b) shows the case for a lens in the perpendicular direction [Fig.~\ref{fig:experimental_setups}(c)].
$\mathbb{S}(\Delta \vec{k},\omega)$ and $\chi(\Delta \vec{k},\Delta\vec{k};\omega)$ are computed at a wavevector
$\Delta \vec{k}$ close to the axis of the lens at $(\theta \sim \pi/2, \phi \sim \pi/4)$.
(b) shows the collective mode at low energies and the
inset shows the single-particle excitations at energies above the gap.
}
  \label{fig:spectrum:rpa:fixpoint}
\end{figure}

The scattering rate of off-resonantly illuminated atoms is generally low. Experimentally, the number of measured photons can be increased by collecting the inelastically scattered photons over a range of
scattering angles using a lens (see Sec.~\ref{sec:optical_componets}). The spectrum, however, changes with the scattering angle, since each $\Delta \vec{k}$ represents a different argument of the dynamical structure factor. In this Section we analyze how much of the spectral structure can be extracted when the measured light is collected by a lens with a large NA and the spectrum is integrated over the corresponding range of scattering angles. We therefore define
\begin{equation}
\label{eq:integratedspectrum}
\mathfrak{s} ^{\text{L}}_{\text{NA}}(\omega) =\int\limits_{\text{L},\text{NA}} d\phi \sin\theta d\theta ~\mathbb{S}(\Delta \vec{k}(\theta,\phi),\omega) \: ,
\end{equation}
where $\text{L}=\text{F},\text{P}$ indicates the experimental configuration:
 $\text{F}$ for the lens with a given NA pointing in the forward direction [Fig.~\ref{fig:experimental_setups}(b)]
and $\text{P}$ for the perpendicular direction [Fig.~\ref{fig:experimental_setups}(c)].

In Fig.~\ref{fig:spectrum:rpa:fixpoint}, we compare
the angular integrated spectrum [Eq.~\eqref{eq:integratedspectrum}] for two different lens sizes with
the physical quantities of interest. These are the angle-resolved spectrum $\mathbb{S}(\Delta \vec{k},\omega)$ and the total susceptibility
$\chi(\Delta \vec{k},\Delta\vec{k};\omega)$, which is the sum of all the longitudinal and transverse susceptibilities. Here
$\mathbb{S}(\Delta \vec{k},\omega)$ and $\chi(\Delta \vec{k},\Delta\vec{k};\omega)$ are computed at a wavevector $\Delta \vec{k}$ close to the axis of each corresponding lens.
In Fig.~\ref{fig:spectrum:rpa:fixpoint}(a) the lens is pointing in the forward direction
($\theta \sim 0, \: \phi \sim 0$) and in Fig.~\ref{fig:spectrum:rpa:fixpoint}(b)
in the perpendicular direction ($\theta \sim \pi/2, \: \phi \sim \pi/4$).

It can be seen that the small lens measurement reproduces quite closely the  spectrum
$\mathbb{S}(\Delta \vec{k},\omega)$, in both forward and perpendicular directions.
However, compared with the large lens,  the small lens leads to the absolute magnitude of the signal being down by nearly
a factor of 200 in the forward direction and about a factor of 10 in the perpendicular direction.
But even the large lens captures well at least the position of the collective mode peak (at low energies)
and single particle peak (at high energies).
The collective mode peak is somewhat broadened by the large lens.
(These peaks are analyzed in detail in the next
subsections.) Note that unlike the total susceptibility $\chi(\Delta \vec{k},\Delta\vec{k};\omega)$,
the  spectrum $\mathbb{S}(\Delta \vec{k},\omega)$ contains the dipole matrix elements
${\sf M}^{g_3g_4}_{g_2g_1}$, Eq.~\eqref{eq:inelastic:angular:40K}, that  skews the
signal towards the forward direction for spin-preserving transitions, and towards the
perpendicular direction for spin-exchanging transitions (see Sec.~\ref{subsec:K40:scattered_intensity:intraband}). Hence one key finding here is that
{\it different lens positions can be used to
select for different types of transitions, to separate out the collective modes versus the
single particle excitations}.

\begin{figure}
  \centering
\includegraphics[width=0.5\textwidth]{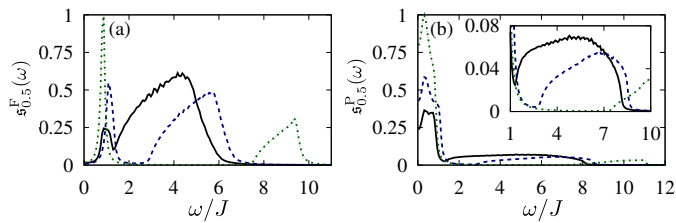}
  \caption{
RPA spectrum collected by a lens for different values of the on-site interaction strength $U$ at $T=0$.
Here the number of sites is 40$\times$40.  The lattice height is $s=25$ and $\kappa=1.5$.
Light is collected with a lens of NA$=0.5$ pointing in
(a) the forward direction [Fig.~\ref{fig:experimental_setups}(b)];
(b) the perpendicular direction [Fig.~\ref{fig:experimental_setups}(c)].
From left to right, black (solid) is $m_\text{RPA}\simeq 0.19$ ($U\approx2.6J$), blue (dashed) is $m_\text{RPA}\simeq 0.25$ ($U\approx4.0J$), and green (dot-dashed) is $m_\text{RPA}\simeq0.30$ ($U\approx8.3J$).
Inset in (b) shows the renormalized single-particle excitations starting at $\omega \approx U/J$.
All the curves have been normalized to the maximum of the $m_\text{RPA}\simeq0.30$ case.
}
  \label{fig:spectrum:rpa:lenses}
\end{figure}

\subsubsection{Single particle excitations} \label{subsubsec:singleparticle}

In Fig.~\ref{fig:spectrum:rpa:lenses}, we plot the RPA spectrum collected by a $\text{NA}=0.5$
lens  in the forward direction [Fig.~\ref{fig:spectrum:rpa:lenses}(a)]
and in the perpendicular direction [Fig.~\ref{fig:spectrum:rpa:lenses}(b)].
A broad peak at high energies can be seen  in both lens directions. This peak is due to gapped single particle excitations already included by MFT.
 The form of the MFT susceptibilities of
Eqs.~\eqref{eq:chi:rhorho:0},~\eqref{eq:chi:zz:0},~\eqref{eq:chi:+-:0} and~\eqref{eq:chi:+-:Q}
indicates that the single particle excitations spectrum is non-zero only for
$ 2 \Delta \leq \omega \leq 2 \sqrt{\Delta^2 + 16 J^2}$.   For $U\gg J$, the
gap can be approximated by $2\Delta = 2 m U \approx U$ as $m$ saturates to $1/2$.
Indeed, it can readily be seen in Fig.~\ref{fig:spectrum:rpa:lenses}
that the gap opens up progressively and more sharply with larger $U$,
and there is signal only between the aforementioned bounds.
To compute numerical values, we give a finite value to the infinitesimal imaginary part  using
$\delta=0.07J$, in the denominators
of the susceptibilities of  Eqs.~\eqref{eq:chi:rhorho:0}, \eqref{eq:chi:+-:0} and \eqref{eq:chi:+-:Q}.
Thus, individual delta functions coming from the susceptibilities are then spread out into Lorentzian
functions, representing the finite frequency resolution of spectral measurements.

\subsubsection{Collective modes} \label{subsubsec:collective}

The other main feature in Fig.~\ref{fig:spectrum:rpa:lenses}(a) and (b)  is the sharp peak at low energies.
This originates from
light scattering off the  spin-exchanging transitions that dominate the transverse
spin susceptibility $\chi^{+-}_{\text{RPA}}(\vec{q},\vec{q};\omega)$.
In turn, such transitions can be induced by the excitation of gapless collective modes, the spin wave excitations above
the AFM ground state (see Sec.~\ref{sec:RPA-Chi} and Fig.~\ref{fig:RPA_chi20}).
Notice that the sharp collective mode at low energy   separates  cleanly in
frequency from the single particle excitations only from moderately large $U/J$ onwards.
A rough criterion for this separation is that the collective mode bandwidth
[$\sim 2 J_H$, see Eq.~\eqref{eq:omega:heisenberg} and the text after] should be smaller
than the single particle gap $2 \Delta$. At small $U/J$, the single particle gap is small
and  the collective mode peak merges with the single particle broad peak, as can be seen for the case of
$U\approx 2.6J$ in Fig.~\ref{fig:spectrum:rpa:lenses}.

\begin{figure}
  \centering
\includegraphics[width=0.5\textwidth]{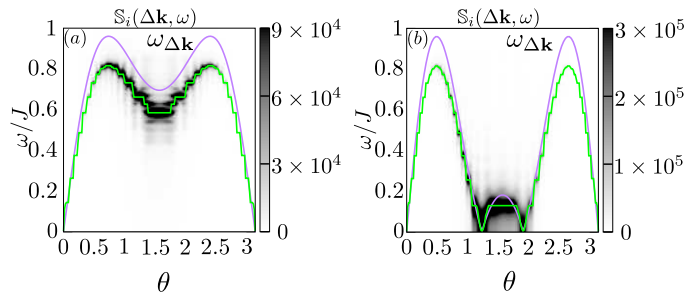}
\caption{RPA scattered spectrum at zero temperature along  $\phi=\pi/4$ for $U=8.3J$ ($m_\text{RPA}\simeq0.30$).
This figure only shows the low energy part of the spectrum that includes the collective modes.
Single particle excitations occur at much higher energies and are not shown.
The gray scale on the left gives the magnitude of the spectrum plotted.
(a) shows $\kappa=1.05$ and (b) $\kappa=1.5$. $\omega_{\Delta \vec{k}}$ (solid line) given by
Eq.~\eqref{eq:omega:heisenberg}  is the spin wave dispersion relation of the Heisenberg model.
The green (step-like) line connects the maximum values of the collective mode peak of the RPA spectrum for each $\theta$ point evaluated.
}
  \label{fig:spectrum:rpa:mobile_lens}
\end{figure}

As discussed in Sec.~\ref{sec:hubbard_model}, for large $U/J$,
the Hubbard model reduces to the Heisenberg model at low energies.
In Fig.~\ref{fig:spectrum:rpa:mobile_lens} we show the angular and frequency dependence of the
scattered light along the $\phi=\pi/4$ line for $U=8.3J$ ($m_\text{RPA}\simeq 0.30$)
and compare it with the spin wave dispersion relation of the
Heisenberg model, Eq.~\eqref{eq:omega:heisenberg}.
As a visual aid, the line connecting the maximum of the spectrum for each $\theta$ point is also
drawn. It is indeed similar to the dispersion relation curve. Such measurement could in principle be
realized with a small NA lens scanning the $\theta$ direction. Note that this figure only
shows the low energy spectrum, cutting off the higher energy
single particle excitations shown in Figs.~\ref{fig:spectrum:rpa:fixpoint} and \ref{fig:spectrum:rpa:lenses}.
 Despite the similar shape, the RPA derived dispersion occurs for smaller frequencies compared with the
Heisenberg model dispersion. This mismatch disappears for values of $U \agt 25 J$.

\section{Concluding remarks}
\label{sec:conclusions}
We have studied off-resonant imaging of AFM correlations in a two-species fermionic atomic gas in a tightly-confined 2D optical lattice. The AFM ordering represents a checkerboardlike alternating density pattern of the two species that effectively doubles the lattice periodicity of each spin component. This can be revealed in emerging magnetic Bragg peaks of elastically scattered light when only one spin component is detected.
The density correlations as well as the longitudinal and transverse spin correlations of the atoms are mapped onto the fluctuations of the scattered light where they can be detected in the inelastically scattered light. We have shown how the standard Feynman-Dyson perturbation theory of interacting many-body systems can then be related to the experimental observables of the intensity and the spectrum of the scattered light. Our specific example concerned RPA of the AFM ordering that corresponds to a partial summation of the diagrammatic perturbation series. The general principle could be adapted to other strongly-correlated states, indicating how off-resonant imaging can provide a powerful diagnostic tool in interacting ultracold atom systems.

\acknowledgments
The authors acknowledge financial support by SEPNET, EPSRC and the Leverhulme trust.
The authors acknowledge the use of the IRIDIS High Performance Computing Facility, and associated support services at the University of Southampton, in the completion of this work

\appendix

\section{Susceptibilities from Mean-field theory and RPA for the AFM via Feynman diagrams}
\label{App:RPA-Feynman}

 \begin{figure*}
\centering
\includegraphics[width=\textwidth]{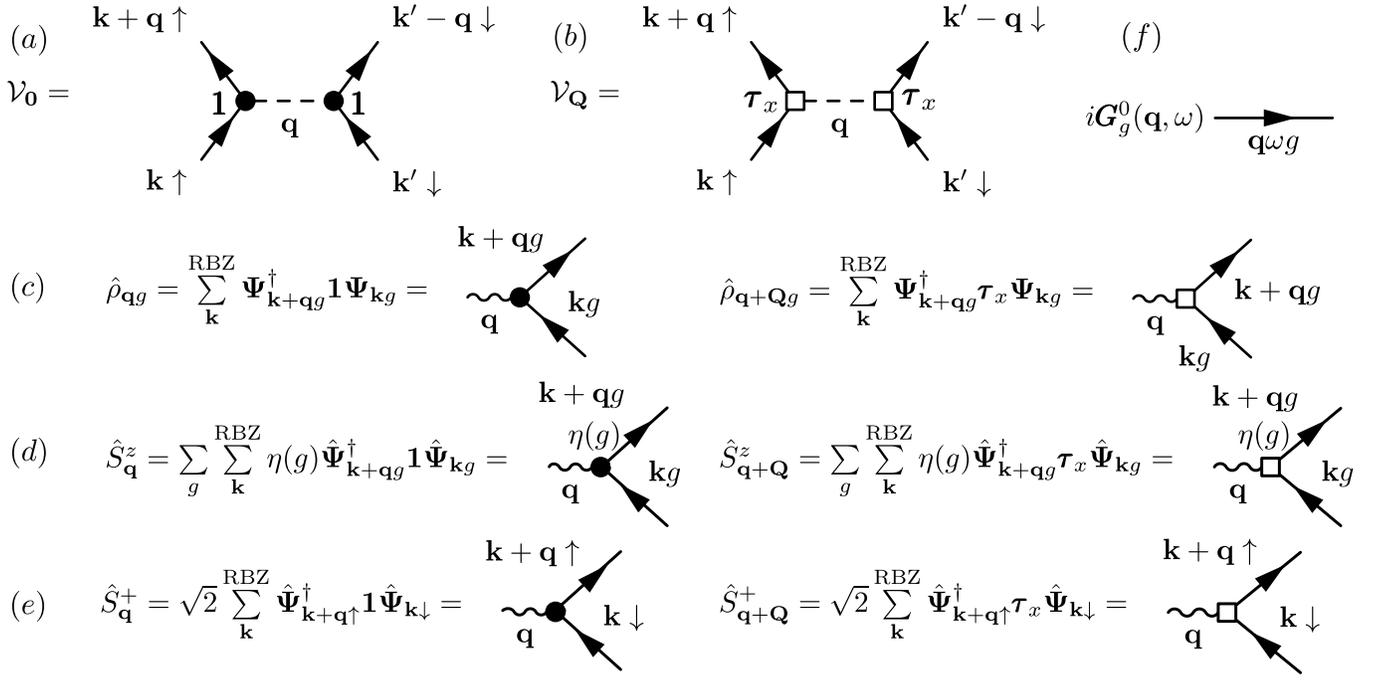}
\caption{Diagrammatic representation of the interaction vertices, spin densities, and Green's function needed for the RPA calculation of the susceptibilities. (a) and (b) interaction vertices [Eq.~\eqref{eq:RPA-H}], (c) density operator [Eq.~\eqref{eq:chargeop}], (d) spin projection on the $z$ direction [Eq.~\eqref{eq:spinSzop}],
(e) spin raising operator [Eq.~\eqref{eq:spinSPlusop}] and (f)  MFT Green's function [Eq.~\eqref{app:rpa:GM^0(q,iomega_n)}].}
\label{fig:diagram-elements}
\end{figure*}
In this Appendix, we outline the calculation of the MFT and RPA susceptibilities for density $\op{\rho}$
Eq.~\eqref{eq:chi:rhorho:RPA}, longitudinal spin $\op{S}^z$ Eq.~\eqref{eq:chi:zz:RPA}
and transverse spin $\op{S}^+ ,  \op{S}^-$ [Eqs.~\eqref{eq:chi:+-:0:RPA} and~\eqref{eq:chi:+-:Q:RPA}] using the diagrammatic
method. MFT susceptibilities (at $T=0$ or finite $T$) can also be calculated directly from the MFT
Hamiltonian [Eq.~\eqref{eq:final-MF-H}] and the Bogoliubov transformation  [Eq.~\eqref{eq:Bog-Schr}]. Here
we derive the $T=0$ MFT susceptibilities first to show briefly how the diagrammatic method works and how to generalize it to RPA.

As mentioned in Sec.~\ref{sec:mean_field_hamiltonian}, we have dropped second order fluctuations to arrive at the MFT Hamiltonian of Eq.~\eqref{eq:bogoliubov:hamiltonian}. The diagrammatic method {\it in the AFM
 ordered state} then approximately reinstates quantum fluctuations as follows:
The full Hamiltonian is a sum of the MFT one $\mathcal{H}$
[Eq.~\eqref{eq:bogoliubov:hamiltonian}] and the fluctuations (the interaction part), using the compact Nambu notation
[Eq.~\eqref{eq:Nambu}],
\begin{align} \label{eq:RPA-H}
\mathcal{H}_{\rm tot} = & \mathcal{H} + \mathcal{V}_{\vec{0}} + \mathcal{V}_{\vec{Q}} \:, \nonumber\\
\mathcal{V}_{\vec{0}} = & \frac{U}{N_s^2} \sum\limits_{\vec{q},\vec{k},\vec{k}'}^{\text{RBZ}}
\vcre{\Psi}{ \vec{k}+\vec{q},\uparrow}\vdes{\Psi}{ \vec{k},\uparrow} \:
  \vcre{\Psi}{ \vec{k}'-\vec{q},\downarrow}\vdes{\Psi}{ \vec{k}',\downarrow} \:, \nonumber \\
\mathcal{V}_{\vec{Q}} = & \frac{U}{N_s^2} \sum\limits_{\vec{q},\vec{k},\vec{k}'}^{\text{RBZ}}
 \vcre{\Psi}{ \vec{k}+\vec{q},\uparrow} {\pmb \tau}_x \vdes{\Psi}{ \vec{k},\uparrow} \:
 \vcre{\Psi}{ \vec{k}'-\vec{q},\downarrow}{\pmb \tau}_x \vdes{\Psi}{ \vec{k}',\downarrow}
 \; .
\end{align}
Here $\mathcal{V}_{\vec{0}}$ corresponds to the interaction with momentum transfer $\vec{q} \in $ RBZ,
while $\mathcal{V}_{\vec{Q}}$ has
momentum transfer $\vec{q} + \vec{Q}  $ that is represented by the Pauli matrix $ {\pmb \tau}_x$ acting
in the Nambu spinor space.

The elements of the diagram technique can now be defined; see Fig.~\ref{fig:diagram-elements}.
The interaction vertices $\mathcal{V}_{\vec{0}}$ and $\mathcal{V}_{\vec{Q}}$
are defined in Fig.~\ref{fig:diagram-elements}(a) and~(b).
The interaction is represented by a dashed line, with an associated factor of $-i U/N_s^2$.
The various density operators (total density, longitudinal and transverse spin) can be written using the
Nambu spinors, as shown in Fig.~\ref{fig:diagram-elements}(c)-(e). Diagrammatically,  a
creation (annihilation) spinor operator $ \vcre{\Psi}{}$ ($\vdes{\Psi}{}$) is represented by an
outgoing (incoming) arrow,
with its momentum and spin label near the arrow. The momentum transfer is represented by a wavy line.
Whenever the momentum transfer $\vec{q}$ is shifted by $\vec{Q}$, a Pauli matrix in Nambu space
${\pmb \tau}_x$  is needed in between $ \vcre{\Psi}{}$ and $\vdes{\Psi}{}$,
and is represented as a open square. When there is no such shift, a unit matrix
$\pmb{1}$  is represented by a filled circle.

In the spin density wave ground state $|\Phi_{\text{sdw}}\rangle$,
we can write the bare Green's function at zero temperature as a 2$\times$2 matrix, using the Nambu spinor defined in
Eq.~\eqref{eq:Nambu},
\begin{equation}
\pmb{G}^0_g(\vec{k},t'-t)
= -i  \langle \Phi_{\text{sdw}}| \mathcal{T} \vdes{\Psi}{\vec{k},g}(t')\vcre{\Psi}{\vec{k},g}(t) |\Phi_{\text{sdw}}\rangle \;.
\end{equation} From now on, we will drop the explicit ground state bra and ket. All expectation values are understood
to be for this ground state.
A matrix Green's function $i \pmb{G}^0_g$ is represented as a straight line with
momentum and spin labels shown in Fig.~\ref{fig:diagram-elements}(f). The arrow
starts from a creation operator to end in an annihilation operator.
$ \pmb{G}^0_g$ can be calculated via a Fourier transform in time, directly
from the MFT  Hamiltonian [Eq.~\eqref{eq:final-MF-H}] and the Bogoliubov transformation
[Eq.~\eqref{eq:Bog-Schr}] to give
\begin{align}
\pmb{G}^0_g(\vec{k}, \omega)
= & \int_{-\infty}^{\infty} dt \: e^{i \omega t} \: \pmb{G}^0_g(\vec{k}, t) \nonumber \\
= & \frac{1}{(\hbar \omega)^2 - E_{\vec{k}}^2 + i \delta}
\begin{pmatrix}
\hbar \omega + \epsilon_{\vec{k}} & \Delta_g \\
 \Delta_g & \hbar \omega - \epsilon_{\vec{k}}
\end{pmatrix} \; .
\label{app:rpa:GM^0(q,iomega_n)}
\end{align}
We will not derive the diagram rules here, as this matrix formalism is a direct generalization of a similar
matrix formalism in the well-documented BCS superconductivity case, which in turn is a matrix generalization of the
usual diagram technique~\cite{mattuck1976guide,schrieffer1999theory}. We will just illustrate the rules with a few examples
in the next subsections.

\begin{figure}
\centering
\includegraphics[width=0.44\textwidth]{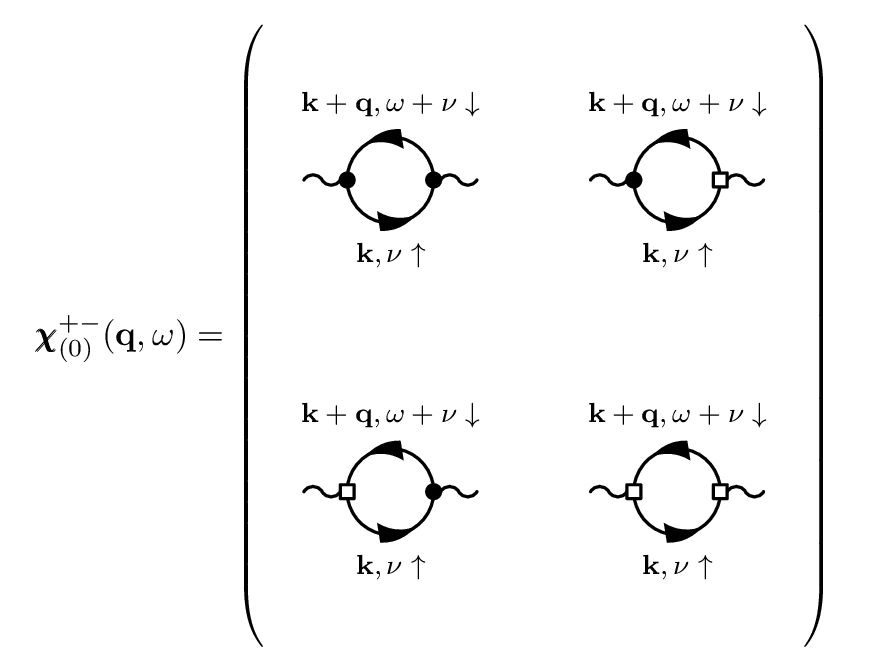}
\caption{Diagrammatic representation of the MFT transverse spin susceptibility matrix, see Eq.~\eqref{eq:susceptibility-Nambu}. The sums over internal momentum $\vec{k}$, frequency $\nu$ have not been written explicitly.}
\label{fig:MF-chi:plusminus}
\end{figure}

\begin{figure}
\centering
\includegraphics[width=0.39\textwidth]{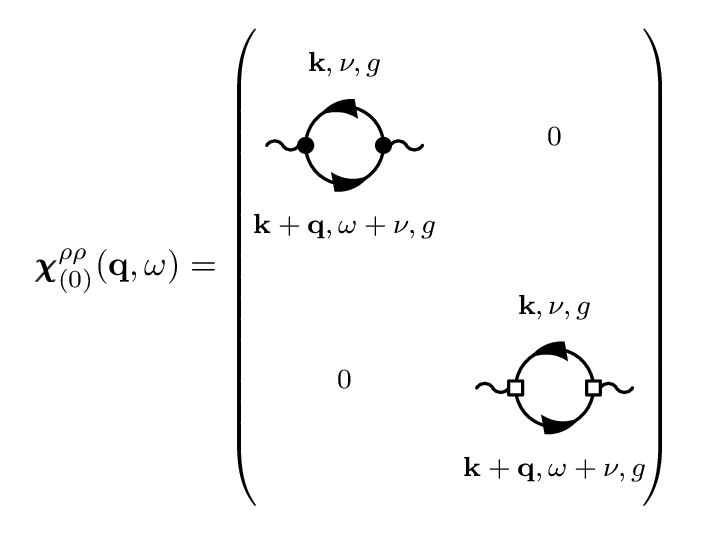}
\caption{Diagrammatic representation of the MFT density-density susceptibility matrix, see Eq.~\eqref{eq:susceptibility-Nambu}. The sums over internal momentum $\vec{k}$, frequency $\nu$, and spin have not been written explicitly.}
\label{fig:MF-chi:rhorho}
\end{figure}

\subsection{Mean-field susceptibilities} \label{App:MF-sus}

The MFT susceptibilities correspond to computing Eq.~\eqref{eq:susceptibility} in the matrix form
Eq.~\eqref{eq:susceptibility-Nambu}  without any interaction vertices
$\mathcal{V}_{\vec{0}}$ and $\mathcal{V}_{\vec{Q}}$. As an example, we calculate the first diagram of the diagonal element of the
susceptibility matrix shown in Fig.~\ref{fig:MF-chi:plusminus},
\begin{align}
\notag
\chi^{+-}_{(0)}(\vec{q}, \vec{q}; t) =& \;
 \frac{2 i}{2 N_s^2}
\\
\times \sum\limits_{\vec{k},\vec{k}'}^{\text{RBZ}}
  \Big\langle \mathcal{T}  &\vcre{\Psi}{ \vec{k}+\vec{q},\uparrow}(t) \vdes{\Psi}{ \vec{k},\downarrow}(t) \:
  \vcre{\Psi}{ \vec{k}'-\vec{q},\downarrow}(0)\vdes{\Psi}{ \vec{k}',\uparrow}(0) \Big\rangle_c \nonumber\\
  =& \; - \frac{i}{ N_s^2} \sum\limits_{\vec{k}}^{\text{RBZ}} \pmb{\text{Tr}} \:
  i \pmb{G}^0_{\downarrow}(\vec{k},t)  i \pmb{G}^0_{\uparrow}(\vec{k}+\vec{q},-t) \; .
\label{eq:app:a4}
\end{align}
The two momenta in the left-hand side of Eq.~\eqref{eq:app:a4} are a consequence of the translation invariance with periodic boundary
conditions that we have imposed on the system (Sec.~\ref{sec:intensity}) and
 $ \pmb{\text{Tr}}$ denotes the trace of the matrix. The overall minus sign comes
from anticommuting the Nambu spinors, and is an example of the diagram rule for a fermion loop leading to a $(-1)$ factor.
In frequency space, this becomes
\begin{equation}
\chi_{(0)}^{+-} (\vec{q},\vec{q};\omega) =  \frac{ i}{ N_s^2} \sum\limits_{\vec{k}}^{\text{RBZ}} \int \frac{d\nu}{2 \pi}
\pmb{\text{Tr}} \:  \pmb{G}^0_{\downarrow}(\vec{k},\nu)   \pmb{G}^0_{\uparrow}(\vec{k}+\vec{q},\omega+\nu) \;.
\end{equation}
Substituting  Eq.~\ref{app:rpa:GM^0(q,iomega_n)} into this and evaluating the frequency integral then leads to Eq.~\eqref{eq:chi:+-:0}. Given that the MFT longitudinal spin susceptibility is equal to the density one [Eq.~\eqref{eq:chi:zz:0}] the only nonzero MFT susceptibilities are those shown in
Figs.~\ref{fig:MF-chi:rhorho} and \ref{fig:MF-chi:plusminus}; in particular,
\begin{equation} \label{eq:chi00:Q}
\chi^{\rho\rho}_{(0)}(\vec{q}, \vec{q}+\vec{Q}; \omega) = \chi^{zz}_{(0)}(\vec{q}, \vec{q}+\vec{Q}; \omega)=0 \;.
\end{equation}
This leads to a simple diagonal structure for the RPA susceptibilities for these quantities, as we shall see next.
However, the matrix structure of $\pmb{\chi}^{+-}_{(0)}$ does lead to a full matrix equation for its  RPA susceptibility.

\begin{figure}
\centering
\includegraphics[width=0.46\textwidth]{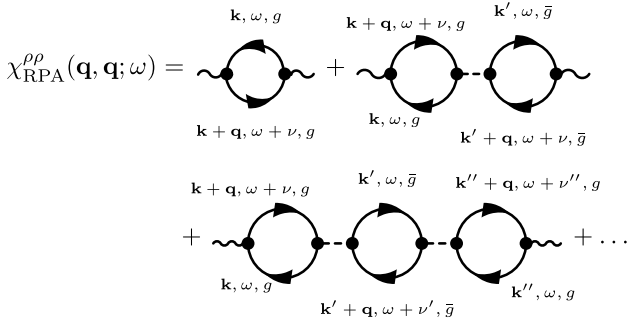}
\caption{Diagrammatic representation of the RPA series for the density susceptibility. The first order term is the MFT bubble.
The sums over internal momenta $\vec{k}, \vec{k}', \vec{k}''$, and frequencies $\nu, \nu',\nu''$ have not been written explicitly.
$\bar g$ stands for the opposite spin of $g$.}
\label{fig:dyson-series}
\end{figure}

\begin{figure}
\centering
\includegraphics[width=0.46\textwidth]{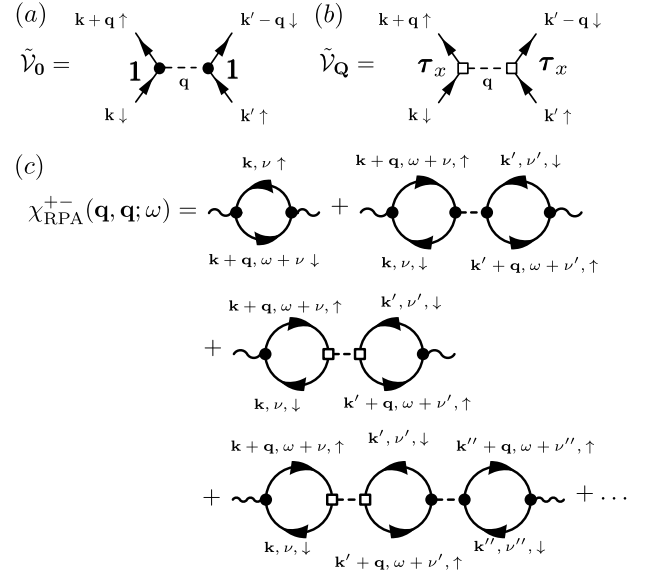}
\caption{Diagrammatic representation of the alternative form of the interaction vertices suited to computing the transverse spin susceptibility  for (a) Eq.~\eqref{eq:Vtilde:0}; (b) Eq.~\eqref{eq:Vtilde:Q}. In
(c), we show the  RPA series for the transverse spin susceptibility. The sums over internal momenta
$\vec{k}, \vec{k}', \vec{k}''$, and frequencies $\nu, \nu',\nu''$ have not been written explicitly.}
\label{fig:dyson-series-splusminus}
\end{figure}

\subsection{RPA susceptibilities} \label{App:RPA-sus}

For the RPA  susceptibilities, the interaction vertices need to be inserted $n$ times for the $n$th order diagram.
We first look at the first order diagrams
 depicted in  the second diagram of Fig.~\ref{fig:dyson-series}.
For example, for the density susceptibility, the first order correction $\chi^{\rho\rho}_{(1)}$ reads
\begin{widetext}
\begin{align} \label{eq:chi_1-firstline}
\chi^{\rho\rho}_{(1)}(\vec{q},\vec{q};t) = & \;  2  \left(\frac{i}{2 N_s^2}\right) \left(\frac{-iU}{ N_s^2}\right)
\sum\limits_{\vec{k},\vec{k}'}^{\text{RBZ}}  \int_{-\infty}^{\infty} dt' \: \left[ \pmb{\text{Tr}} \:
  \pmb{G}^0_{\uparrow}(\vec{k},t-t')   \pmb{G}^0_{\uparrow}(\vec{k}+\vec{q},t'-t) \;
 \pmb{\text{Tr}} \:  \pmb{G}^0_{\downarrow}(\vec{k}',t')   \pmb{G}^0_{\downarrow}(\vec{k}'+\vec{q},-t')  \right. \nonumber\\
   &  \left. \qquad \qquad \qquad \qquad \qquad \qquad  \qquad
   +  \pmb{\text{Tr}} \: \pmb{G}^0_{\uparrow}(\vec{k},t-t') {\pmb \tau}_x  \pmb{G}^0_{\uparrow}(\vec{k}+\vec{q},t'-t) \;
 \pmb{\text{Tr}} \: \pmb{G}^0_{\downarrow}(\vec{k}',t')   \pmb{G}^0_{\downarrow}(\vec{k}'+\vec{q},-t') {\pmb \tau}_x
    \right] \\
  = & \: - U   \int_{-\infty}^{\infty} dt' \:
  \chi^{\rho\rho}_{(0)}(\vec{q},\vec{q};t-t') \chi^{\rho\rho}_{(0)}(\vec{q},\vec{q};t')  \; . \label{eq:chi_1-secondline}
\end{align}
\end{widetext}
In Eq.~\eqref{eq:chi_1-firstline}, the extra factor of $2$ comes from the spin sum in
Fig.~\ref{fig:dyson-series}, and
using the fact that each  bubble of a specific spin is spin independent. Importantly, the second line of
Eq.~\eqref{eq:chi_1-firstline} does not contribute because this is proportional to
$\int dt' \chi^{\rho\rho}_{(0)}(\vec{q}, \vec{q}+\vec{Q}; t-t') \chi^{\rho\rho}_{(0)}(\vec{q}, \vec{q}+\vec{Q}; t')$, which is zero by Eq.~\eqref{eq:chi00:Q}. Hence, $\pmb{\chi}^{\rho\rho}_{(0)}$ is a diagonal matrix in Nambu space: effectively, there is no matrix
structure needed.

This example can be generalized to higher order contributions and we find in the frequency domain, in the full matrix form,
\begin{align}
\notag
\pmb{\chi}^{\rho\rho}_{\text{RPA}}(\vec{q},\omega)=& \:\pmb{\chi}^{\rho\rho}_{(0)}(\vec{q},\omega)-\pmb{\chi}^{\rho\rho}_{(0)}(\vec{q},\omega)U\pmb{\chi}^{\rho\rho}_{(0)}(\vec{q},\omega)
\\\notag
&+\pmb{\chi}^{\rho\rho}_{(0)}(\vec{q},\omega)U\pmb{\chi}^{\rho\rho}_{(0)}(\vec{q},\omega)U\pmb{\chi}^{\rho\rho}_{(0)}(\vec{q},\omega)+...
\\=&  \: \pmb{\chi}^{\rho\rho}_{(0)}(\vec{q},\omega)\left[1+U\pmb{\chi}^{\rho\rho}_{(0)}(\vec{q},\omega)\right]^{-1} .
\label{eq:dyson-series-rhorho}
\end{align}

For $\op{S}^z$, we sum the same set of  diagrams in Fig.~\ref{fig:dyson-series},
except that each of the spin density operator $\op{S}^z$ has a spin-dependence [Eq.~\eqref{eq:spinSzop}],
which leads to a $(-1)$ extra factor. Thus, we obtain
\begin{equation}
%\label{eq:chi:zz:RPA}
\pmb{\chi}^{zz}_{\text{RPA}}(\vec{q},\omega)=\pmb{\chi}^{zz}_{(0)}(\vec{q},\omega)\left[1-U\pmb{\chi}^{zz}_{(0)}(\vec{q},\omega)\right]^{-1}.
\end{equation}

For the transverse spin susceptibility,  the computation can be simplified in the following way:
instead of writing the interaction term as
$\sim U \hat{\rho}_{\uparrow} \hat{\rho}_{\downarrow}$ as for Eq.~\eqref{eq:RPA-H}, we can instead write it as $\sim  -U\hat{S}^+ \hat{S}^- $. More precisely, we write
\begin{align}
\label{eq:Vtilde:0}
\tilde{\mathcal{V}}_{\vec{0}} = & -\frac{U}{N_s^2} \sum\limits_{\vec{q},\vec{k},\vec{k}'}^{\text{RBZ}}
\vcre{\Psi}{ \vec{k}+\vec{q},\uparrow}\vdes{\Psi}{ \vec{k},\downarrow} \:
  \vcre{\Psi}{ \vec{k}'-\vec{q},\downarrow}\vdes{\Psi}{ \vec{k}',\uparrow} \:\;,  \\
\label{eq:Vtilde:Q}
\tilde{\mathcal{V}}_{\vec{Q}} = & -\frac{U}{N_s^2} \sum\limits_{\vec{q},\vec{k},\vec{k}'}^{\text{RBZ}}
 \vcre{\Psi}{ \vec{k}+\vec{q},\uparrow} {\pmb \tau}_x \vdes{\Psi}{ \vec{k},\downarrow} \:
 \vcre{\Psi}{ \vec{k}'-\vec{q},\downarrow}{\pmb \tau}_x \vdes{\Psi}{ \vec{k}',\uparrow}
 \; .
\end{align}
The new forms of the interaction vertices
are depicted in Fig.~\ref{fig:dyson-series-splusminus}(a), (b). Using this new interaction form,
the RPA series of diagrams, depicted in Fig.~\ref{fig:dyson-series-splusminus}(c),  have the same structure as for
the longitudinal spin or density susceptibility cases.
In particular, the first order term for the transverse spin susceptibility
(second and third diagrams of Fig.~\ref{fig:dyson-series-splusminus}(c))
becomes
\begin{widetext}
\begin{align} \label{eq:chi_+-_1-firstline}
\chi^{+-}_{(1)}(\vec{q},\vec{q};t) = & \;  2  \left(\frac{i}{2 N_s^2}\right) \left(\frac{+iU}{ N_s^2}\right)
\sum\limits_{\vec{k},\vec{k}'}^{\text{RBZ}}  \int_{-\infty}^{\infty} dt' \left[
\pmb{\text{Tr}} \; \pmb{G}^0_{\downarrow}(\vec{k},t-t')  \pmb{G}^0_{\uparrow}(\vec{k}+\vec{q},t'-t) \;
\pmb{\text{Tr}} \; \pmb{G}^0_{\downarrow}(\vec{k}',t')   \pmb{G}^0_{\uparrow}(\vec{k}'+\vec{q},-t') \right. \nonumber\\
   &  \left. \qquad \qquad \qquad \qquad \qquad \qquad  \qquad
   +  \pmb{\text{Tr}} \; {\pmb \tau}_x  \pmb{G}^0_{\downarrow}(\vec{k}',t')  \pmb{G}^0_{\uparrow}(\vec{k}'+\vec{q},-t') \;
 \pmb{\text{Tr}} \; \pmb{G}^0_{\uparrow}(\vec{k}+\vec{q},t'-t)  \pmb{G}^0_{\downarrow}(\vec{k},t-t') {\pmb \tau}_x
    \right]  \\
  = & \:  U   \int_{-\infty}^{\infty} dt' \: \left[
  \chi^{+-}_{(0)}(\vec{q},\vec{q};t-t') \chi^{+-}_{(0)}(\vec{q},\vec{q};t')
  + \chi^{+-}_{(0)}(\vec{q},\vec{q}+\vec{Q};t-t') \chi^{+-}_{(0)}(\vec{q}+\vec{Q},\vec{q};t')
  \right] \; .    \label{eq:chi_+-_1-secondline}
\end{align}
\end{widetext}
Note here that the extra factor of 2 in the first line of Eq.~\eqref{eq:chi_+-_1-firstline}
is due to the definition of $\op{S}^+, \op{S}^-$ in Eq.~\eqref{eq:spinSPlusop}, respectively.

Other components of $\pmb{\chi}^{+-}_{(1)}$ can similarly be evaluated, to give the full matrix equation
for the first order term
\begin{align}
\pmb{\chi}^{+-}_{(1)}(\vec{q}, t ) =
U \int_{-\infty}^{\infty} dt' \;
\pmb{\chi}^{+-}_{(0)} (\vec{q}, t - t' ) \; \pmb{\chi}^{+-}_{(0)} (\vec{q}, t' ) \: .
\end{align}
Now it can be seen that the higher order terms  will  have the same bubble series geometry
{\it in a full matrix equation form}. In the frequency domain the summation of the series yields
\begin{align}
\pmb{\chi}^{+-}_{\text{RPA}} ({\vec{q}}, \omega ) =
\pmb{\chi}^{+-}_{(0)} ({\vec{q}}, \omega ) \left[
\pmb{1} - U \pmb{\chi}^{+-}_{(0)} ({\vec{q}}, \omega ) \right]^{-1} \;.
\end{align}
This is the matrix form for Eq.~\eqref{eq:chi:+-:0:RPA:matrix}.

\section{Mean-field finite temperature susceptibilities}  \label{app:mfa_susceptibilities}

In this Appendix we give the full finite temperature MFT result for the correlation function needed
in the structure factor of  Eq.~\eqref{eq:static_structure_factor} that goes into the inelastic spectrum Eq.~\eqref{eq:intensity-inelastic-general}.
Instead of using the imaginary time Matsubara formalism, we evaluate to the correlation functions directly in the MFT from the Hamiltonian \eqref{eq:final-MF-H} as follows: First, for the
connected correlation function in Eq.~\eqref{eq:static_structure_factor},  the
original fermions belonging to the full Brillouin Zone are transformed into the two-band fermions of the
RBZ via  the Bogoliubov transformation  Eq.~\eqref{eq:Bog-Schr}. Next, the resulting correlation function
can be simplified because of the simple quadratic form of the MFT Hamiltonian of
Eq.~\eqref{eq:final-MF-H}: any given term
$\ev{\op{c}^\dagger_{\alpha_4 \vec{k}+\vec{q}, g_4}\op{c}_{\alpha_3 \vec{k}, g_3}\op{c}^\dagger_{\alpha_2 \vec{k}'-\vec{q}', g_2}\op{c}_{\alpha_1 \vec{k}', g_1}}_c$
will be zero unless the effective band index ($\alpha_i, \; i=1,\ldots,4$), the  spin index ($g_i$), and the
momenta all match up pairwise. Now, we have
$\ev{ \op{c}^\dagger_{\alpha \vec{k}, g}\op{c}_{\alpha \vec{k}, g}}= n_{\alpha\vec{k}g} $,
where $n_{1\vec{k}g} = f(E_{\vec{k}g}) $ and  $n_{2\vec{k}g} = 1 - f(E_{\vec{k}g}) $,
and $f(E) = [\exp(E/k_B T)+1]^{-1}$ denotes the Fermi-Dirac distribution.
Hence, collecting all the factors, we get the MFT static structure factor [see Eq.~\eqref{eq:static_structure_factor}]
\begin{widetext}
\begin{equation}
\label{eq:corr:inelastic}
            S^{g_3g_4}_{g_2g_1}(\Delta\vec{k})
 =  \delta_{g_4,g_1}\delta_{g_2,g_3}
\frac{1}{N_s^4}
\sum\limits_{\alpha,\beta=1,2}\sum\limits_{\vec{q}\vec{q}'}^{RBZ}
g_{\alpha,\beta}^{g_4g_2}
(\bar{\Delta \vec{k}},\vec{q},\vec{q}') n_{\alpha\vec{q}g_4}\left(1 -n_{\beta\vec{q}'g_2}\right) \; ,
\end{equation}
where
\begin{align}
\notag
g_{\alpha,\beta}^{g_4g_2}(\Delta \vec{k},\vec{q},\vec{q}')=
         &\frac12\mathfrak{u}^*_{\bar{\Delta\vec{k}}-\vec{q}}\mathfrak{u}_{\bar{\Delta\vec{k}}-\vec{q}'}
         \left[
               1+ \frac{\Delta_{g_4}\Delta_{g_2}+\epsilon_{\vec{q}}\epsilon_{\vec{q}'}}{(-1)^{\alpha+\beta}E_{\vec{q}}E_{\vec{q}'}}
         \right]
 %\\\notag&
+   \mathfrak{u}^*_{\bar{\Delta\vec{k}}-\vec{q}+\vec{Q}}\mathfrak{u}_{\bar{\Delta\vec{k}}-\vec{q}'}
         \left[
           \frac{\Delta_{g_4}}{(-1)^{\alpha+1}E_{\vec{q}}}
           +\frac{\Delta_{g_2}}{(-1)^{\beta+1}E_{\vec{q}'}}
         \right]
\\
&+
         \frac12\mathfrak{u}^*_{\bar{\Delta\vec{k}}-\vec{q}+\vec{Q}}\mathfrak{u}_{\bar{\Delta\vec{k}}-\vec{q}'+\vec{Q}}
         \left[
               1+\frac{\Delta_{g_4}\Delta_{g_2}-\epsilon_{\vec{q}}\epsilon_{\vec{q}'}}{(-1)^{\alpha+\beta}E_{\vec{q}}E_{\vec{q}'}}
         \right]\,,
\label{eq:g_function}
\end{align}
\end{widetext}
and $\alpha$ and $\beta$ are the effective band indices.
The product of the occupation numbers $ n_{\alpha\vec{q}g_4}\left(1 -n_{\beta\vec{q}'g_2}\right) $ indicates the Pauli blocking where the scattering from the initial state $n_{\alpha\vec{q}g_4}$ to an already occupied final state $n_{\beta\vec{q}'g_2}$ is forbidden. At $T=0$, only the scattering from the filled lower band to the empty upper band is allowed, while at the finite temperature also other processes are possible.

%%%%%%%%
%%%%%%%%%%%%%%%% 	BIBLIOGRAPHY
%%%%%%%%

%\addcontentsline{toc}{section}{References}
%\bibliographystyle{revtex}
%\bibliography{./bibliography}
%merlin.mbs apsrev4-1.bst 2010-07-25 4.21a (PWD, AO, DPC) hacked
%Control: key (0)
%Control: author (8) initials jnrlst
%Control: editor formatted (1) identically to author
%Control: production of article title (-1) disabled
%Control: page (0) single
%Control: year (1) truncated
%Control: production of eprint (0) enabled
%

\end{document}